\documentclass[useAMS,usenatbib,babel,superscriptaddress]{mnras}

\usepackage[english,english]{babel}
\usepackage{amsmath}
\usepackage{amssymb,amsfonts,textcomp}
\usepackage{empheq}
\usepackage{array}
\usepackage{supertabular}
\usepackage{hhline}
\usepackage{soul}
\usepackage[usenames]{color}
\usepackage{algpseudocode}
\usepackage{times}
\usepackage{siunitx}
\DeclareSIUnit\hred{\emph{h}}
\usepackage{physics}
\usepackage{wasysym}
\usepackage{xspace}
\usepackage{xfrac}
\usepackage{hyperref}
\usepackage{cleveref}
\usepackage{enumerate}
\usepackage{relsize}
\crefname{section}{Section}{Sections}                 %
\crefname{equation}{equation}{equations}              %
\crefname{figure}{Fig.}{Figs.}                        %
\crefname{table}{Table}{Tables}                       %
\crefname{appendix}{Appendix}{Appendix}               %
\crefrangelabelformat{equation}{(#3#1#4)--(#5#2#6)}

\Crefname{section}{Section}{Sections}                 %
\Crefname{equation}{Equation}{Equations}              %
\Crefname{figure}{Figure}{Figures}                    %
\crefname{table}{Table}{Tables}                       %
\Crefname{appendix}{Appendix}{Appendix}               %

\algrenewcommand\textproc{}%

\def\gtrsim{\lower.5ex\hbox{$\; \buildrel > \over \sim \;$}}
\usepackage{graphicx}

\newcommand{\pd}{\partial}
\newcommand{\rr}{\vb*{r}}
\newcommand{\vvec}[1]{\vb*{#1}}
\newcommand{\dirac}{\delta_{\rm D}}
\newcommand{\heaviside}{\Theta_{\rm H}}
\newcommand{\disperse}{\mbox{{\sc \small DisPerSE}}}

\newcommand{\ie}{\emph{i.e.}\xspace}
\newcommand{\wrt}{with respect to\xspace}
\newcommand{\eg}{e.g.\xspace}
\newcommand{\deltac}{\delta_{\mathrm{c}}}
\newcommand{\me}{\mathrm{me}}
\renewcommand{\cp}{\mathrm{cp}}
\newcommand{\ce}{\mathrm{ce}}
\renewcommand{\TH}{{\mathrm{TH}}}

\def \prd {{\it Phys. Rev. D}}

\def \apj {{\it ApJ}}
\def \apjl {{\it ApJ Let.}}
\def \apjs {{\it ApJ Sup.}}

\def \aap {{\it A\&A}}

\def \mnras {{\it MNRAS}}
\def \nat {{\it Nature}}

\def \jcap {{\it JCAP} }

\definecolor{grey}{rgb}{0.75,0.75,0.75}
\definecolor{Orange}{rgb}{1.0,0.5,0.15}
\definecolor{brown}{rgb}{0.7,0.25,0.0}
\definecolor{pink}{rgb}{1.0,0.5,0.5}
\definecolor{darkerred}{rgb}{0.8,0,0}
\definecolor{darkerblue}{rgb}{0,0,0.8}
\definecolor{Blue}{rgb}{0,0.08,0.65}
\definecolor{Red}{rgb}{0.65,0.08,0.05}
\definecolor{Green}{rgb}{0.15,0.45,0.25}

\begin{document}

\author[C. Cadiou  et al.]{
\parbox[t]{\textwidth}{C.~Cadiou$^{1,2}$\thanks{E-mail: c.cadiou@ucl.ac.uk}, C.~Pichon$^{2,3,4}$, S.~Codis$^{2,4}$,  M.~Musso$^{5}$,  D.~Pogosyan$^{3,6}$,\\
Y. Dubois$^{2}$, J.-F.~Cardoso$^{2}$, S.~Prunet$^{7}$
  }
\vspace{1em}\\
$^{1}$ Department of Physics and Astronomy, University College London, London WC1E 6BT, UK\\
$^{2}$ CNRS and Sorbonne Universit\'e, UMR 7095, Institut d'Astrophysique de Paris, 98 bis Boulevard Arago, 75014 Paris, France\\
$^{3}$ Korea Institute of Advanced Studies (KIAS) 85 Hoegiro, Dongdaemun-gu, Seoul, 02455, Republic of Korea\\
$^{4}$ IPHT, DRF-INP, UMR 3680, CEA, Orme des Merisiers Bat 774, 91191 Gif-sur-Yvette, France\\
$^{5}$ East African Institute for Fundamental Research (ICTP-EAIFR), KIST2 Building, Nyarugenge Campus, University of Rwanda, Kigali, Rwanda\\
$^{6}$ Department of Physics, University of Alberta, 11322-89 Avenue, Edmonton, Alberta, T6G 2G7, Canada\\
$^{7}$ Canada-France-Hawaii Telescope, 65-1238 Mamalahoa Highway, Kamuela, HI 96743, USA
}

\title[Critical event theory in a multi-scale landscape.]{
When do cosmic peaks, filaments or walls merge? \\
A theory of  critical events in a multi-scale landscape.
}

\maketitle

\begin{abstract}
The merging rate of cosmic structures is computed, relying on the \emph{Ansatz} that they can be predicted in the initial linear density field from the coalescence of critical points with increasing smoothing scale, used here as a proxy for cosmic time.
Beyond the mergers of peaks with saddle points (a proxy for halo mergers), we consider the coalescence and nucleation of all sets of critical points,  including wall-saddle to filament-saddle and wall-saddle to minima
(a proxy for filament and void mergers respectively), as they impact the geometry of galactic infall, and in particular filament disconnection.
Analytical predictions of the one-point statistics are validated against multiscale measurements in 2D and 3D realisations of Gaussian random fields (the corresponding code being  available upon request) and compared qualitatively to cosmological $N$-body simulations at early times ($z\geq 10$) and large scales ($\geq \SI{5}{Mpc\per\hred}$).
The rate of filament coalescence is compared to the merger rate of haloes and
the two-point clustering of these events is computed, along with their cross-correlations with critical points. %
These correlations are qualitatively consistent with the preservation of the connectivity of dark matter haloes, and the impact of the large scale structures on assembly bias.
The destruction rate of haloes and voids as a function of mass and redshift is quantified down to $z=0$ for a $\Lambda$CDM cosmology.
The one-point statistics in higher dimensions are also presented, together with consistency relations between critical point  and critical event counts. 
\end{abstract}


\section{Introduction}

The large-scale structures of our observable Universe are routinely observed through the distribution of galaxies, neutral gas or dark matter.
As such, galaxies and their haloes are both probes of the large-scale density field (from the point of view of cosmology), and the subject of interest (from the point of view of galaxy formation).
It is now accepted that the large-scale structures are key to understand galaxy formation, for example by driving angular momentum acquisition through cosmic cold streams \citep{dekel_galaxy_2006,agertz_disc_2009,pichonetal11,danovichetal11,Dubois2011} and by galaxy mergers, which efficiently disrupt galaxies into ellipticals \citep[\eg][]{toomre_GalacticBridgesTails_1972,naab_StatisticalPropertiesCollisionless_2003,2007A&A...476.1179B}. This scale coupling is also relevant to cosmology, as it influences for example lensing observations through spin alignments \citep{Cri++01,codis2015}.
In the era of precision cosmology, any attempt to infer cosmological parameters from observations of galaxies and haloes should therefore take into account the influence of the surrounding large-scale structures.
Since the details of the buildup of cosmic structure and galaxies are encoded in the initial matter density field and are coupled, one could, in principle, predict their joint evolution from the initial conditions.
Considering that an initial Gaussian random field with small density perturbations
leads to the formation of both cosmic structures and galaxies, some descriptive statistics of this field can be used to jointly predict the final fate of galaxies, halos and the cosmic web. More specifically -- and this will be the topic of this paper -- we should be able to identify special sets of points via a multi-scale analysis of the initial conditions (as a means of compressing the relevant information content of this field) and use them to predict the fate of cosmic structures.

The topology of the initial density field at a given smoothing scale is encoded in the positions and heights of all its critical points, namely extrema (maxima and minima) and saddle points (filament-type and wall-type saddle points).
For instance, peaks in the initial conditions will later form the nodes of the cosmic web \citep{BBKS}, while bridges in between, in the middle of which is  found a saddle-point,  will subsequently collapse due to the tidal anisotropies to form filaments \citep{bkp96,Rossi:2013fk}. Conversely, voids will develop from the initial minima \citep{sheth_hierarchy_2004}, and walls around wall-type saddle points. 
Beyond the strong focus on extrema, 
\cite{pogo09}  developed a theoretical framework, the skeleton,  to understand the structure of the cosmic web as a whole (walls and filaments) in terms of gradient lines joining peaks and voids through saddle points. 
In this context, computational geometry  allows us to quantify the strength of  topological 
 pairing between critical points \citep{sousbie10,vandeWeygaert2013} through persistence \citep{Edelsbrunner2000, Pranav2016} which measures their relative heights, and defines a  scale-free hierarchy amongst filaments, walls and voids of the cosmic web.   

Focusing now on haloes,
within the paradigm of the spherical gravitational collapse, 
one can draw a  relationship between the time of collapse of the initial overdense patch and the scale at which it must be smoothed so as to pass a theoretically given overdensity threshold \citep{press_formation_1974}, and therefore map features of the initial linear density field to later-time non-linear structures.
In practice, not only does the patch need to pass a given density threshold as a function of smoothing but additional constraints must be added, notably to avoid double counting (the so-called cloud-in-cloud problem). This requires enforcing a first crossing condition to ensure that no larger scales than the one considered has collapsed, which makes the core of the excursion set approach \citep{PH90,Bondetal1991, Jedamzik95,MaggioreRiotto2009,MussoSheth2012}. Better agreement with actual collapsed haloes can be achieved with a modified stochastic threshold that incorporates the effects of tidal forces on top of spherical collapse \citep{bond1996,SMT01}.
When studying halo statistics in cosmological models with 
cold dark matter (CDM, \citet{1984Natur.311..517B})
that exhibit the hierarchical clustering, 
it is also of  interest to investigate the substructures (hence smaller scales) within a given patch so as to study its assembly history. \cite{LaceyCole1993} showed that the properties of the excursion set trajectories carry information on the matter accretion history of the forming haloes, allowing us to split this accretion into a smooth component on the one hand and mergers on the other hand.
In this sense, the fate of a given region is encoded in its initial conditions and is captured by the multi-scale properties of the corresponding Gaussian random field.

Going one step further, \cite{Manrique1995,Manrique1996} brought together the virtues of the two approaches (peaks and excursion sets) in the so-called confluent system formalism, where excursion set trajectories are not randomly located in space and concentric, but insist on peaks and follow their position as the smoothing scale changes. This approach was later perfected and made more analytically manageable \citep{Paranjape2012,ESP2013}, including the effect of tidal shear \citep{2016arXiv161103619C}.

The very notion of special points in the position-smoothing space is hence crucial in the context of modelling the evolution of haloes but also of the cosmic web as a whole.
The drift of critical points with smoothing defines the so-called skeleton tree \citep{hanami} which
captures the variation of this topology with
smoothing scale, hence time.
One can identify special scales at which two such points coalesce, hence producing merger \emph{events}, as they are located in time as well as in space, of different types, corresponding to mergers of haloes, filaments, walls or voids.
In that paper the focus was on the coalescence of filament saddles with maxima, which the author named sloping saddles  (as they are vanishing saddle points on the slope of peaks), identified as proxies for merging events.  These are known to play a significant role in triggering AGN feedback,
which impacts gas inflow and therefore galactic morphology \citep{Dubois2016}.
Coalescence of other critical points also
impact the geometry of the cosmic web (in particular the filaments)
which defines preferred directions along which galaxies are fed cold gas and
acquire their spin. They also impact wall disappearance, hence void statistics \citep{1993ApJ...410..458D}.

The focus should now therefore be on special points in the 3+1D\footnote{The field smoothed at all scales has a 3D spatial component and a 1D smoothing scale component, hence the short-hand notation ``3+1D''.}
position-smoothing space, where these  paired critical points merge, \ie when the persistence level tends to zero as a function of smoothing.
Using the  above-mentioned mapping between scale and cosmic time provided by the spherical collapse model, we will rely on the \emph{Ansatz} that these mergers in the initial linear density field can be matched
to structurally important special moments that  modify the topology
of the evolved non-linear density field, while a more careful evaluation of the accuracy of this assumption will be carried out in future works.
For instance, when two haloes merge,
the topology of the excursion set of the density field (\ie the region above a given threshold) is changed, because it decrements the number of components above the threshold.

Mapping the geometry of the Gaussian random field onto the knowledge of only
these singular events  is a very efficient  and useful compression
of the information encoded in the field. It is efficient because  it compresses the information about a 3D random field
into a finite set of points in 3+1D.
It is useful because i) these points bear significance in terms of cosmology or galaxy formation,  and ii) we will be able to 
characterise the corresponding point process in terms of the properties of the underlying initial Gaussian
field -- therefore statistically in terms of the underlying power spectrum.

Hence, in this paper we will present a ``critical event theory'' to  capture not only the evolution of the halo hosting the galaxy via its merger tree, but also the evolution of the spatial structures that fed it, which are known to affect the acquisition of secondary galactic properties (such as their angular momentum) and may thus contribute to assembly bias.
We will include the coalescence of minima with wall-type saddles
and wall-type saddles with filament-type saddles corresponding respectively to the
merging of two walls (with a void disappearing in between) and two filaments (with disappearance of a wall).  We will finally study the  clustering properties of all  these critical events in the multi-scale landscape, as a means to relate their sequence and geometry to the events relevant to the evolution of galaxies.

 Our astrophysical motivations are the following. 
 Study the generalised history of accretion: what kind of mergers happen when, and where?
Quantify the conditional rate of filament and wall  disappearance in conjunction to that of  an existing larger scale critical point.
Understand the  origin of void disappearance and its usefulness as a cosmic probe for dark energy.
Connect the multiscale landscape of initial conditions to the  properties of dark matter haloes. %
Study how the anisotropic  large-scale modes bias its assembly history.
Beyond astrophysics,  we aim to quantify the statistical properties 
of zero persistence points in a multiscale landscape and to provide tools to identify 
such points.

This paper proceeds as follows.
\Cref{sec:theory} forecasts critical events through the coalescence of critical points in the multi-scale density landscape.
We present a formal definition of critical events and re-derive the condition for a critical event in an arbitrary frame
We derive their one-point statistics in 2 and 3 dimensions (in the main text) and higher dimensions (in the appendix) as a function of their height and kind.
We also propose an extension of the theory for mildly non-Gaussian fields.
\Cref{sec:theory2pt} presents the two-point statistics of critical events in simplified setups.
We present numerical integrations of the auto- and cross-correlation functions of different kind of critical events and of their number densities in the presence of a large-scale tide.
We also derive analytically the clustering of peak critical events in their simplest configuration.
\Cref{sec:measurements} compares the predictions of the one-point statistics to realisations of Gaussian random fields for validation for different linear matter power spectra.
We also present measurements of the auto- and cross-correlation functions in the general case.
\Cref{sec:discussion} discusses possible applications in astrophysics and beyond.
We present predictions of the destruction rate of halos and voids as a function of cosmic time, we present consistency relations with the evolution of the cosmic connectivity.
We develop how the framework can be applied to the problem of assembly bias.
We also make qualitative comparisons to $N$-body simulations
\Cref{sec:conclusion} presents our conclusions.
A summary of the notations and conventions used throughout the paper is provided in \cref{tab:notations-definitions}.
  
\Cref{sec:NDevents} presents the counts in arbitrary dimensions and illustrates them in up to 6D.
\Cref{sec:net_mergers} explores the duality between critical points and critical events.
\Cref{sec:dual-interpretation} discusses alternative interpretations of critical events from low to high densities.
\Cref{sec:nucl} describes the local behaviour of critical point lines near their coalescence.
\Cref{sec:generation} presents algorithms to generate Gaussian random fields satisfying a set of given `events' at some scale and position.
\Cref{sec:crit-event-numb-different-def} generalises some results using alternative definitions to relate critical events to mergers in physical space.
\Cref{sec:jointpdf} gives the joint PDF of a Gaussian random field up to the third derivative of the field.
\Cref{sec:detection} explains how the critical events are measured in random field maps and cubes.

\begin{table*}
  \caption{Summary of notations and definitions.}
  \label{tab:notations-definitions}
  \begin{tabular}{lll}
    Notation & Equation & Note \\
    \hline\hline
    \textbf{Definitions} & & \\
    $\delta(R)$ & \eqref{eq:delta-definition} & Density field smoothed over a scale $R$\\
    $\sigma_i^2(R)$ & \eqref{eq:defsigi} & Variance of the $i$-th derivative of the density field\\
    $x,x_{i},x_{ij},x_{ijk}$ & \eqref{eq:scalefr} & Density, density first, second and third derivatives, normalised by their variance\\
    $R_0,R_*,\tilde{R}$ & \eqref{eq:defR0} & Typical separation between zero-crossings, critical points and inflection points \\
    $\gamma, \tilde{\gamma}$ & \eqref{eq:gammadef} & Cross-correlation coefficients between the field and its derivatives\\
     $\mathbf{H}, H=\det(\mathbf{H})$ & \eqref{eq:hessian-definition}, \eqref{eq:hessian-det-definition} & Density Hessian matrix and determinant \\
    \hline\hline
    \textbf{Notations} & & \\
    $\mathrm{p,f,w,v}$ & & Peak, filament-saddle, wall-saddle and void (minima) critical points\\
    ${\cal P, F, W}$ & & Peak, filament and wall critical events\\
    \hline\hline
    \textbf{Critical point definitions} & & \\
    $n_\cp$ &  & Total number density \\
    $n_{\cp}^{(j)}$& & Total number density of kind $j\in\{\mathrm{p,f,w,v}\}$\\
    $n_{\cp}^{(j)}(\nu)$& & Number density of kind $j\in\{\mathrm{p,f,w,v}\}$ at height $\nu$\\
    \hline\hline
    \textbf{Critical event definitions} & & \\
    $n_\ce$ & \eqref{eq:ce_definition}, \eqref{eq:eventcount_covariant}& Total number density \\
    $n_{\ce,+},n_{\ce,-}$ & \eqref{eq:nce+-_def} & Total number density of nucleation ($+$) and destruction ($-$) \\
    $n_{\me}$ & \eqref{eq:nme_def} & Total net merger rate (critical event net density) \\
    $n_{\me}^{(j)}$ & \eqref{eq:eventcount} & Total net merger rate of kind $j\in \{\cal P, F, W\}$ \\
    $n_{\me}^{(j)}(\nu)$ & \eqref{eq:nmerger_nu} & Net merger rate of kind $j\in \{\cal P, F, W\}$ at height $\nu$ \\
  \end{tabular}
\end{table*}

\section{Theory: 1-pt statistics of critical events}
\label{sec:theory}

In this paper we consider the overdensity at position $\vvec{r}$, $\delta(\vvec{r})= \rho(\vvec{r})/\bar\rho -1 $, to be a homogeneous and
isotropic Gaussian random field of zero mean and linear power spectrum $P_k$, smoothed on scale $R$.
In this section, we will focus on one-point statistics associated with merger rates of the field
critical points as the smoothing scale increases.
In~\cref{sec:charac-features} we define different quantities used throughout the paper to describe the relevant features of the field.
\Cref{sec:definition-crit-events} presents the number density of critical
events. %
\Cref{sec:ce_by_type} introduces critical events of different types (peak, filament and wall mergers)
and calculates their total and differential densities at given height. %
\Cref{sec:theory2D} sketches the corresponding  theory for projected maps, while \cref{sec:non-gaussianity} presents its extension to non-Gaussian fields.

In this section, limiting assumptions are introduced as late as possible:
all the results are general up to \cref{sec:definition-crit-events}, which introduces the requirement for a Gaussian filter.
In \cref{sec:ce_by_type} and after, and unless stated otherwise, it is assumed that the field is a Gaussian random field smoothed by a Gaussian filter.
We also remind the reader that a summary of the notations and definitions used throughout the paper can be found in \cref{tab:notations-definitions}.

\subsection{Characteristic features of a field}
\label{sec:charac-features}
Let us first introduce the dimensionless quantities for the density field, smoothed over a scale $R$ by a filter function $W$
\begin{equation}
  \delta(\vvec{r},R) = \int \frac{\dd[3]{k}}{(2\pi)^3} \delta(\vvec{k}) W(kR)
  \mathrm{e}^{i\vvec{k}\vdot\vvec{r}}\,.
  \label{eq:delta-definition}
\end{equation}
We will consider the statistics of this field and its derivatives in this paper. For practical purposes, let
us introduce the dimensionless quantities
\begin{equation}
\hskip -0.05cm
x \equiv \frac{\delta}{\sigma_0 },\, x_{k}\equiv \frac{\nabla_k \delta}{ \sigma_1},
\,  x_{kl}\equiv \frac{\nabla_k \nabla_l \delta}{\sigma_2},\,
x_{klm}\equiv \frac{\nabla_k \nabla_l \nabla_m \delta}{\sigma_3 },
\label{eq:scalefr}
\end{equation}
which are normalised by their respective variance 
\begin{equation}
\sigma_i^2(R)\equiv \frac{1}{2\pi^2}\int_0^\infty \dd{k} k^2 P_k(k) k^{2i} W^2(k R)\,,
\label{eq:defsigi}
\end{equation}
so that we have 
$\langle x^2\rangle = \sum_k\langle x_k x_k\rangle = \sum_{k,l}\langle x_{kl}x_{kl}\rangle = \sum_{k,l,m}\langle x_{klm}x_{klm} \rangle = 1$.
Note that here and in the rest of the paper, we have dropped the explicit dependence of the quantities of \cref{eq:scalefr} on the smoothing scale.

Following closely \cite{pogo09}, let us introduce the characteristic scales of the field
\begin{equation}
R_{0} = \frac{\sigma_{0}}{\sigma_{1}}, \quad R_{*} = \frac{\sigma_{1}}
{\sigma_{2}}, \quad \tilde R  = \frac{\sigma_{2}}{\sigma_{3}}\,.
  \label{eq:defR0}
\end{equation}
These scales are ordered as $R_0 \ge R_* \ge \tilde R  $.
The first two have well-known meanings of typical separation between zero-crossing of the field and mean distance between extrema respectively \citep{BBKS}.
The third one, $\tilde R$, is by analogy the typical distance between inflection points.

Let us define a set of spectral parameters that depend on the shape of
the underlying power spectrum.
Out of the three scales introduced above, two dimensionless ratios may be   constructed
that are intrinsic parameters of the theory
\begin{equation}
\displaystyle
\gamma   \equiv \frac{R_{*}}{R_{0}}=\frac{\sigma_1^2}
{\sigma_0\sigma_2} , \quad
{\tilde \gamma}  \equiv \frac{\tilde R}{R_{*}}=\frac{\sigma^2_2}
{\sigma_1 \sigma_3} \,.
\label{eq:gammadef}
\end{equation}
From the geometrical point of view $\gamma$ specifies
how frequently one encounters  a maximum between two zero-crossings
of the  field, while $\tilde \gamma$ describes, on average, how many
inflection points are between two extrema. From  a statistical perspective,
$\gamma$ and $\tilde\gamma$ are cross-correlation coefficients between the field and its
derivatives at the same point
\begin{equation}
\gamma=-\frac{\langle \delta \Delta \delta \rangle}{\sigma_0 \sigma_2} , \quad
{\tilde \gamma} =
-\frac{\langle \grad \delta \cdot \laplacian \grad\delta \rangle}{\sigma_1 \sigma_3}\,.
\quad \label{eq:defgam}
\end{equation}
These scales and scale ratios fully specify the correlations between the field and its derivative (up to third order) at the same point. For power-law power spectra of index $n$, $P_k(k) \propto k^n$, with Gaussian smoothing at the scale $R$ in 3D,
$R_0= R \sqrt{2/(n+3)}$, $R_* = R\sqrt{2/(n+5)}$ and $\tilde R = R\sqrt{2/(n+7)}$ while $\gamma = \sqrt{(n+3)/(n+5)}$ and $\tilde \gamma = \sqrt{(n+5)/(n+7)}$.
See also \cref{sec:spectral-params-d} for their generalisation to any dimension.

\subsubsection{Critical points of the random field}
Critical \emph{points} of the 3D field at fixed smoothing scale
are defined as places where the spatial gradient of the field vanishes: $\grad \delta = 0$. This provides a number of conditions exactly equal to the dimensionality of the space, and thus is in general satisfied only at isolated points.
The type of critical point
is given by the signs of the eigenvalues $\sigma_2\lambda_i$ of the Hessian of the field
\begin{equation}
  \mathrm{\mathbf{H}} \equiv \grad \grad \delta,
  \label{eq:hessian-definition}
\end{equation}
which we will always consider sorted $\lambda_1 \le \lambda_2 \le \lambda_3$.

Local extrema of the field are critical points whose eigenvalues have all the same sign, negative for maxima, and positive for minima. Other critical points are saddles of different types: in 3D filamentary saddles have  $\lambda_1 \le \lambda_2 < 0 < \lambda_3$ and wall-like saddles have $\lambda_1 < 0 < \lambda_2 \le \lambda_3$.
Requiring ever more eigenvalues to be positive, we go from maxima to filamentary saddles to wall saddles and to minima,
each type differing from the neighbours by the sign of one eigenvalue.  Correspondingly, the Hessian determinant
\begin{equation}
  H\equiv\det(\grad \grad \delta)=\sigma_2^3{\lambda_1 \lambda_2 \lambda_3},
  \label{eq:hessian-det-definition}
\end{equation}
changes sign at every step of this progression.

In Euclidian space the average Euler characteristic is zero. This means that the alternating sum of critical points is null $
n_{\cp}^\mathrm{p}-n_{\cp}^\mathrm{f}+n_{\cp}^\mathrm{w}-n_{\cp}^\mathrm{v} = 0$, with $n_{\cp}^\mathrm{p,f,w,v}$ the mean number densities of peaks, filament-type saddle, wall-type saddles and voids respectively. A more formal definition is given in \cref{sec:net_mergers}.
Thus, the density of all the critical points with $H>0$ is equal to the density of ones with $H<0$. For any dimension, this mathematically reads $\sum_{H_i > 0} n_\cp^{(i)} = \sum_{H_i < 0} n_\cp^{(i)}$.

\subsubsection{Critical event definition}
\label{sec:crit-event}

\begin{figure}
  \centering
  \includegraphics[width=\columnwidth]{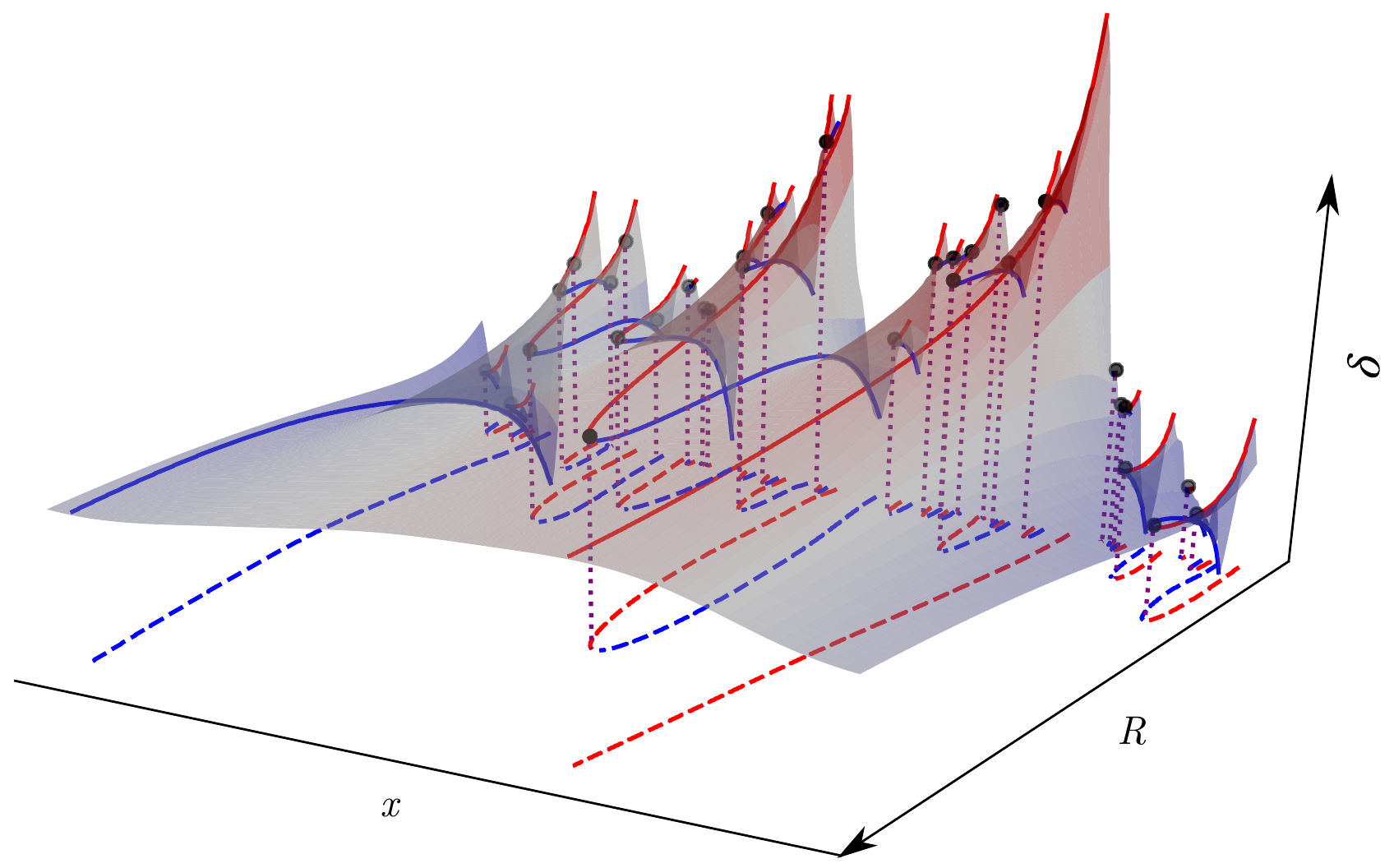}
  \caption{1+1D landscape of a 1D field smoothed at a scale $R$. Solid lines indicate maxima (red) and minima (blue). Smoothing length $R$ is the smallest at the backplane and increases toward the viewer, critical point lines end at critical events (black dots). The critical point lines are projected on the $\delta=-1$ plane (red and blue dashed lines). Vertical purple lines indicate the projection of critical events onto the $\delta=-1$ plane and illustrate that critical events are found at the location where two critical points merge.
  An interactive version can be found in the online supplemental material and \href{http://www2.iap.fr/users/pichon/Critical/}{online}.
 }%
  \label{fig:scheme_critical_events}
\end{figure}

\begin{figure}
  \centering
  \includegraphics[width=\columnwidth]{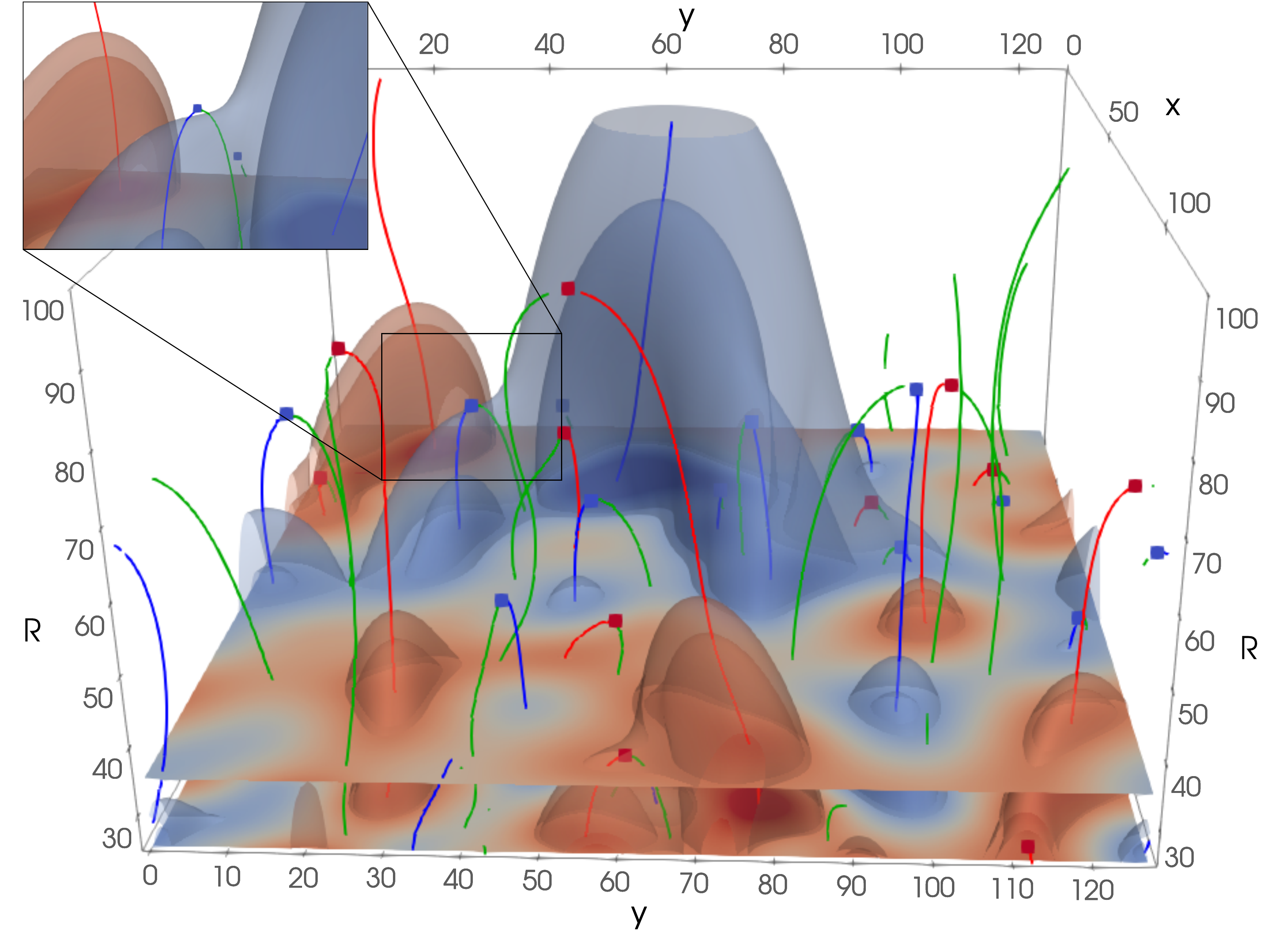}
  \caption{2+1D landscape of a 2D field smoothed at a scale $R$. The density field (blue to red map) is smoothed at increasing (upward) $R$. For each scale, the critical points (red lines: peaks, green lines: saddle points, blue lines: minima) are found. At the tip of each branch a critical event is found (red: peak-saddle critical events, blue: saddle-minima). 
  Isocontours of density in 2+1D are shown as transparent surfaces (blue for negative density and red for positive density).
  An interactive version can be found in the online supplemental material and \href{http://www2.iap.fr/users/pichon/Critical/}{online}.
 Note that the critical points at coordinate $\sim\!(50,20)$ (see inset) or $\sim\!(110,55)$ are indeed clearly  sloping saddles: one of the eigenvalues of the Hessian vanishes
 as the curvature changes. 
  }%
  \label{fig:scheme_critical_events_2D}
\end{figure}

Let us now define critical \emph{events}. These events -- that generalise the notion of sloping saddles in \cite{hanami} -- are defined in the 3+1D position-smoothing space as locations where, besides $\grad \delta=0$,  the Hessian determinant $H$ also vanishes\footnote{We warn against possible confusion
that critical events are not a
generalisation of critical points to the position-smoothing space. The additional condition imposed is not $\partial \delta/\partial R=0$ but $H=0$.}.
Because we  impose these four conditions in a four-dimensional space, the solution is a set of points in position-smoothing space, which will be interpreted as points in space-time, hence the denomination \emph{events}.
These events in 3+1D space correspond to mergers of the trajectories traced by critical \emph{points}  as the smoothing scale $R$ changes. Since in general at each critical event only one eigenvalue of $\mathbf{H}$ vanishes, only the tracks of critical points of neighbouring types can merge.

\Cref{fig:scheme_critical_events} shows the critical events for a 1+1D field. These events are found at the tip of
critical point lines and represent the disappearance of a pair of critical points of neighbouring kind
(\eg a maximum and a minimum in case of one spatial dimension, a maximum and a saddle point for higher dimensions).
At a critical event the topology of the field at fixed $R$ slice is changed by removing a pair of critical
points.  The inverse process where a critical point pair is created and two trajectories emerge from a critical
event\footnote{In this paper, we will always be speaking about mergers or creation as smoothing increases, \ie consider trajectories traced by critical points in the direction of increasing $R$. In general, one could also consider un-smoothing the field as a proxy for 
some control parameter in the context of  bifurcation theory.}
is also possible (for a Gaussian filter, only in more than 1D), although, as we will show further, much less probable.

\begin{figure*}
  \centering
  \includegraphics[width=1.75\columnwidth]{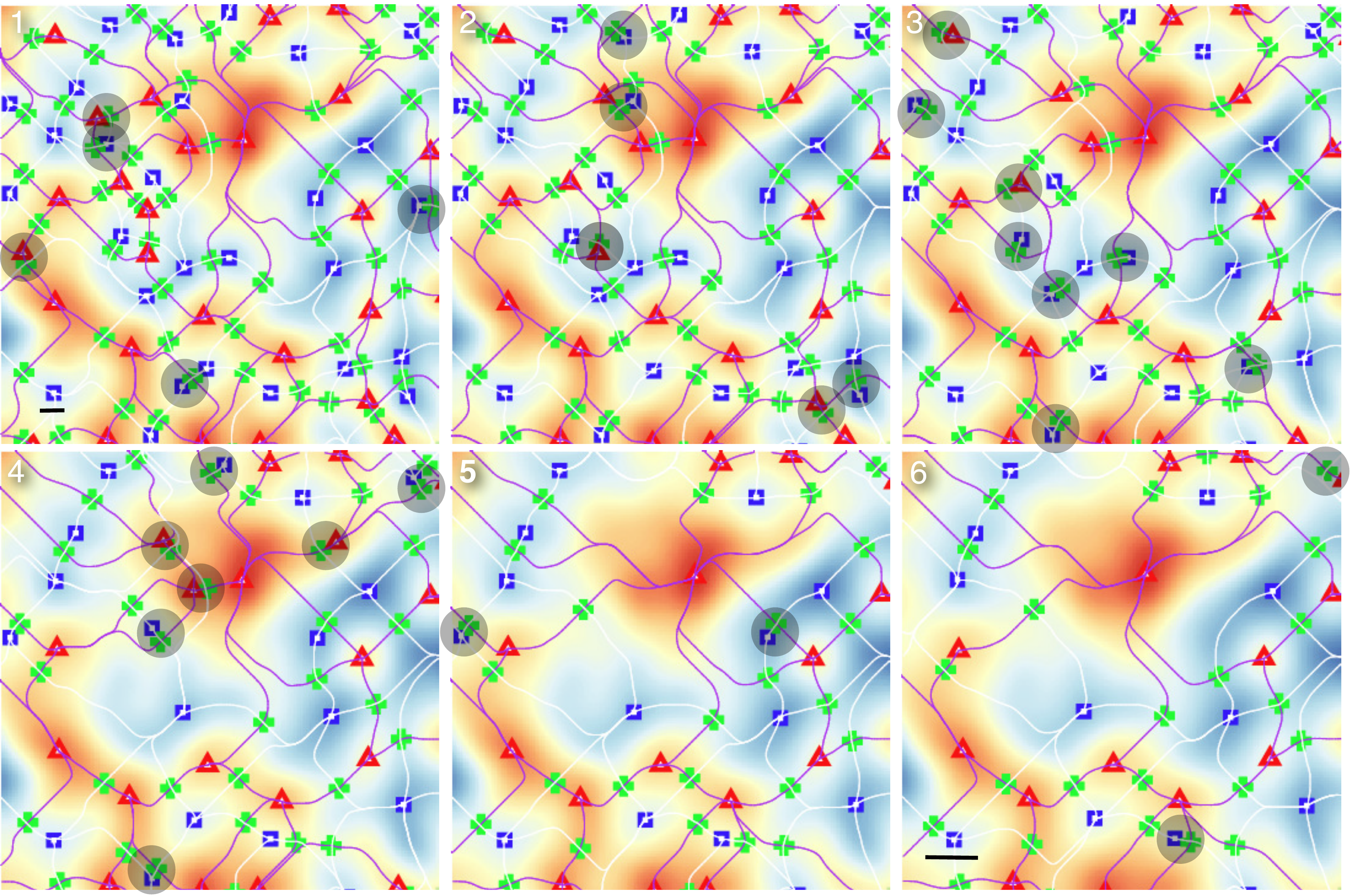}
  \caption{From left to right and top to bottom,
  a smoothing sequence of a Gaussian random field, whose density is colour coded from blue to red as a function of height (analogous to the slices shown in \cref{fig:scheme_critical_events_2D}). The skeleton tracing the ridges \citep{pogo09} is shown in purple,
  while the anti-skeleton tracing the trough is shown in white. The saddles
  shown as green crosses lie at the intersection. The maxima are shown as red triangles
  and the minima as blue squares. As one smooths the field, these critical points
  drift towards each other along the skeletons, until they vanish in (zero persistence) pairs.  The upcoming coalescence are  identified  with grey circles.
  Note that as saddle points vanish, the two corresponding skeletons do too. Note also that the direction of coalescence is typically set by the skeleton's just before coalescence.
  In this two-dimensional example, the ratio of peak+saddle to void+saddle event is one. The black segment  in the bottom left of the first and last image represents the amount of smoothing.
  This paper is concerned with studying the one and two-point statistics of these grey circles. Note that these events are indeed proxy for  mergers of the peaks of the underlying field: for instance, between snapshot 3 and 5 the central four peaks have merged into one. Similarly, between 1 and 4 the central four voids have merged into one.
  We provide an interactive tool to follow such events in \href{http://www2.iap.fr/users/pichon/Critical/critical_point_2D.html}{2D} and \href{http://www2.iap.fr/users/pichon/Critical/critical_point_3D.html}{3D}.
   }%
  \label{fig:timesequence}
\end{figure*}

Let us illustrate the concept of critical events using an analogy with a mountainous landscape, the latter being restricted to 2D space. A mountainous landscape is made of peaks analogous to proto-haloes. A peak is linked to some of its  neighbours via mountain passes, that form a proto-filamentary structure. Following the ridge from one peak to another one is analogous to following a filamentary structure between two proto-haloes. 
With the action of time, the mountains will erode until eventually no peak will subsist -- this is analogous to the smoothing operation.
In the process, a disappearing peak will see its height (the density) decrease with time.
If the peak is not prominent enough, it will eventually be smoothed to the point where it no longer is a peak but a shoulder on another peak's slope.
Just before the peak disappears, it is still linked to its neighbour via a pass.
When the peak disappears so does the pass -- indeed a pass is always located between \emph{two} peaks; when one disappears, so does the pass.
This particular event is what we defined as a critical \emph{event}.
It encodes the moment when two critical points (here a peak and a saddle point) annihilate.
This can also be interpreted as the moment a peak disappears on the slope of its nearest neighbour -- the two peaks merged and the most prominent subsisted.
Critical events have hence a dual interpretation. \cref{fig:scheme_critical_events_2D,fig:timesequence} shows an illustration in this specific case of a 2+1D field using a 3D visualisation and a sequence of 2D renderings at various smoothing length.
Critical events can be equivalently defined as pairs of critical points with vanishing persistence \citep{Edelsbrunner2000}\footnote{Recalling that topology defines a special relationship between specific sets of critical points, which create and destroy topological components of the excursion, persistence -- the height difference between such points -- is a measure of  the robustness of the newly created component. Hence, vanishingly low persistence pairs correspond to vanishingly short-lived components when scanning the excursion, or changes in topology when smoothing. Note that  
persistence is traditionally used in computational geometry to de-noise data, rather than to probe multiple scales at once.}.

In the following, we will rely on the \emph{Ansatz} that critical points (peaks, filament saddles, wall saddles and minima) in the linear density field can be mapped into late time structures of the cosmic web (haloes, filaments, walls and voids respectively), with increasing smoothing scales probing later times. Under this assumption, critical events (where critical points merge) can be interpreted as mergers of cosmic structures. While this assumption sounds generally reasonable, a word of caution is required. For instance, the formation redshift, $z$, of a halo is usually related to the height $\delta$ of the corresponding peak through the relation $\delta=\deltac/D(z)$, with peaks of vanishing height forming haloes asymptotically late in the future. This automatically excludes local maxima of negative height from the picture. Thus, critical events where local maxima of negative height disappear should never be associated to halo mergers, nor should those where local minima of positive height disappear be associated to void mergers. Similarly, critical events leading to the creation of critical points (unlikely, but possible nucleation) have no obvious late-time counterparts. We will come back to these details later.

\begin{figure}
  \centering
  \includegraphics[width=\columnwidth]{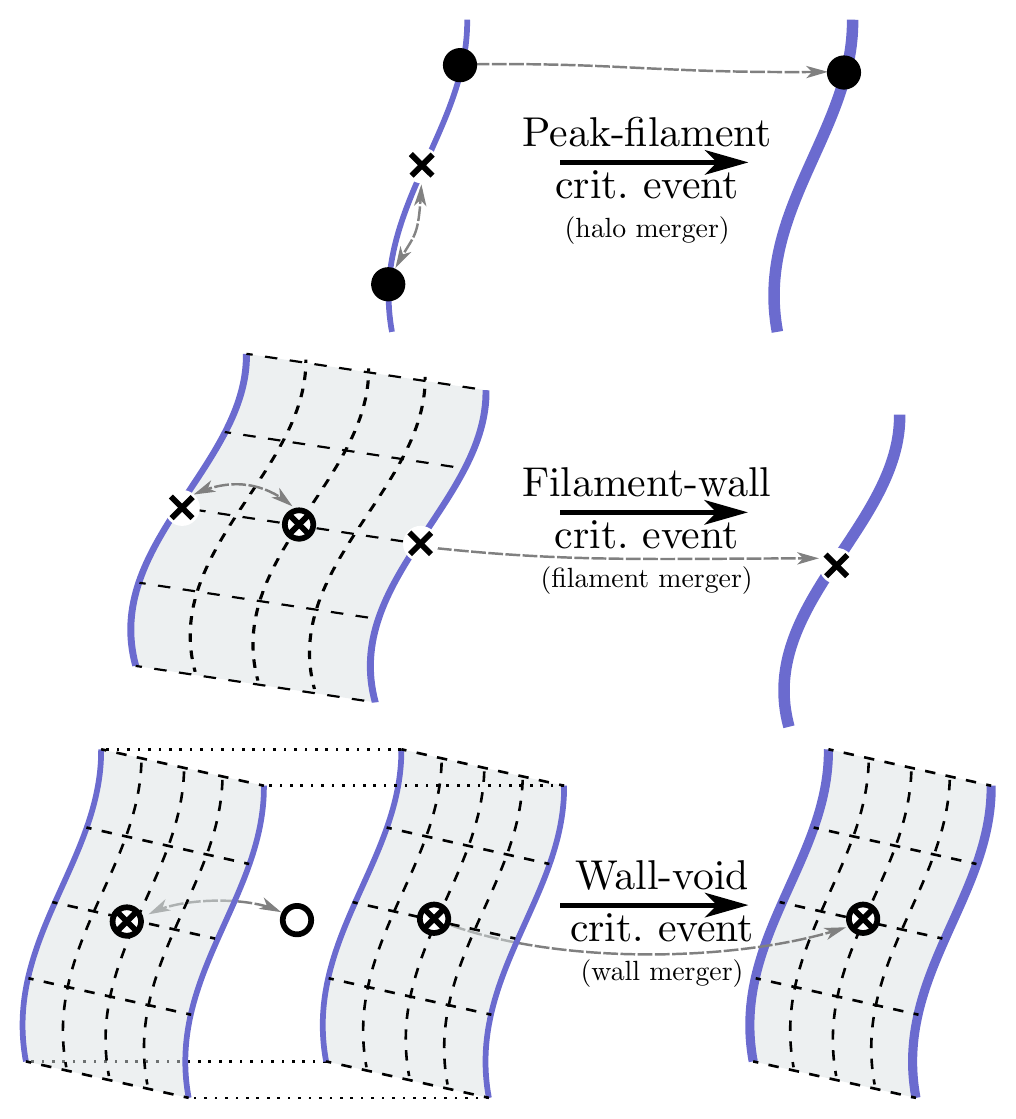}
  \caption{Illustration of critical events in a 3D random field and their physical meaning. $\newmoon$ symbols are peaks, $\times$ symbols are filament-type saddle points (filament centres), $\otimes$ symbols are wall-type saddle points (wall centres) and $\fullmoon$ symbols are minima (void centres). \emph{Top:} Peak-filament critical events encode the merger of two haloes and the disappearance of their shared filament. After the merger, only one peak subsists and the filament disappears. \emph{Middle:} Filament-wall critical events encode the merger of two filaments and the disappearance of their shared wall. After the merger, only one filament subsists. \emph{Bottom:} Wall-void critical events encode the merger of two walls and the disappearance of their joint void (surrounded by the two walls and the dotted lines). After the merger, only one wall-type saddle-point subsists and the void has disappeared. Halo mergers are encoded by peak-filament critical events, filament mergers. Alternatively, one could have chosen to describe these events as resp. filament, wall and void disappearances,  while describing the excursion from the low density end. This is illustrated on \cref{fig:scheme_critical_events_3D_dual}.
  }
  \label{fig:scheme_critical_events_3D}
\end{figure}

Since the primordial density field is a 3D field, the density landscape is made of peaks (proto-haloes), saddle-points (proto-filaments and proto-walls) and minima (proto-voids). Critical events record the merger of peaks into proto-filaments (PF critical events), of proto-filaments into proto-walls (FW critical events) and of proto-walls into proto-voids (WV critical events).
This is illustrated in \cref{fig:scheme_critical_events_3D}. PF critical events (top panel) encode the merger of two haloes separated by a filament. After the merger, the most prominent peak subsists, while the other proto-halo and the proto-filament have disappeared. FW critical events (centre panel) encode the merger of two filaments separated by a wall. After the merger, the most prominent proto-filament subsists, while the other proto-filament and the proto-wall have disappeared. WV critical events (bottom panel) encode the merger of two walls separated by a void. After the merger, the most prominent wall subsists, while the other proto-wall and the proto-void have disappeared.
Note that here we have interpreted the merger from the viewpoint  of the densest surviving structure (\eg the surviving peak of a peak-filament merger), but a dual interpretation is possible that instead takes the viewpoint of the least dense structure. This is further discussed in \cref{sec:dual-interpretation} but is kept out of the main text for the sake of conciseness. In the rest of the paper and unless stated otherwise (as in \eg \cref{sec:Mzmerger}), we will always use the former interpretation.

\subsection{Critical event number density}
\label{sec:definition-crit-events}

In this section, we will present the derivation of the 
mean
number density of critical events in
3+1D position--smoothing space. The averaging is performed over ensemble of field realisations on 3D spatial
slices and the resulting mean density is smoothing dependent.
In \cref{sec:general-formula-critevent}, we demonstrate how one can express the critical event constraint as a function of the local properties of the field and its derivatives.
We also
describe  in more details the  link between the 3+1D density of critical events and the rate of change
with smoothing of the 3D spatial density of critical points and introduce the concept of
\emph{net merger event density} (see also  \cref{sec:net_mergers}).
We then perform in \cref{sec:expression-hessian}
the computation of the critical event density in the eigenframe of the Hessian of the field
 where it takes a simpler form.

\subsubsection{General formulation}
\label{sec:general-formula-critevent}
As defined in \cref{sec:crit-event}, each critical event is a solution $(\vvec{r}_\ce,R_\ce)$ of the set of constraint equations $\grad \delta = 0$ and $H=0$, the latter implying that one eigenvalue of the Hessian vanishes.
In the direction of the null eigenvector, the field behaves as at a flat (critical) inflection point.
Following \cite{hanami}, the number density of critical events in position-smoothing space is given by
\begin{equation}
  n^{\rm 3D}_\ce\equiv\bigg\langle \sum_\ce\dirac^{(3)}(\vvec{r}-\vvec{r}_\ce)  \dirac(R-R_\ce) \bigg\rangle\,,
  \label{eq:ce_definition}
\end{equation}
where $\vvec{r}_\ce$ is the position of a critical event (\ie a critical point with a degenerate direction) in real space and $R_\ce$ its associated smoothing scale, 
and $\delta_{\rm D}$ is the Dirac function.  
The brackets in \cref{eq:ce_definition} denote the 3+1D spatial averaging over volume $V$ and scale range $\Delta R$, $\langle \dots \rangle = (V \Delta R)^{-1} \int_{\Delta R} \dd{R} \int_V \dots \dd[3]{\vvec{r}} $.

In the following, we will use $\pd_R$ to denote derivatives with respect to scale $R$. Since critical events are characterised by $H$ and $\grad \delta $, let us rewrite \cref{eq:ce_definition} in terms of the properties of the field, using the coordinate transformation from $(\rr, R)$ to $(\grad \delta, H)$.
As pointed out by \cite{Musso2019}, this involves the 3+1D Jacobian of the transformation
\begin{align}
  \label{eq:jacobian}
  \hskip -0.1cm
  J(H,\grad\delta) & =
  \begin{vmatrix}
    \pd_R H & \grad H \\
    \pd_R \grad \delta & \grad\grad \delta
  \end{vmatrix}\,,\nonumber
  \\
  &= H \left( \pd_R H - \pd_R \grad \delta \cdot \mathbf{H}^{-1} \cdot \grad H \right).
\end{align}
The latter expression for the $3+1$ decomposition of the Jacobian formally requires the Hessian $\mathbf{H}$ to be invertible, which is not the case at the critical event. Still, the Jacobian is well-defined even there since the product $H \mathbf{H}^{-1}$ remains finite in the $H\to0$ limit.
Interestingly, the term $\partial_R H$
does not contribute to $J$ since it enters the result only multiplied by the vanishing $H$.

The fully covariant formulation of the number density of critical events, which generalises \cite{hanami}, is then
\begin{equation}
  \label{eq:eventcount_covariant}
  n_{\ce}^{\rm 3D} = \left\langle \abs{J}\ \dirac^{(3)}(\grad\delta) \dirac(H)\right\rangle,
\end{equation}
where the brackets now indicate the expectation value over the joint distribution of the field and its successive derivatives up to second order,
as well as derivatives of the field gradient with respect to $R$,
${ P}(x,x_i,x_{ij}, \pd_R x_i)$.

The statistics of  $\partial_R \grad\delta$ variables depends on the choice of filtering function and may be non-local.
Its treatment is significantly simplified when filtering with a Gaussian window, in which case the change in the value of the field with
$R$ is given by a local quantity via the diffusion-type equation
\begin{equation}
   \pd_R \delta = R \laplacian\delta,
   \label{eq:diffusion}
\end{equation}
so we can replace the problem by averaging over the one-point distribution of the field and its derivatives up to the third order,
${ P}(x,x_i,x_{ij}, x_{ijk})$. This distribution involves 20 variables, see \cref{sec:jointpdf}
for the PDF for Gaussian random fields.
For the calculations that follow, we will use a Gaussian filtering model and \cref{eq:diffusion}.

\begin{figure}
  \centering
  {\includegraphics[trim={0cm 2.8cm 0cm 2cm},clip,width=\columnwidth]{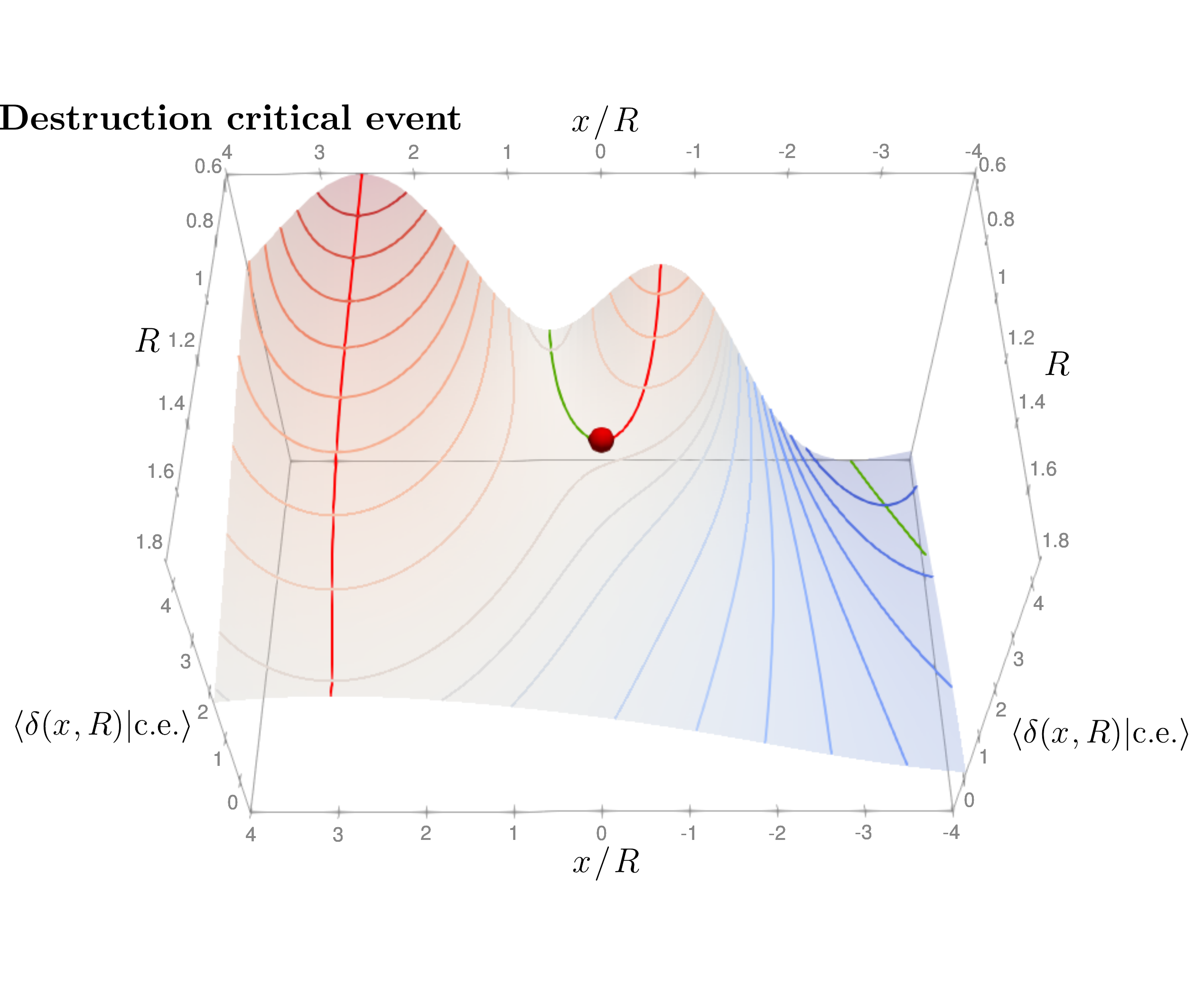}}
  {\includegraphics[trim={0cm 2.5cm 0cm 2cm},clip,width=\columnwidth]{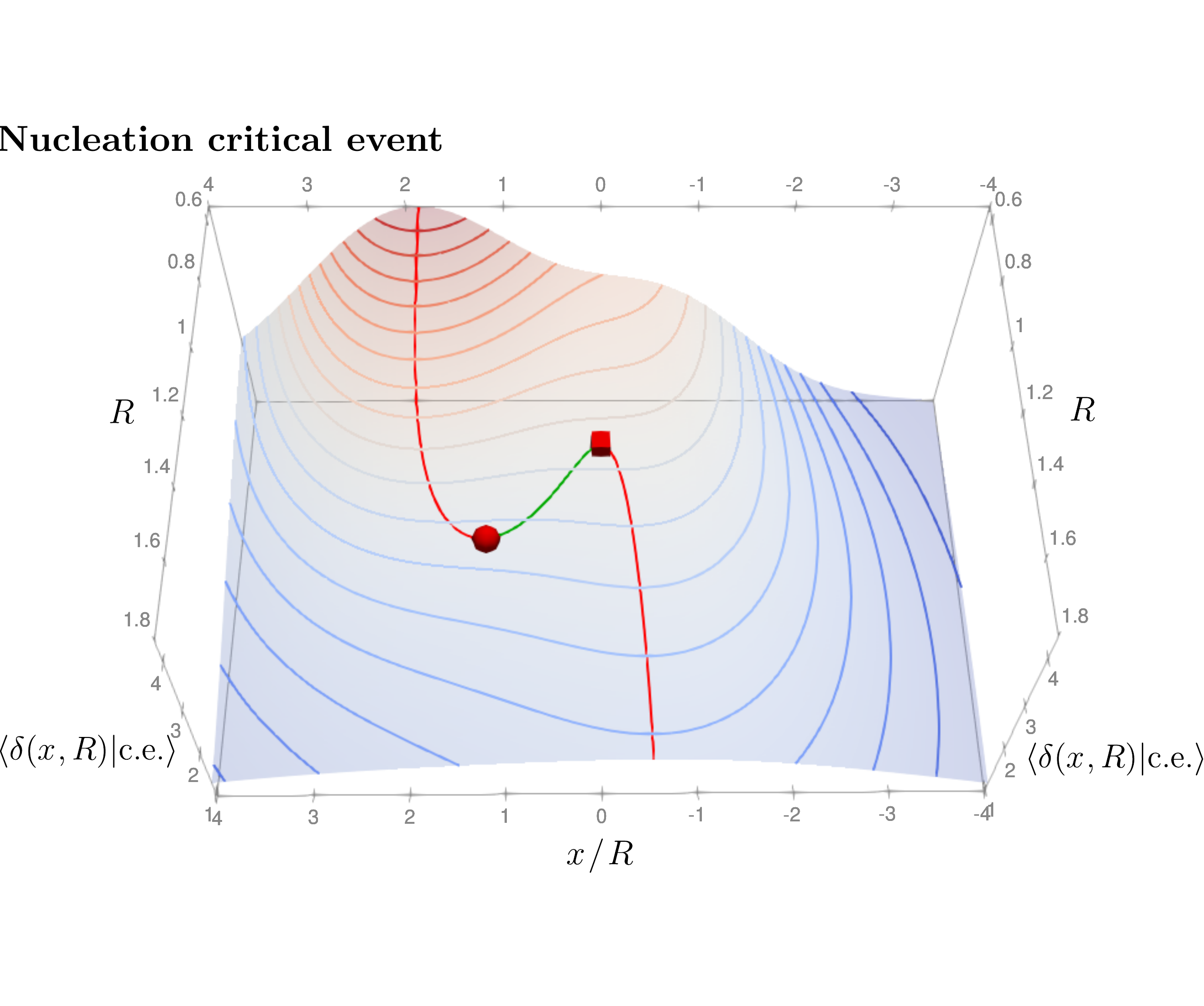}}
  \caption{
    2D slice, in the $(x,R)$ plane, of the conditional mean density in 3+1D position-smoothing space, under constraint of a destruction critical event (red sphere, \emph{top panel}) and a nucleation critical event (red box, \emph{bottom panel}) at $R=1, x=0$. The slice position is chosen to contain the events. 
    Green lines show filament-type saddle points at each $R$ and red lines show peaks.
    Density isocontours are represented as coloured lines (from red, high density to blue, low density).
    The top panel is reminiscent of \cref{fig:scheme_critical_events}, since 3D merger events closely resemble 1D merger events along filaments, 
    while the bottom panel is unique to dimensions larger than one for Gaussian smoothing.
    An interactive version can be found in the online supplemental material and \href{http://www2.iap.fr/users/pichon/Critical/}{online}.    
    While nucleation events such as that shown on the bottom panel can occur, they statistically seem to remain short-lived and are less frequent than destruction ones.
    }
  \label{fig:nucleation-destruction}
\end{figure}

It is important to stress now that the 3+1D number density of critical events given by \cref{eq:eventcount_covariant}
is not equivalent to the rate of change with smoothing $R$ of the 3D density of critical point pairs.
Indeed, at a critical event, one pair of critical points of adjacent
topological types (\eg maximum and filamentary saddle) coalesce, but as a local analysis in
\cref{sec:nucl} demonstrates, this event can describe either the merging or the creation of the pair,
depending on the sign of Jacobian $J$. Namely, the partial number densities
\begin{equation}
  n_{\ce,\pm}^{\rm 3D} \equiv
  \left\langle \abs{J}\heaviside(\pm J)\ \dirac^{(3)}(\grad\delta)
  \dirac(H)\right\rangle,
  \label{eq:nce+-_def}
\end{equation}
such that $n_{\ce}^{\rm 3D}= n_{\ce,-}^{\rm 3D} + n_{\ce,+}^{\rm 3D}$,
count separately critical events where a pair of critical points is created ($+$, also called a nucleation) or destroyed ($-$).
The two kinds are illustrated on \cref{fig:nucleation-destruction}, which was generated using the code detailed in \cref{sec:generation} for two likely configurations. 
Note however that nucleation critical events are $\sim$ 30 times less probable than the destruction critical event (see \cref{fig:total-vs-net}) for $n_\mathrm{s} < -1$ ($\gamma \lessapprox 0.8$).
The quantity that is equal to the rate of change of the density of critical points with smoothing is, therefore, obtained by
removing the absolute value from the Jacobian in \cref{eq:eventcount_covariant}, as shown in \cref{sec:net_mergers}, which shows that 
the rate of change of the number density of critical points with smoothing obeys
\begin{equation}
\label{eq:dndRmaintext}
\frac{\mathrm{d} n_{\cp}}{\mathrm{d} R} = 2 \left\langle J\; \dirac^{(3)}(\grad \delta)\dirac(H) \right\rangle
\equiv -2 \, n^{\rm 3D}_\me~,
\end{equation}
where we introduced  the `net merger rate' (taken with minus sign) as
\begin{equation}
  n^{\rm 3D}_\me  %
  = n_{\ce,-}^{\rm 3D} - n_{\ce,+}^{\rm 3D}\,.
\label{eq:nme_def}
\end{equation}
While in this paper, the term `net merger rate' has been chosen for the sake of readability, we must emphasize that this quantity measures rates in position-smoothing scale space.
Care should be taken to relate these mergers to mergers in space-time, as will be discussed in \cref{sec:Mzmerger}.

\subsubsection{Expression in the frame of the Hessian}
\label{sec:expression-hessian}

One possible method to yield an analytical expression of \cref{eq:dndRmaintext} is to re-express it in the frame of the Hessian, where the Jacobian becomes sparse and can be computed in terms of the field variables.
We shall denote the field variables in the eigenframe of the Hessian with a tilde. In this frame the diagonal
components of the Hessian itself are given by the eigenvalues $\tilde{x}_{ii}=\lambda_i$ with off-diagonal ones being zero $\tilde{x}_{i\ne j}=0$.
The Jacobian is by construction invariant under rotation, so we can rewrite it in the Hessian eigenframe
without loss of generality.
Developing $H$ into $ \sigma_2^3 \lambda_{1}\lambda_{2}\lambda_{3}$ and assuming, for instance, that  direction 3 is the degenerate one, the Jacobian can be rewritten as follows
\begin{align}
  \frac{J(H,\grad\delta)}{\sigma_1 \sigma_2^4 \sigma_3}
  & = \lambda_{1} \lambda_{2}
  \begin{vmatrix}
    \pd_R  \lambda_{3} &  \tilde x_{331} &  \tilde x_{332} &  \tilde x_{333} \\
    \pd_R  \tilde x_1 &  \lambda_{1} & 0 & 0 \\
    \pd_R  \tilde x_2 & 0 &  \lambda_{2} & 0 \\
    \pd_R  \tilde x_3 & 0 & 0 & 0
  \end{vmatrix}\,, \label{eq:J2}\\
   & = - (\lambda_{1} \lambda_{2})^2 \tilde x_{333} \pd_R \tilde x_3 \,,
\label{eq:J3}
\end{align}
where the factorisation of $\lambda_{1} \lambda_{2}$  in \cref{eq:J2} is a consequence of
$ \lambda_{3}$ being zero, 
which also nulls the last component of the last row.
Using \cref{eq:diffusion} to re-express the derivative \wrt smoothing in terms of the Laplacian of the field, we find the number density of critical events in \cref{eq:eventcount_covariant} to be\footnote{ One factor of $\lambda_{1} \lambda_{2}$ drops between \cref{eq:J3} and \eqref{eq:eventcount0-0}
because of $\dirac(H)$ in \cref{eq:eventcount_covariant}. We also note that $\lambda_1 \lambda_2 \ge 0$ when $\lambda_3=0$.}
 \begin{align}
\label{eq:eventcount0-0}
   n_{\ce}^{\rm 3D} \!=\!\frac{ R}{{\tilde R}^{2}\, R_*^3  }%
   &
  \left\langle   \abs{\mbox{$\sum_l$} \tilde x_{3ll}} \abs{\tilde x_{333}}
  \dirac^{(3)}(\tilde x_i) \lambda_{1}\lambda_{2}
  \dirac(\lambda_{3})  \!\right\rangle ,
\end{align}
where $\dirac^{(3)}(\tilde{x}_i)$ is understood as the product of the Dirac delta functions of all components of the gradient of the field.
$R_*$ and $\tilde R$ are the typical inter-critical point and inter-inflection point separation introduced in \cref{eq:defR0}.

Let us stress that in \cref{eq:eventcount0-0} the averaging is performed over the distribution of the fields expressed in the frame of the Hessian matrix \citep{doroshkevich70} that  differs functionally from the distribution in an arbitrary frame.  For computational purposes it is useful to avoid this complication. We achieve this by noticing that in the integral over the Hessian space, the transition to  the eigenframe can be introduced using
the Dirac delta functions on off-diagonal elements of the Hessian coupled with the Jacobian of the transformation $\propto (\lambda_3-\lambda_1)(\lambda_3-\lambda_2)(\lambda_2-\lambda_1)$ times $2\pi^{2}$ due to integration over angles of Hessian orientation assuming statistical isotropy.
Namely, \cref{eq:eventcount0-0} can be cast in the form of an average over the distribution of field variables in an arbitrary frame as
 \begin{align}
\label{eq:eventcount0}
   n_{\ce}^{\rm 3D}  &= \frac{ 2 \pi^2 R}{{\tilde R}^{2}\, R_*^3  }
     \Big\langle   \abs{\mbox{$\sum_l$} x_{3ll}} \abs{x_{333}}
  \dirac^{(3)}(x_i) x_{11}^2 x_{22}^2 (x_{22}\!-\!x_{11})
  \notag
  \\ &\quad \times \displaystyle
   \heaviside(-x_{22}) \heaviside(x_{22}\!-\!x_{11}) \dirac(x_{33}) \delta^{(3)}_{\rm D}(x_{i\neq k})
  \Big\rangle\,.
\end{align}
We can use this expression as is to compute the average $n_{\ce}^{\rm 3D}$ over any isotropic distribution given in an arbitrary coordinate frame, since the Hessian eigenframe condition is now enforced explicitly by $\dirac^{(3)}(x_{i\neq k})$ which denotes again a product of Dirac delta functions of all the off-diagonal components of the Hessian matrix, while the Heaviside functions $\heaviside$ enforce the sorting of the Hessian's diagonal elements. Thus, we have
dropped the tilde sign from the variables.  For compactness, we have given the integrand in non-rotation invariant form, having
used the presence of $\dirac(x_{33})$ in the integral that describes condition of the vanishing third Hessian eigenvalue.

The novelty of \cref{eq:eventcount0} compared to the classical BBKS formula is the weight $ \abs{\sum_i x_{3ii}} \abs{x_{333}} $, which
requires the knowledge of the statistics of the 3rd order derivatives of the field. The expectations in \cref{eq:eventcount0} can be evaluated  with the joint statistics of the field and its successive derivatives,
${ P}(x_{113},x_{223},x_{333},x_{11},x_{22})$,
which now only involves five of the variables listed above to average over.

Following the same derivation, one can also compute the net merger rate
 \begin{align}
\label{eq:eventcount00}
  n^{\rm 3D}_{\me} &= \frac{ 2 \pi^2 R}{{\tilde R^{2}}\, R_*^3  }
  \left\langle  \left( x_{333}\mbox{$\sum_l$} x_{3ll}  \right)
  \dirac^{(3)}(x_i) x_{11}^2 x_{22}^2 (x_{22}\!-\!x_{11})
  \right.  \notag
  \\ &\quad \times \displaystyle
  { \heaviside(-x_{22}) \heaviside(x_{22}\!-\!x_{11}) \dirac(x_{33}) \delta^{(3)}_{\rm D}(x_{i\neq k})}
  \Big\rangle\,.
\end{align}
Let us stress here that \cref{eq:eventcount0,eq:eventcount00} describe different quantities
that were defined in \cref{eq:eventcount_covariant,eq:nme_def}, respectively.
In \cref{eq:eventcount00} and in the rest of the paper, the quantity of interest will be the net merger rate.

 Note
 that   \cref{eq:eventcount0-0}  closely resembles
 the equation giving the flux of critical lines per unit surface presented in \cite{pogo09}, up to the delta function on the third eigenvalue (and the corresponding Jacobian). It involves the product of the transverse curvatures, because the larger those curvatures the larger the flux of such lines per unit  transverse surface.
 The extra   third eigenvalue delta function reflects that we also now require  that along the filament's direction the curvature should be flat, whereas they marginalised over all possible longitudinal curvature.
 The similarity implies that  critical points essentially slide along critical lines as one smooths the field, see \cref{fig:timesequence}: in some loose sense
 the 3D event count can be  approximately recast into a 1D event count along the ridges.

\subsection{Gaussian number density of critical events per type}
\label{sec:ce_by_type}

In this section, the number counts are extended to distinguish different critical event types and count them as a function of density.
\Cref{sec:number_counts} presents the number count of the different types of critical events.
\Cref{sec:critical-event-versus height} presents their number count as a function of their density. Throughout the section, the field will be assumed to be a Gaussian random field.

\subsubsection{Different  critical events and their mean number density}
\label{sec:number_counts}

In the previous example we chose the largest eigenvalue $\lambda_3$ to be vanishing at the critical event, which corresponds to
the coalescence of a peak-filamentary saddle pair.  Thus, we did not count all possible critical events in \cref{eq:eventcount0,eq:eventcount00}.
While the coalescence of peaks with filaments \citep[PF critical events, the sloping saddles of][]{hanami} are clearly central to the theory of mass assembly,
the coalescence of filament-saddles with wall-saddles (FW critical events) and of wall-saddles with voids  (WV critical events) are also likely to affect the topology of galactic infall.
FW critical events correspond to the case when the middle  eigenvalue $\lambda_2$ vanishes, while WV critical events are the ones with the lowest eigenvalue $\lambda_1$ being zero. %

Let us therefore compute the net merger rate for each type of mergers ($\cal P\equiv$ PF, $\cal F\equiv$ FW and $\cal W \equiv$ WV) using Gaussian assumption about the density field. For an isotropic Gaussian field, odd- and even-order derivatives of the field at the same point are completely statistically independent.
Therefore, \cref{eq:eventcount00}, generalised
to the case where any eigendirection can be chosen as a degenerate one,
can be split into odd- and even-order derivative terms as
\footnote{From now on to simplify the notation we will drop
the superscript 3D from the critical event densities where it does not lead to confusion.}
\begin{equation}
  n_{\me}^{(j)} = \frac{R}{\tilde{R}^2R_*^3} \; {C_\mathrm{odd}} \;
  C_{j,\mathrm{even}} \,,
\label{eq:eventcount}
\end{equation}
where
\begin{align}
  \label{eq:Cjeven}
  C_{j,\mathrm{even}} &=
  \bigg\langle 2 \pi^2 %
  {\dirac(x_{jj})}   \heaviside(x_{33}\!-\!x_{22}) \heaviside(x_{22}\!-\!x_{11}) \nonumber \\
 & \hskip 0.5cm \times  \dirac^{(3)}(x_{k\neq l})
 \bigg|{\sum_{kl}}\tfrac{1}{2}\varepsilon^{jkl} x_{kk}^2 x_{ll}^2 ({x_{kk}\!\!-\!x_{ll}})\bigg| \bigg\rangle\,,
\end{align}
with $\varepsilon^{jkl}$ being the completely anti-symmetric Levi-Civita tensor and $j = {3,2,1}$ for peak, filament and walls.
In turn the term that involves the odd-order derivatives of the field,
\begin{equation}
  \label{eq:Codd}
 C_\mathrm{odd} =
  \bigg\langle %
  {\sum_l x_{jll}}{x_{jjj}} \dirac^{(3)}(x_i)\bigg\rangle,
\end{equation}
is actually independent on $j$ due to isotropy. In the rest of the paper, we will also make use of the notation $\cal P, F, W$ instead of $j=3,2,1$ (for peak, filament and wall mergers resp.) in formulas with an astrophysical interpretation.

The factors $C_\mathrm{odd}$ and $C_{j,\mathrm{even}}$ that constitute $n_{\me}^{(j)}$ are readily evaluated. In 3+1D they are
\begin{equation}
  \label{eq:C2even3D}
  C_{1,\mathrm{even}} = C_{3,\mathrm{even}} =
  \frac{29-6\sqrt{6}}{18\sqrt{10\pi}}\,,
   \quad
  C_{2,\mathrm{even}}
  = %
  \frac{2}{\sqrt{15\pi}}\,,
\end{equation}
while common to all merger event types,
\begin{equation}
  C_\mathrm{odd}=\frac{1}{5}\left({\frac{3}{2\pi}}\right)^{3/2}
(1\!-\!\tilde{\gamma}^2)\,.
  \label{eq:alpha-term}
\end{equation}
$C_\mathrm{odd}$ can also be computed in arbitrary dimensions as shown in \cref{sec:NDcounts}.

We note that for Gaussian fields, the computation of the total critical event density $n_{\ce}$,
as well as partial densities of creation and destruction events $n_{\ce,\pm}$,
differ from the computation of $n_{\me}$ only in the $C_\mathrm{odd}$ term that can also be found analytically for these quantities. The corresponding values are given in \cref{sec:crit-event-numb-different-def}.

In addition, let us note that the quantities $n_{\me}^{(j)}$ correspond to the following changes of the critical point densities 
\begin{align}
\mathrm{d} n_{\cp}^\mathrm{p} / \mathrm{d} R = & -n_{\me}^\mathcal{P}, \quad
\mathrm{d} n_{\cp}^\mathrm{f} / \mathrm{d} R = -(n_{\me}^\mathcal{P} + n_{\me}^\mathcal{F}), \nonumber \\
\mathrm{d} n_{\cp}^\mathrm{v}/ \mathrm{d} R = & -n_{\me}^\mathcal{W}, \quad \mathrm{d} n_{\cp}^\mathrm{w} / \mathrm{d} R
= -(n_{\me}^\mathcal{F} + n_{\me}^\mathcal{W}) ~.
\label{eq:me_to_sp}
\end{align}
Here, superscripts $\mathrm{p,f,w,v}$ denote peaks, filament-saddle, wall-saddle and minima respectively.
Thus, for instance, $n_{\me}^\mathcal{P}$ and the change in the density of peaks $\mathrm{d} n_{\cp}^{\mathrm{p}} / \mathrm{d} R $ both evaluate to
\begin{equation}
 n_{\me}^\mathcal{P}
=
 \frac{3R}{R_{\star}^{3}\tilde R^{2}}
 (1-\tilde \gamma^{2})
 \frac{29\sqrt {15} -18 \sqrt{10}}{1800\pi^{2}}~.
\end{equation}
This coincides with the result of \cref{sec:nce_versus_nsp}, obtained by direct differentiation of $n_{\cp}^{\mathrm{p}}$.

From \cref{eq:C2even3D} we can compute the ratio of filament to peak mergers $r_\mathrm{\!{\cal F}/\!{\cal P}} \equiv n_{\me}^\mathcal{F}/n_{\me}^\mathcal{P} = C_{2,\mathrm{even}}/C_{3,\mathrm{even}}$. Interestingly, the merger event ratio is independent of the spectral index of the field and is given by
\begin{equation}
  r_\mathrm{{\cal F}\!/\!{\cal P}} = \frac{24\sqrt{3}}{29\sqrt{2}-12\sqrt{3}}\approx 2.05508~
  ,
  \label{eq:crit-event-ratio-th}
\end{equation}
which is nothing but the ratio between the mean number of wall-type saddles and peaks minus 1. This relation can be readily obtained also from \cref{eq:me_to_sp} by noting that the relative fraction of different critical points is smoothing-independent, and thus should be, after some algebra, the ratio
of their rates of change, \eg, $(n_{\me}^\mathcal{P}+n_{\me}^\mathcal{F})/n_{\me}^\mathcal{P} = n_{\cp}^\mathrm{f}/n_{\cp}^\mathrm{p}$ so that %
$r_\mathrm{{\cal F}\!/\!{\cal P}}=n_{\cp}^\mathrm{f}/n_{\cp}^\mathrm{p}-1$.
\Cref{eq:crit-event-ratio-th} also shows that there are about twice more filaments disappearing in filament merger events (${\cal F}$ events) than
in halo merger events (${\cal P}$ events).
Similarly, we can compute $r_\mathrm{{\cal F}\!/\!{\cal W}}$ to deduce that there are twice as many walls disappearing due to filament mergers (${\cal F}$ events) as due to wall mergers (${\cal W}$ events).
\Cref{sec:ratio-critical-events-ND} also presents these ratios in dimension 4 to 6.

\subsubsection{3D differential event counts of a given height}
\label{sec:critical-event-versus height}

As argued by \cite{press_formation_1974,BBKS}, the initial mean density profile of a proto-object
contains information about its future evolution (\eg the time of collapse).
In this section, we therefore extend our previous results by computing the net merger rate in 3+1D space as a function of the field height (the overdensity).
While the {\it density-integrated} net merger rates, computed in the previous section, are directly connected to the derivatives of the {\it density-integrated} number density of critical points through \cref{eq:me_to_sp}, the net merger rates at fixed $\nu=\delta/\sigma_0$\footnote{Note that here and in the following, $\nu$ refers to specific values that the random field $x$ may take.} do not verify such a simple relation, since the field height $\nu$ is not preserved along the 3+1D trajectory of an individual critical point.
In other words, the field height of the critical event is not simply related to the height of its two progenitors.
This gives us an additional source of change in the critical point number density at fixed $\nu$.
Thus, $n_{\me}(\nu)$ is a new statistics, not equivalent to $\mathrm{d} n_{\cp}(\nu)/ \mathrm{d} R$,
that focuses specifically on the contribution of mergers to the change of the critical point number density at a given $\nu$.
Studying $n_{\me}(\nu)$ allows us to make the distinction between mergers of important critical points and less significant ones.
In particular, if we identify astrophysical objects by a threshold in $\nu$, we will be able to study the mergers of that particular population.

The differential net merger density as a function of height is obtained  by introducing $ \dirac(x-\nu)$ in the expectation of \cref{eq:eventcount}.
Under the assumption of a Gaussian random field, the field only correlates with its even-order derivatives (second in this case).
Imposing the height of the critical events considered here therefore only modifies the term $C_{j,\mathrm{even}}$ while $C_{\mathrm{odd}}$ is left unchanged, following
\begin{align}
\label{eq:Ceven-nu}
\!C_{j,\mathrm{even}} (\nu)&= \bigg\langle \vphantom{\prod_{k\neq j}} \!{\dirac(x-\nu)} {\dirac(x_{jj})}
\heaviside(x_{33}\!-\!x_{22}) \heaviside(x_{22}\!-\!x_{11})  \notag\\
& \hskip -0.2cm \times 2 \pi^2 \dirac^{(3)}(x_{k\neq l})
\bigg|{\sum_{kl}}\tfrac{1}{2}\varepsilon^{jkl} x_{kk}^2 x_{ll}^2 (x_{kk}\!-\!x_{ll})\bigg| \bigg\rangle\,.
\end{align}
The net merger density of kind $j$ at height $\nu$, $n_{\me}^{(j)}(\nu)$ then reads
\begin{equation}
  \label{eq:nmerger_nu}
  n_{\me}^{(j)}(\nu) \equiv \frac{R}{\tilde{R}^2R_*^3}  C_{j,\mathrm{even}}(\nu) C_\mathrm{odd}.
\end{equation}
Interestingly, $C_{j,\mathrm{even}} (\nu)$ appears to have an analytical expression once rotational invariants are used to evaluate the expectations. Following the formalism described  in \cite{pogo09b}, we introduce the variables
\begin{align}
J_{1}&=I_{1}\,,\quad
J_{2}=I_{1}^{2}-3I_{2}\,,\\
J_{3}&=\frac {27}{2} I_{3}-\frac 9 2 I_{1}I_{2} +I_{1}^{3}\,,\quad
\zeta=\frac{x + \gamma J_1}{\sqrt{1 - \gamma^2}}\,,
\label{eq:defJ1J2J3}
\end{align}
that are linear combinations of the density field $x$ and rotational invariants of its second derivatives, namely the trace $I_{1} \sigma_2=\tr \mathbf{H}=\sigma_2(\lambda_{1}+\lambda_{2}+\lambda_{3})$, minor $I_{2} \sigma_2^2=1/2((\tr \mathbf{H})^{2}-\tr \mathbf{H}\cdot {\mathbf H})=\sigma_2^{2}(\lambda_{1}\lambda_{2}+\lambda_{2}\lambda_{3}+\lambda_{3}\lambda_{1})$ and determinant $I_{3} \sigma_2^ 3=\det \mathbf{H}=\sigma_2^3\lambda_{1}\lambda_{2}\lambda_{3}$ of the Hessian matrix $\mathbf{H}$.
The distribution of these variables is given by
\begin{equation}
{ P}(\zeta, J_{1},J_{2},J_{3})=\frac{25\sqrt{10\pi}}{24\pi^{2}}\exp\!\left(-\tfrac{1}{2}\zeta^{2}-\tfrac{1}{2} J_{1}^{2}-\tfrac{5}{2} J_{2} \right),
\end{equation}
where $J_{3}$ is uniformly distributed between $-J_{2}^{3/2}$ and $J_{2}^{3/2}$ and $J_{2}$ is positive.
Using these rotational invariants, one can rewrite \cref{eq:Ceven-nu} for each type of critical event as
\begin{align}
\label{eq:Ceven-nu-3}
C_{3,\mathrm{even}} (\nu) =  \left\langle \vphantom{\Theta_\mathrm{H}(-\sqrt{\!J_{2}}\!-\!J_1)} \right.  &
\!\tfrac{1}{3}\!\left( J_1^2-J_2 \right)^2 \dirac(x\!-\!\nu) \dirac(I_{3}) \times  \nonumber \\
& \times \left. \Theta_\mathrm{H}(J_1\!+\!2\sqrt{\!J_{2}})\Theta_\mathrm{H}(-\sqrt{\!J_{2}} \! -\! J_1)
\right\rangle ,\\
\label{eq:Ceven-nu-2}
C_{2,\mathrm{even}} (\nu) =  \left\langle \vphantom{\Theta_\mathrm{H}(-\sqrt{\!J_{2}}\!-\!J_1)} \right. &
\!\tfrac{1}{3}\!\left( J_2 -J_1^2 \right)^2 \dirac(x\!-\!\nu) \dirac(I_{3}) \times  \nonumber \\
& \times \left. \Theta_\mathrm{H}(J_1\!+\!\sqrt{\!J_{2}}) \Theta_\mathrm{H}(\sqrt{\!J_{2}}\!-\!J_1)
\right\rangle , \\
\label{eq:Ceven-nu-1}
C_{1,\mathrm{even}} (\nu) =   \left\langle \vphantom{\Theta_\mathrm{H}(-\sqrt{\!J_{2}}\!-\!J_1)} \right. &
\!\tfrac{1}{3}\!\left( J_1^2-J_2 \right)^2 \dirac(x\!-\!\nu) \dirac(I_{3}) \times  \nonumber \\
& \times \left. \Theta_\mathrm{H}(J_1\!-\!\sqrt{\!J_{2}})\Theta_\mathrm{H}(2\sqrt{\!J_{2}}\!-\!J_1)
\right\rangle,
\end{align}
with
 \begin{align}
  \dirac(I_{3})&= \frac {27}{2}\dirac\left(J_{3}-\frac{ 3J_{1}J_{2}-J_{1}^{3}}{2}\right),\\
  \dirac(x-\nu)&= \frac {1}{\sqrt{1-\gamma^{2}}}\dirac\left(\zeta-\frac{\nu + \gamma J_1}{\sqrt{1 - \gamma^2}}\right).
  \end{align}
The condition that the determinant $I_3$ is null due to specific $\lambda_{j}$ being zero is enforced by restricting the range of $J_{1}$
according to the product of Heaviside functions as specified in \cref{eq:Ceven-nu-3,eq:Ceven-nu-2,eq:Ceven-nu-1}.
The integration in \cref{eq:Ceven-nu-3,eq:Ceven-nu-2,eq:Ceven-nu-1} can be done analytically and an exact expression for $C_{j,\rm even}(\nu)$  follows
\begin{align}
\label{eq:C1even_nu}
C_{3,\mathrm{even}} (\nu)&\!=\!
\sum_{i=5,6,9}c_{3,i}\,\exp\left(-{\frac{\nu ^2}{2 \left(1-5\gamma ^2/i\right)}}\right)\notag
,\\
C_{2,\mathrm{even}} (\nu)&\!= \!c_{2,6}\,\exp\left(-\frac{\nu^{2}}{2(1 - 5\gamma^2/6)}\right),\notag\\
C_{1,\mathrm{even}} (\nu)&\!=\! C_{3,\mathrm{even}}(-\nu),
\end{align}
with
\begin{align}
c_{3,5}&=\frac{3 \sqrt{5} \gamma  \nu \sqrt{1-\gamma ^2} \left(275 \gamma ^4+30 \gamma ^2 \left(2 \nu ^2-23\right)+351\right)}
{\pi \sqrt{2 \pi} \left(9-5 \gamma
   ^2\right)^4}\,,\notag\\
c_{3,6}&=-\frac{\text{erf}\left(\frac{\gamma  \nu }{\sqrt{2 (1-\gamma^2)(6-5\gamma^2)}}\right)+1}{\sqrt{5} \pi  \sqrt{6-5 \gamma ^2}}\,, ~~~
c_{2,6}=\frac{2}{\sqrt{5}\pi\sqrt{6 - 5\gamma^2}}\,, \notag
\\
c_{3,9}&=\frac{  \text{erf}\left(\frac{\sqrt 2\gamma  \nu }{ \sqrt{(1-\gamma^2)(9-5\gamma^2)}}\right)+1}{4 \pi \sqrt 5 \left(9-5 \gamma ^2\right)^{5/2}} \times \notag\\
 \notag &\hskip -0.2cm\!\!\left(\!\frac{3600 \gamma ^4 \nu ^4}{\left(9\!-\!5
   \gamma ^2\right)^2}\!\!+\!\frac{120 \gamma ^2\! \left(27\!-\!35 \gamma ^2\right) \!\nu ^2}{9\!-\!5
   \gamma ^2}\!\!+\!575 \gamma ^4\!-\!1230 \gamma ^2\!+\!783\!\right)\!.
\end{align}
The resulting net merger rate as a function of their height $\nu$ is plotted in \cref{fig:PDF-3D}, bottom panel, for
different values of the spectral index $n_\mathrm{s}$. Note that $n_{\me}^{(j)}(\nu)$ scales like $1/R^4 $ but is also a function of
$R$ via the spectral parameters $\gamma$ and $\tilde \gamma$.
A comparison to measurement in numerically drawn random fields will be presented later in \cref{fig:crit-event-PDF-GRF}.
Note that the mean density of net peak mergers, given by \cref{eq:nmerger_nu} for $j=3$ and \cref{eq:C1even_nu}, is equivalent to formula C30 in \cite{hanami}.

\subsection{2D event counts and differential counts}
\label{sec:theory2D}
Given its astrophysical interests when considering 2D maps in various contexts, 
let us also briefly present the analogues of \cref{eq:eventcount}
for 2+1D fields. It reads
\begin{align}
  n_\me^{{\cal P}, \rm 2D}(\nu, R)
  \!&=\! \frac{ {-}2 \pi R}{\tilde{R}^2R_*^2}
  {\left\langle { (x_{211}+x_{222})}{x_{222}} \dirac^{}(x_1)\dirac^{}(x_2)\right\rangle}\times  \notag 
 \\
  &\hskip -0.2cm {\left\langle   \heaviside({ -}x_{ 11}) {\dirac(x_{ 22})} \dirac^{}\!(x_{12}) {\dirac(x-\nu)}{x_{ 11}}\right\rangle} \,, \label{eq:eventcount2D}
\end{align}
where the even part  ${ -}2 \pi \!\left\langle  \heaviside({ -}x_{ 11}) {\dirac{}\!(x_{ 22})} \dirac^{}\!(x_{12}) {\dirac{}\!(x-\nu)}{x_{ 11}} \right\rangle$ is nothing but
\begin{equation}
C_{\rm even}(\nu)=\left\langle I_{1}^{2} \heaviside({ -}I_{1}) {\dirac(I_{2})} {\dirac(x-\nu)} \right\rangle,
\end{equation}
once written in terms of the trace $I_{1}$ and determinant $I_{2}$ of the Hessian matrix.

\begin{figure}
  \includegraphics[width=0.98\columnwidth]{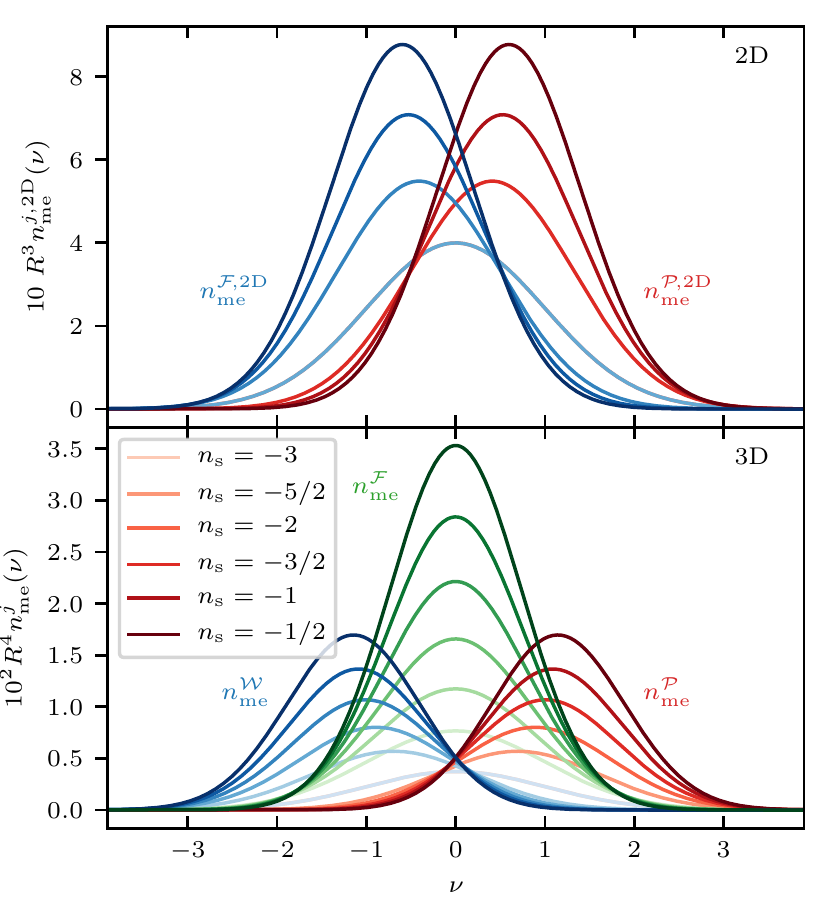}
  \caption{ 
    The PDFs of critical events net merger rates of the various types ${\cal P,F}$ in 2D for $n_\mathrm{s}$ from $-2$ to $-1/2$ (\emph{top panel}) and ${\cal P,F,W}$ in 3D for $n_\mathrm{s}$ from $-3$ to $-1/2$ (\emph{bottom panel}), as labelled.
    Note that the dominant change with spectral index is in the amplitude which scales like $1/({\tilde R}^2 R_\star^d)$.
    The rest of the shape variation comes from the  weaker $\gamma$ and $\tilde \gamma$ dependence of $C_{\rm odd}$
    and $C_{\rm even}$.
    In 2D, the ${\cal P, F}$ merger rates coincide for $n_\mathrm{s}=-2$ as the field and its second derivatives become uncorrelated ($\gamma = 0$).
    }
  \label{fig:PDF-2D}
  \label{fig:PDF-3D}
\end{figure}

After some algebra, given the knowledge of the 2D PDF written in \cref{sec:jointpdf},  we get for the peak merger rate
\begin{multline}
  n_\me^{{\cal P}, \rm 2D}(\nu, R)
  =
  \frac{ 2\pi R C_{\rm odd}^{\rm 2D}}{\tilde{R}^2R_*^2} \bigg[
  \frac{ 4\gamma\nu\sqrt{1-\gamma^{2}}}{\left(3\!-\!2 \gamma ^2\right)^{2}}
  \exp({-\frac{1}{2}\frac{\nu ^2}{1- \gamma ^2}}) \\
  +  \frac{\sqrt{8\pi}
   ((1-\gamma^2)(3-2\gamma^2)+\gamma^2\nu^2)}
 {\left(3\!-\!2 \gamma ^2\right)^{5/2}
 } \\
  \times \mathrm{erfc} \bigg(\frac{-\gamma
   \nu }{\sqrt{2(1-\gamma^2)(3-2\gamma^2) }}\bigg)
  \exp(\!-\frac{3\nu ^2}{6-4\gamma ^2}\!)\bigg]\,,
\label{eq:differentialcount2D}
\end{multline}
with
\begin{equation}
C_{\rm odd}^{\rm 2D}= \frac 3{8\pi} (1-\tilde{\gamma }^2)\,. \notag
\label{eq:eventcount2Dfinal}
\end{equation}
The wall-filament merger rate is obtained by swapping $\nu$ to $-\nu$ in \cref{eq:differentialcount2D}.
The two rates are plotted in  \cref{fig:PDF-2D}, top panel, and validated
against Gaussian random fields later in \cref{fig:crit-event-PDF-GRF-2D}.
The net merger rate, $n_\me^{{\cal P},{\rm 2D}}(R) = 2 C_{\rm odd} R /(3\sqrt{3} {\tilde{R}^2R_*^2})$
 follows by integration over $\nu$.\footnote{The code to reproduce the figures can be found in the online supplemental material and \href{http://www2.iap.fr/users/pichon/Critical/}{online}.}

\Cref{sec:NDevents} presents also the differential counts in dimension 4 to 6,
together with asymptotic expressions in the large dimension limit for the integrated count ratios.
As expected, for any dimension the number counts per unit log-volume 
are scale invariant (up to the slow variation in the spectral parameters), \ie $R^d n^{j,d\mathrm{D}}_{\me}(\nu,R)$ for any $j\in\{1,\dots,d\}$, is a function of $\gamma$,
$\tilde \gamma$ and $\nu$ only.

\subsection{Beyond Gaussian statistics} \label{sec:non-gaussianity}
Let us finally compute the one-point statistics for weakly non-Gaussian fields.
Following \cite{Gay2012}, the Edgeworth expansion around a Gaussian kernel of the joint statistics of the field $x$ and its derivatives, ${ P}(x,x_i,x_{ij},x_{ijk})$ involves the hierarchy of cumulants and reads
\begin{equation}
{ P}(\vvec{x})= { P}_{\rm G}(\vvec{x})\left(1+ \sum_{n=3}^\infty \sigma_{0}^{n-2} \frac{\langle {\mathbf{H}}_n(\vvec{x})  \rangle}{\sigma_0^{2n-2}} \cdot \mathbf{H}_n(\vvec{x}) \right)\,,
\label{eq:defPDFNG}
\end{equation}
where $\vvec{x}=(x,x_i,x_{ij},x_{ijk})$, $\mathbf{H}_n$ is a vector of orthogonal polynomials\footnote{Not to be confused with the Hessian matrix $\mathrm{\mathbf{H}}$ used elsewhere in the paper} \wrt the Gaussian kernel ${ P}_{\rm G}$,  obeying
$\mathbf{H}_n=(-1)^n \partial^n { P}_{\rm G}/\partial \vvec{x}^n/{ P}_{\rm G} $. At tree level in perturbation theory \citep{Bernardeau2002}, $\langle \mathbf{H}_n(\vvec{x})  \rangle/\sigma_{0}^{2n-2}$ is independent of  the variance at redshift $z$, $\sigma_0(z)$, below $n=6$.
Cumulants such as $\langle x_1^2 x_{113} \rangle$ entering \cref{eq:defPDFNG} could, in the context of a given cosmological model,
involve a parametrisation of modified gravity (via \eg a parametrisation of the perturbation theory kernel $F_2(\vvec{k}_1,\vvec{k}_2)$), and/or primordial non-Gaussianities (via \eg the local non-Gaussianity parameter $f_{\rm NL}$), 
and  enable us to study the first stages of the non-linear evolution of the Universe under the action of gravity.
From this expansion, or relying on the connection between event ratio and
connectivity discussed in Appendix~\ref{sec:self},
we can for instance compute the non-Gaussian correction to the ratio of critical events, defined in \cref{eq:crit-event-ratio-th} as
\begin{equation}
\label{eq:kappa1NL}
 \frac{ r_\mathrm{{\cal F}\!/\!{\cal P}}}
 { r_\mathrm{{\cal F}\!/\!{\cal P},G}}=1\!+c_r\left(8\left\langle J_{1}^{3}\right\rangle\!-\!10\left\langle J_{1} J_{2}\right\rangle\!-\!21\left\langle  J_{1} q^{2}\right\rangle\right),
\end{equation}
where $c_r = \left(29 \sqrt{2}\!+\!12
   \sqrt{3}\right)\!/210/\!\sqrt{\pi}$, while
 $q^{2}=\sum_{i} x_{i}^{2}=|\nabla \delta|^{2}/\sigma_{1}^{2}$ is the modulus square of the gradient, and  $J_{1}$ and $J_{2}$ are defined  in \cref{eq:defJ1J2J3} via the trace and minor of the Hessian.
   These extended skewness parameters  are  isotropic moments of the underlying bispectrum which, when gravity drives the evolution,
     scale with $\sigma$ at tree  order in perturbation theory (\eg
   $\left\langle J_{1}^{3}\right\rangle/\sigma_0$ is independent of $\sigma_0$).
   The correction to one entering \cref{eq:kappa1NL} is negative (approximately equal to $-\sigma_0(1/7\!-\log(R)/5)$ for a $\Lambda$CDM spectrum smoothed over $R$ \si{Mpc\per\hred}),
   suggesting that gravitational clustering reduces the relative number of peak mergers
   compared to filament mergers.
When astronomers constrain the equation of state of dark energy
using the cosmic evolution of voids  disappearance, they effectively estimate
$\sigma$ (\emph{via} its dependence  in the cumulants) in \cref{eq:kappa1NL}.
Conversely, for primordial non-Gaussianities,
the extended skewness parameters from pure gravitational origin must be updated accordingly \citep[see][]{Gay2012,Codis2013}.
For instance, $\langle J_1 q^2 \rangle=\langle J_1 q^2 \rangle_{\rm grav}\!-\!2f_{\rm NL}\sqrt{1\!+\! f_{\rm NL}^2}/(1\!+\!4 f_{\rm NL}^2)$.

\begin{figure}
  \centering
  \includegraphics[width=\columnwidth]{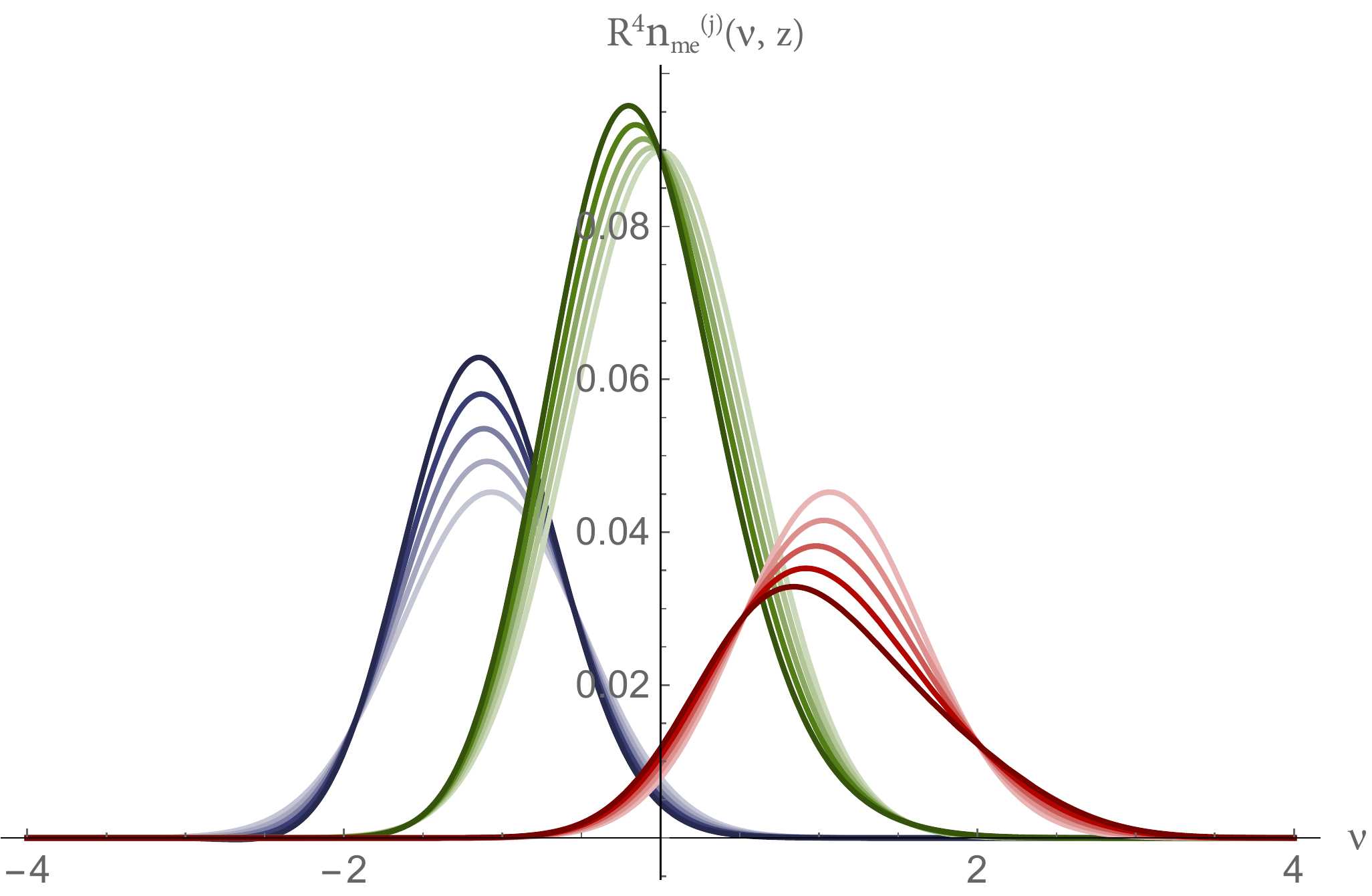}
  \caption{Predicted cosmic evolution
    of the product of extrema counts as a proxy for the event counts ($\cal W$ in blue, $\cal F$ in green and  $\cal P$ in red) for the variances
    $\sigma_{0}(z)=0,0.04,0.08,$ $0.12,0.16$ (from light to dark) and an underlying scale invariant power spectrum of index  $n=-1$.
    The $\cal F$ counts have been rescaled by a constant $205/332$ factor to better match the actual counts.
    The predicted trends with $\sigma_{0}$ are
    in qualitative agreement with the measured counts presented in \cref{fig:nbody-critevents}.
     }
  \label{fig:prediction-critevents-product}
\end{figure}

Since the  computation of the expectation \eqref{eq:eventcount} with the Edgeworth expansion~\eqref{eq:defPDFNG}
 is beyond the scope of this paper, let us investigate an alternative proxy for the event rate.
\Cref{fig:prediction-critevents-product} makes use of the perturbative prediction of \cite{Gay2012} to first order in $\sigma$
for the gravitationally-driven non-Gaussian   differential extrema counts  to
compute the product of such counts as a proxy for the events, namely
$n_{\rm me}^{\cal P}(\nu,z)\propto n_{\rm cp}^{\rm p}(\nu,z)\!\times\!n_{\rm cp}^{\rm f}(\nu,z)$,  
$n_{\rm me}^{\cal F}(\nu,z)\propto n_{\rm cp}^{\rm f}(\nu,z)\!\times\!n_{\rm cp}^{\rm w}(\nu,z)$
and
$n_{\rm me}^{\cal W}(\nu,z)\propto n_{\rm cp}^{\rm w}(\nu,z)\!\times\!n_{\rm cp}^{\rm v}(\nu,z)$.
This \emph{Ansatz} is reasonable, since for a merger to occur, two critical points of the same height must  exist beforehand.
 We  use the Gaussian PDF as a reference, to recalibrate the relative
amplitude of the filament to peak merger counts. Since \cite{Gay2012} provide fits to the critical point PDFs as a function of $\sigma_0$, it is straightforward to compute
their product.

From \cref{fig:prediction-critevents-product}, we see that
gravitational clustering shifts the peak event counts to lower contrast.
Less trivially, the filament merger rates also shift towards negative contrasts.
From these PDFs we can re-compute the cosmic evolution of the ratio of critical events
which appears to closely follow
$r_{\cal P\!/\!F}= 7/34(1-\sigma_0/7)$ (for $n=-1$), in good agreement
with \cref{eq:kappa1NL}, suggesting that this approximation indeed
captures the main features of gravitational  clustering.

\section{Theory: 2-pt statistics of critical events}
\label{sec:theory2pt}
Let us now present a method to compute  the two-point statistics of critical events.
Such statistics are of interest,  as they allow us 
to qualitatively understand the upcoming sequencing of processes of importance for 
galaxy formation, for example to study the cosmic evolution of the
connectivity of peaks, or to understand how large scale tides bias
mass accretion (the so-called assembly bias).  
\Cref{sec:clustering} presents the two-point statistics of merger events in 3D,
while \cref {sec:1Dcorrel} provides analytical approximations
assuming mergers occur along a straight filament.
\Cref{sec:condcluster} computes the conditional merger rates
subject to larger scale tides.
We match these predictions to simulations in \cref {sec:measurements}  below.

\subsection{Clustering of critical events in \texorpdfstring{$R,\vvec{r}$}{R, r} space}
\label{sec:clustering}

We cannot generally assume that the orientations of two critical events are aligned \wrt the separation vector, so the covariant condition for critical event of type $j\in\{{\cal P, F,W }\}$,
 ${\rm cond}_j$, is given by the argument of the expectation in \cref{eq:dndRmaintext} %
multiplied by a requirement on the sign of  the two non-zero eigenvalues.
For instance
\begin{multline}
 {\rm cond}_{\cal P}(\vvec{x}) \!=\! {J}\, \dirac^{(3)}(x_i) \dirac(H) \times \\
 \heaviside(\!-\!{\rm tr}( x_{ik}) ) \heaviside( {\rm tr}^2(x_{ik})  \!-\!{\rm tr} (x_{il} x_{lk}))
 \,, \notag
\end{multline}
where the two Heaviside conditions ensure that the trace is negative and the minor
positive so that the two eigenvalues are negative. Note that we use an implicit sum on repeated indices here.
From the joint two-point count of critical events, we can define  the relative clustering of critical events of kind $i,j$ smoothed at scales $(R_x, R_y)$ and located at positions $(\rr_x, \rr_y)$, $\xi_{ij}(\vvec{s})$  as
\begin{align}
 1+\xi_{ij}(\vvec{s}) =\frac{\langle  {\rm cond}_i(\vvec{x})\times {\rm cond}_j(\vvec{y}) \rangle}{\langle  {\rm cond}_i(\vvec{x}) \rangle\langle  {\rm cond}_j(\vvec{x}) \rangle}\,,
\label{eq:jointeventcount}
\end{align}
where $\vvec{x}=\{x,x_i,x_{ij},x_{ijk}\}$ (resp. $\vvec{y}$) is the set of fields at location $\rr_x$ (resp $\rr_y$), and
\begin{equation}
\vvec{s}\equiv \sqrt{2} \left(\frac{\vvec{r}_x-\vvec{r}_y}{\sqrt{R_x^2+ R_y^2}} \right)\,,
  \label{eq:separation-def}
\end{equation}
the event separation which we define as the spatial separation between the two points in units of the quadratic mean smoothing length.
We chose 
this definition as we expect the correlation lengths to be proportional to the smoothing scale, hence events at different scales can only be meaningfully stacked if distances are expressed in terms of the smoothing length. Because we focus on a Gaussian smoothing, it is natural to associate the two smoothing scales using a quadratic mean as the product of two Gaussian kernels with scales $R_x, R_y$ is equivalent to smoothing at a single scale $R=\sqrt{(R_x^2+R_y^2)/2}$.
Evaluating the expectation in \cref{eq:jointeventcount} requires full knowledge of the joint statistics of the field  ${ P}(\vvec{x},\vvec{y})$ (involving 40 variables, see \cref {sec:twptPDF}).

We rely on Monte-Carlo
methods in {\sc Mathematica} in order to evaluate numerically \cref{eq:jointeventcount}. Namely, we draw random
numbers from the conditional probability  that $\vvec{x}$ and $\vvec{y}$ satisfy
the joint PDF, subject to the condition that $x_j=y_j=0$, 
$x=\nu_1$ and $y=\nu_2$.
For each draw $(\vvec{x}^{(\alpha)},\vvec{y}^{(\alpha)})$, $\alpha=1,\dots,N$, we drop or keep the sample, depending on the type of critical event given by the signs of ${\rm tr}( x_{ij})$ and  ${\rm tr}^2(x_{ij})  \!-\!{\rm tr} (x_{il} x_{lj})$;
if it is kept, we evaluate ${J(\vvec{x})} \delta^{(\epsilon)}_{\rm D}\!\left(H(\vvec{x})\right) {J(\vvec{y})} \delta^{(\epsilon)}_{\rm D}\!\left(H(\vvec{y})\right) $ where
$\delta^{(\epsilon)}_{\rm D}$ is a normalised Gaussian of width $\epsilon$, which in the limit of  $\epsilon\rightarrow0$ would correspond to a Dirac function imposing here that the
two determinants are zero.
For small enough $\epsilon$, we then have
\begin{align}
\langle  &{\rm cond}_i(\vvec{x}){\rm cond}_j(\vvec{y}) \rangle
\! \approx\! \frac{ \!{ P}_m(x\!=\nu_1,\!y\!=\!\nu_2,x_{l}\!=\!y_{l}\!=\!0)\!}{N}
\\
&\!\sum_{k\in {\cal S}_{ij}} \!   {J(\vvec{x}^{(k)})} \delta^{(\epsilon)}_{\rm D}\left(H(\vvec{x}^{(k)})\right) {J(\vvec{y}^{(k)})} \delta^{(\epsilon)}_{\rm D}\left(H(\vvec{y}^{(k)})\right),
\nonumber
\end{align}
where $N$ is the total number of draws, $P_m$ the marginal probability for the field values and its gradients,  and ${\cal S}_{ij}$ is the subset of the indices of draws
satisfying the constraints $i,j$ on the Hessians.
The same procedure can be applied to evaluate the denominator of \cref{eq:jointeventcount}, which then yields an estimation of  $\xi_{ij}(s,\nu_1,\nu_2)$. This algorithm is embarrassingly
parallel.

\begin{figure}
\includegraphics[width=0.98\columnwidth]{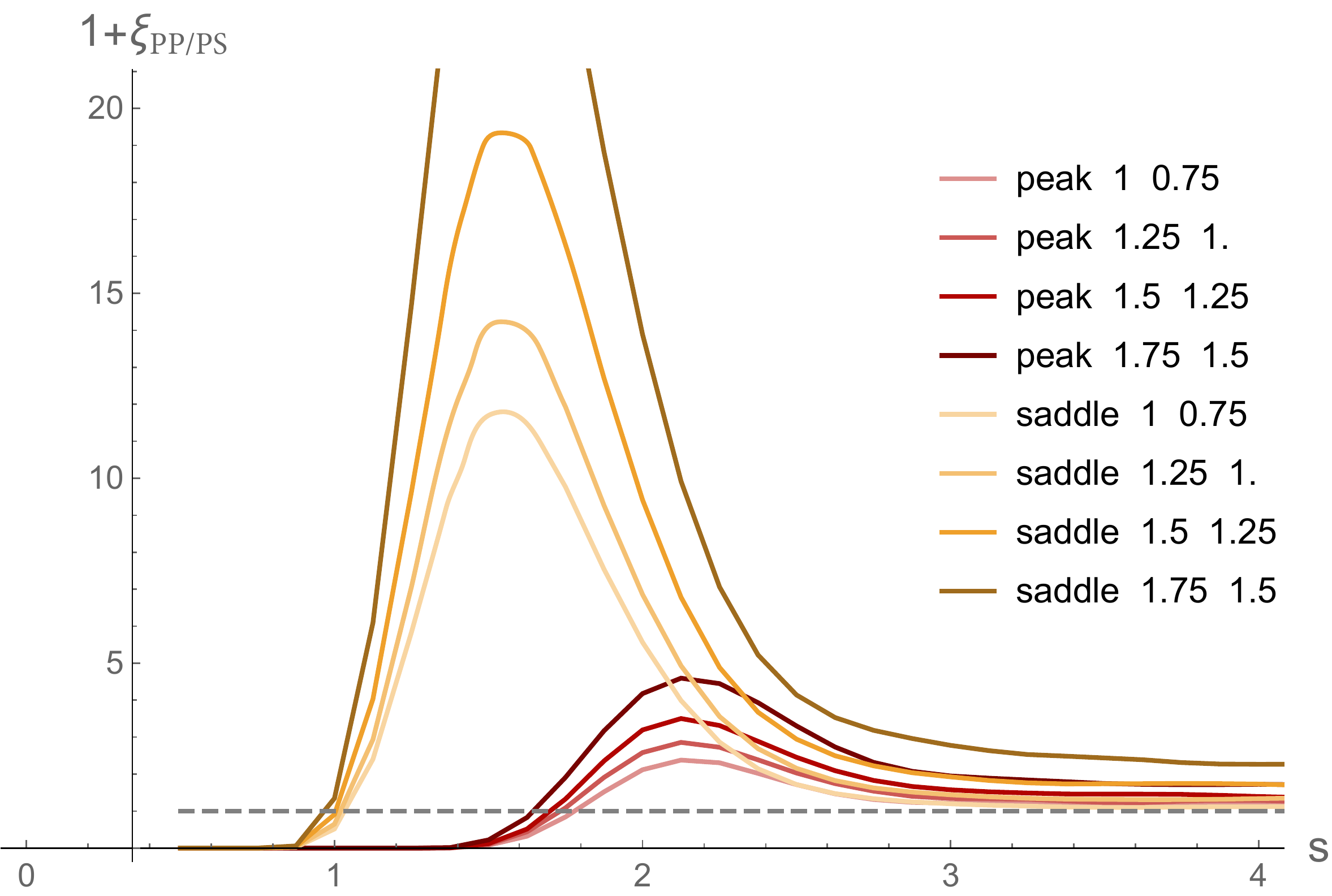}
\caption{
The auto-correlation of peak merger $\xi_{\cal P\!P}$  (in shades of red, as labelled in
terms of the height of the two critical points) and the cross-correlation of peak and filament merger $\xi_{\cal P\!F}$ (in shades of yellow, as labelled) as a function of separation $s$.
As expected, the saddle mergers are clustered closer to the higher peak compared to the
peak mergers.
\label{fig:cluster}}
\end{figure}

The result of the numerical integration is presented in \cref {fig:cluster}, which shows  the auto-correlation of peak merger $\xi_{\cal P\!P}$ on the one hand, and the cross-correlation of peak and filament merger $\xi_{\cal P\!F}$
on the other hand at fixed merger height, as labelled.
Here we used $\epsilon=0.002$.
Note that because \cref{eq:jointeventcount} is a ratio, the prefactors in the counts involving scales
all cancel out.

\begin{figure}
\center\includegraphics[width=0.9\columnwidth]{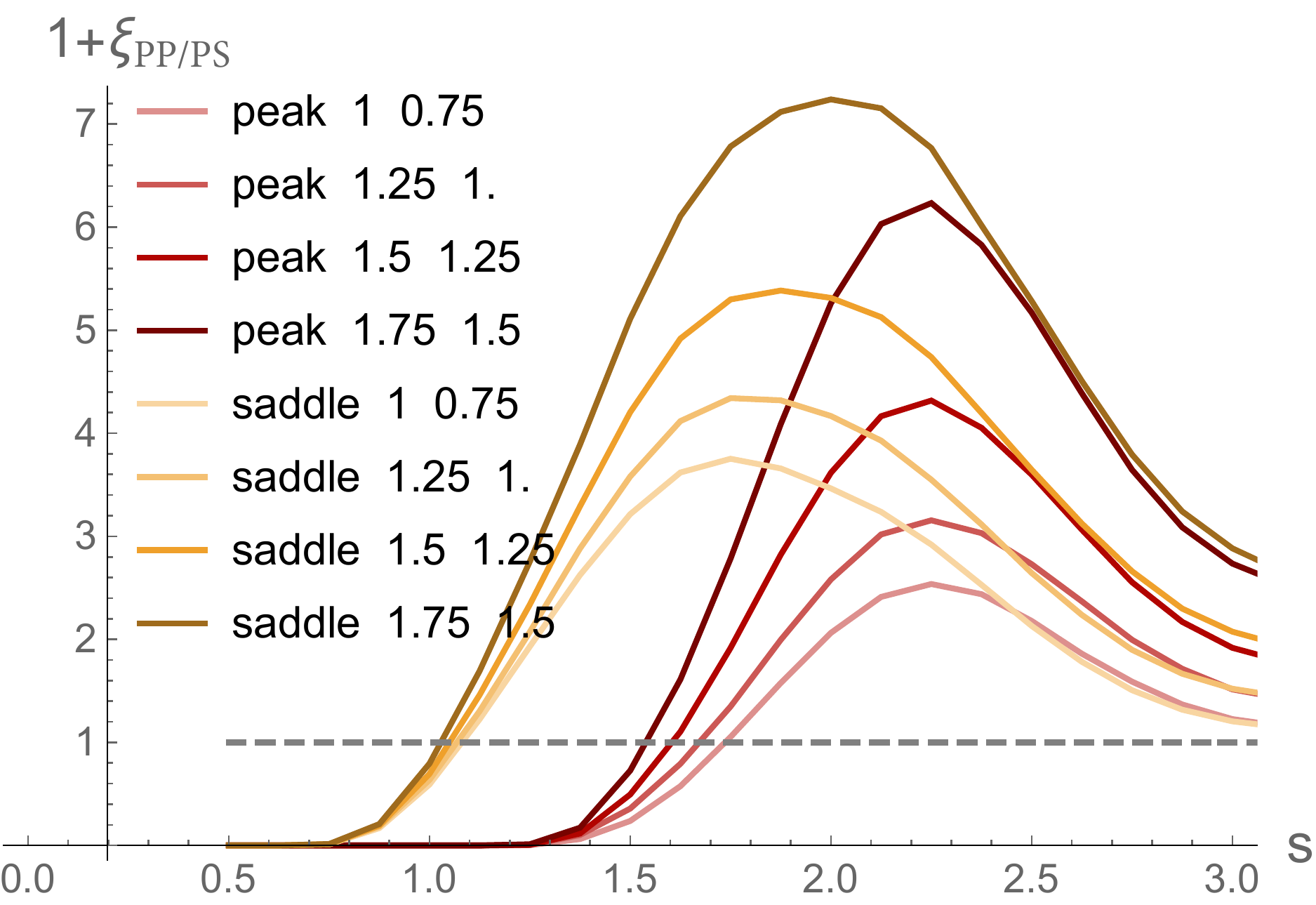}
\caption{
Same as \cref{fig:cluster} for the two-point correlation of events in 2D fields with scale invariant power spectrum of index $n_s=-1$. Next filament mergers will occur before 
 next peak mergers. The rarer the event the more delayed, and the higher the clustering amplitude.  
\label{fig:cluster-2D}}
\end{figure}

\subsection{Correlation of peak mergers along filament}
\label{sec:1Dcorrel}

Let us briefly present the two-point statistics of high density peak mergers while assuming for simplicity  that the mergers occur along the same (straight) filament (discussed in \cref{sec:definition-crit-events}),
as it is instructive and simpler. In this approximation we can resort to one dimensional statistics.
In the high density limit, we may drop the Heaviside constraint on the sign of the
eigenvalues since all high density critical points tend to be automatically maxima.
Then  the (1D) correlation function of peak mergers, $1+\xi_{\nu_1\nu_2}(s)$ of height $\nu_1$ and $\nu_2$  becomes
\begin{equation}
\frac{\langle  \dirac(\!x\!-\!\nu_1\!) x_{111}^2   \dirac(x_1)  \dirac(x_{11})\ \dirac(\!y\!-\!\nu_2\!) y_{111}^2  \dirac(y_1)
 \dirac(y_{11})
\rangle}{ \langle \dirac(\!x\!-\!\nu_1\!) x_{111}^2 \dirac(x_1)  \dirac(x_{11}) \rangle
              \langle   \dirac(\!y\!-\!\nu_2\!) y_{111}^2 \dirac(y_1)  \dirac(y_{11}) \rangle}\notag
\end{equation}
where the expectation   is over the Gaussian
PDF whose covariance for the field $(x,x_1,x_{11},x_{111},y,y_1,y_{11},y_{111})$ obeys
\begin{equation}
\left(
\begin{array}{cccccccc}
 1             & 0                & -\gamma       & 0                &  \gamma _{00} &  \gamma _{01}    &  \gamma _{02} &  \gamma _{03}    \\
 0             & 1                & 0             & -\tilde{\gamma } & -\gamma _{01} &  \gamma _{11}    &  \gamma _{12} &  \gamma _{13}    \\
 -\gamma       & 0                & 1             & 0                &  \gamma _{02} & -\gamma _{12}    &  \gamma _{22} &  \gamma _{23}    \\
 0             & -\tilde{\gamma } & 0             & 1                & -\gamma _{03} &  \gamma _{13}    & -\gamma _{23} &  \gamma _{33}    \\
  \gamma _{00} & -\gamma _{01}    &  \gamma _{02} & -\gamma _{03}    &  1            &  0               &  -\gamma      &  0               \\
  \gamma _{01} &  \gamma _{11}    & -\gamma _{12} &  \gamma _{13}    &  0            &  1               &  0            & -\tilde{\gamma } \\
  \gamma _{02} &  \gamma _{12}    &  \gamma _{22} & -\gamma _{23}    &  -\gamma      &  0               &  1            &  0               \\
  \gamma _{03} &  \gamma _{13}    &  \gamma _{23} &  \gamma _{33}    &  0            & -\tilde{\gamma } &  0            &  1               \\
\end{array}
\right)\,,
\end{equation} 
where for instance $\gamma_{02}(s)= \langle  x(\rr_x) y_{11}(\rr_y) \rangle$.
The dominant contribution in the large threshold $\nu_1,\nu_2\gg 1$, large separation
$s\gg 1$ regime reads
\begin{equation}
\xi^0_{\nu_1\nu_2}(s)=\frac{\nu _1 \nu _2   \left(\gamma _{00}(s)+\gamma  \left(2 \gamma _{02}(s)+\gamma  \gamma
   _{22}(s)\right)\right)}{\left(1-\gamma ^2\right)^2}\,,
\end{equation}
which as expected  scales like the underlying correlation, $\gamma _{00}(s)$, boosted
by the bias factor $\nu_1 \nu_2$ \citep{Kaiser1984}.
In that limit, the next order correction to the correlation function involving the third derivative of the field reads
\begin{equation}
 \xi^1_{\nu_1\nu_2}(s)=
\frac{2  \left(\tilde{\gamma }^2 \gamma _{11}(s)+2 \tilde{\gamma } \gamma _{13}(s)+\gamma
   _{33}(s)\right)^2}{\left(1-\tilde{\gamma }^2\right)^2}\,,
\end{equation}
where  $\tilde \gamma$-weighted linear combination of  the auto-correlation  of $\nabla \Delta \delta$
and the cross-correlation of  $\nabla \laplacian \delta$ and  $\nabla  \delta$ appear,  evaluated at events separated by $s$.
The assumption of successive mergers of peaks occurring along a straight filament is of course very simplified,  and prevents us from considering cross-correlations between
peak mergers and \eg filament mergers.

\subsection{Conditional merger rates in the vicinity of larger tides}
\label{sec:condcluster}
In the context of galaxy formation, it is of interest to quantify  conditional merger rates subject to  tides imposed by the large scale structure  to explain geographically the origin of assembly bias. To do so one must compute the conditional event counts, subject to a given large scale critical point at some distance $\vvec{s}$
from the running point $\vvec{x}(\vvec{r}_x)$.
The critical point can  be \eg a peak of a given geometry and height,
if one is concerned with  the impact of clusters on mergers trees of dark matter haloes in their
vicinity \citep{Hahn2009,Ramakrishnan2019}, or it could be a saddle point, as a proxy for a larger scale filament,
when studying how haloes growth stalls in such vicinity \citep{Borzyszkowski2017,Musso2018}.
In turn this involves the joint expectation
\begin{equation}
\langle  {\rm cond}_j(\vvec{x})\,\dirac(y_i) |\!\det y_{ij}|  \rangle\,.
\label{eq:eventcountcond}
\end{equation}

Evaluating \cref{eq:eventcountcond} requires the full knowledge of the joint statistics of the field at $\vvec{x}(\rr_x)$ and $\vvec{y}(\rr_{y})$, ${ P}(x,x_i,x_{ij},x_{ijk},y,y_i,y_{ij})$ (involving 30 variables).
The correlations of the PDF involves the covariance of the field and its derivatives computed at two smoothing scales, $R$ and $R_c$ corresponding to the proxy for the timeline of the haloes
 and the large scale structure respectively.
We can then marginalise over all variables, subject to \eg imposing the height, $\nu_c$ and shape, ${\mu}^c_i$ of the large scale critical point
\begin{equation}
\langle  {\rm cond}(\vvec{x})\dirac(y_i) |\!\det y_{ij}| \dirac(x-\nu)\dirac(y-\nu_c)\heaviside(-\lambda_i)  \dirac(\mu_i-\mu^c_i) \rangle \notag
\end{equation}
where $\lambda_i$ are the eigenvalues of $x_{ij}$ and $\mu_i$ are the eigenvalues of $y_{ij}$.
The conditions imposed by the mergers and the properties of the peaks and large scale environment reduce the number of integrals from
30 to 21.
\Cref{sec:generation} describes how to sample conditional event counts using constrained realisations of
Gaussian random fields.

\begin{figure}
  \centering
  \includegraphics[width=0.8\columnwidth]{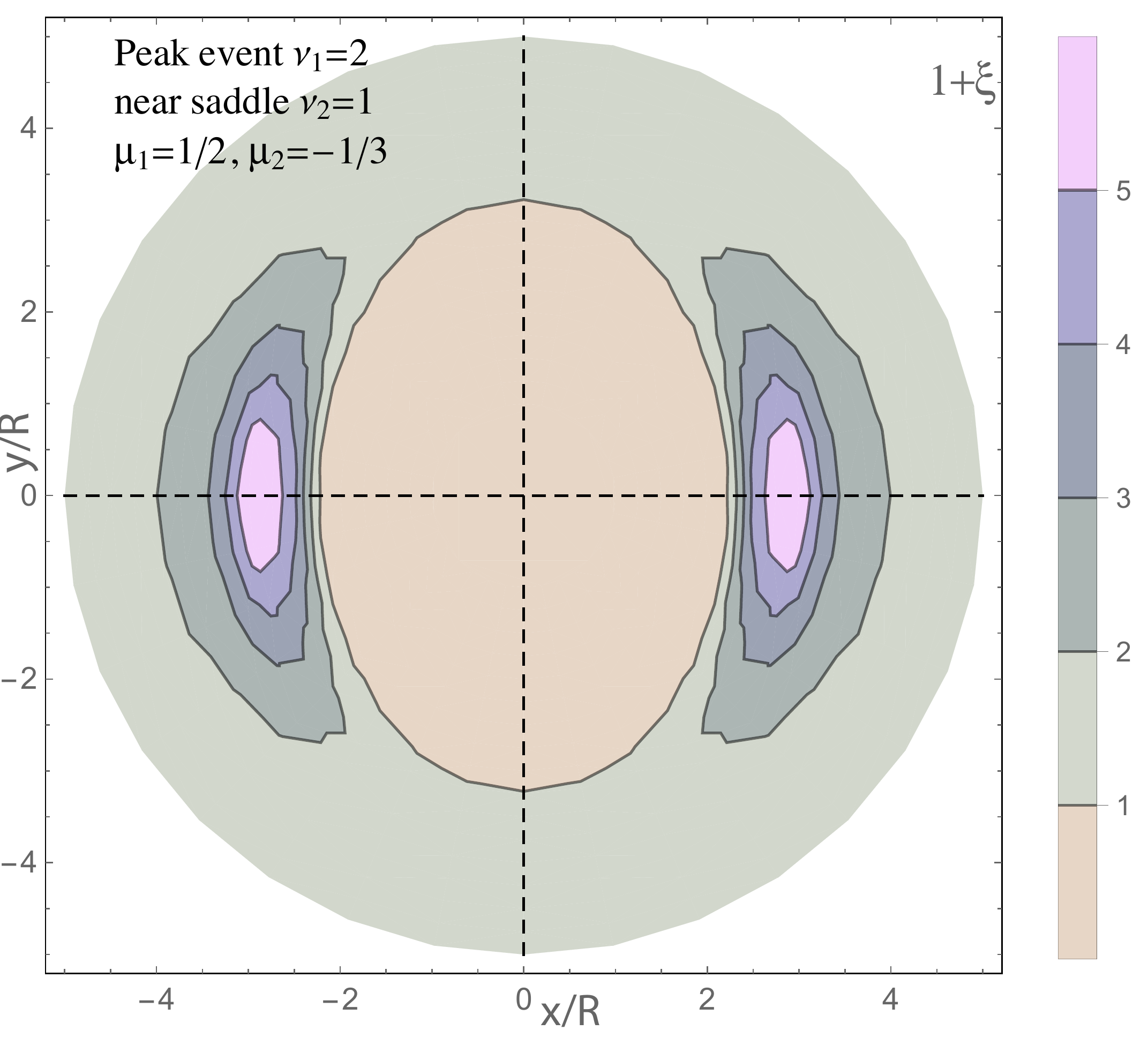}
  \includegraphics[width=0.8\columnwidth]{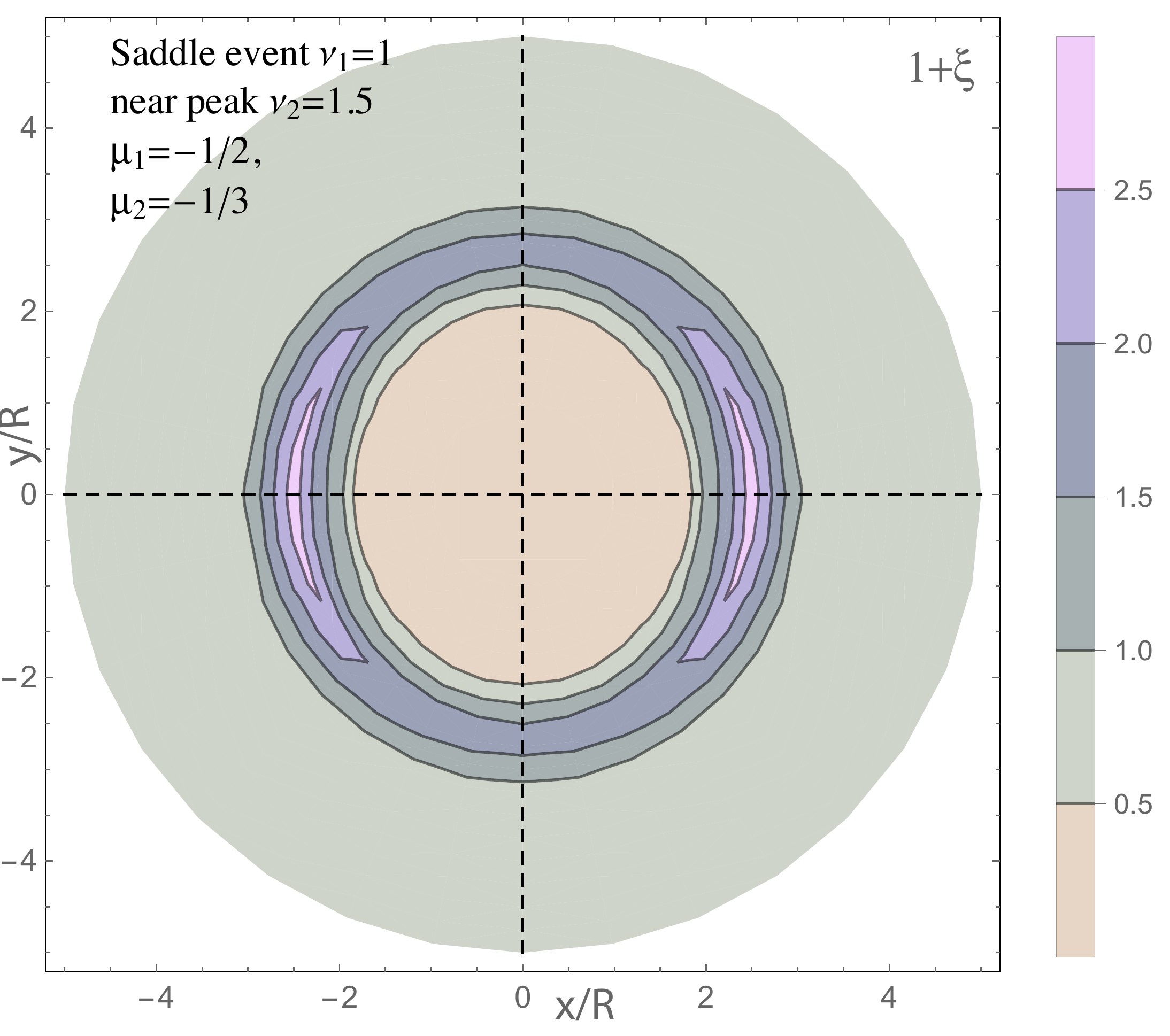}
  \caption{ Theoretical prediction for the 
  conditional excess probability, $1+\xi$  of peak merger events in the frame of a 2D critical point  at the origin as labelled. 
  The critical point defines a local exclusion zone whose geometry is set by its fixed eigenvalues.
    For simplicity, we have chosen $R_c=R$, while the underlying power spectrum index is $-1$.
See \protect\cref{fig:filament-crit-event-counts} for  {\sl measured} 3D counterparts.
  }
  \label{fig:conditional-saddle-2D}
\end{figure}

For the sake of simplicity, let us restrict computation to the conditional merger rates in 2D.
\Cref{fig:conditional-saddle-2D} presents  the excess probability of
having a  peak/filament merger  at some distance $r$ and orientation $\theta$ \wrt the frame
set by a given critical point.  Two configurations and types of events are considered.  As expected, the tides impact merger rates. 
While it is beyond the scope of this paper to explore systematically 
all possible geometries and relative heights, let us stress that such two-point functions 
are  physically very informative: for instance,  the bottom panel is an indication of the  early disappearance of filaments perpendicular to a wall  embedding a filament,
which seems qualitatively consistent with what is observed in $N$-body simulations. 

\section{Measurements for Gaussian random fields}
 \label{sec:measurements}
Let us validate the theory while counting
critical events within  realisations of  Gaussian random fields.
We then  bin them to estimate  their one and two-point statistics.
\subsection{Method}
\label{sec:measurements-method}
For each power-law power spectrum $P_k(k)=k^{n_\mathrm{s}}$, with spectral index $n_\mathrm{s} = -2, -1.5, -1, -0.5$, we have generated 250 Gaussian random fields.
We have also generated 400 Gaussian random fields with a $\Lambda$CDM power spectrum using \texttt{mpgrafic} \citep{prunetetal08} in a \cite{Planck2018} cosmology generated using the \cite{eisenstein_power_1999} fitting formula.
Each realisation will henceforth be called a `cube'.
Each cube has a size of $256^3$ pixels and a physical extent of \SI{100}{Mpc\per\hred}.\footnote{The box size is only relevant in the $\Lambda$CDM case, as the power-law cases are scale invariant.}
Each cube has been smoothed using a Gaussian filter with scale ranging from \SI{1}{Mpc\per\hred} to \SI{20}{Mpc\per\hred} (\SI{2.56}{px} to \SI{51.2}{px}).
The smoothing was operated in Fourier space, assuming periodic boundary conditions. At each scale, all critical points are detected (maxima, minima and saddle points) using the method detailed in \cref{sec:critical_points_detection}. The critical events are then detected by matching cubes of different smoothing scales using the method detailed in \cref{sec:critical_events_detection}.

Additionally, we have generated 200 $2048^2$ cubes with a power-law power spectrum with spectral index $n_\mathrm{s} = -1$ and a physical box size of \SI{1000}{Mpc\per\hred} which we smoothed with a Gaussian filter with scale ranging from \SI{1}{Mpc\per\hred} to \SI{20}{Mpc\per\hred}.

\subsection{Critical events counts}

In this section we present the number density of critical events measured in cubes with a power-law power spectrum and compare the theoretical predictions of \cref{sec:critical-event-versus height} to measurements in cubes.

We first measured the ratio of the number of critical events of different kind.
We found $r_{\cal F\!/\!P}=r_{\cal F\!/\!W} \approx 2.1$, regardless of the smoothing scale or the underlying power spectrum.
This excess of about 2\% in the ratio originates from a slight over-detection of saddle points with respect to local extrema.
Theory predicts this ratio to be $N_\mathrm{saddle}/N_\mathrm{peak}\approx 3.055$ in 3D \citep[see \eg\@][equation 2]{codis2018} while the measured value is $3.1$.
In the rest of the paper, we have corrected the excess number density of $\mathcal{F}$ critical events so that the number density ratio matches the prediction.

\begin{figure}
  \centering
  \includegraphics[width=1.0\columnwidth]{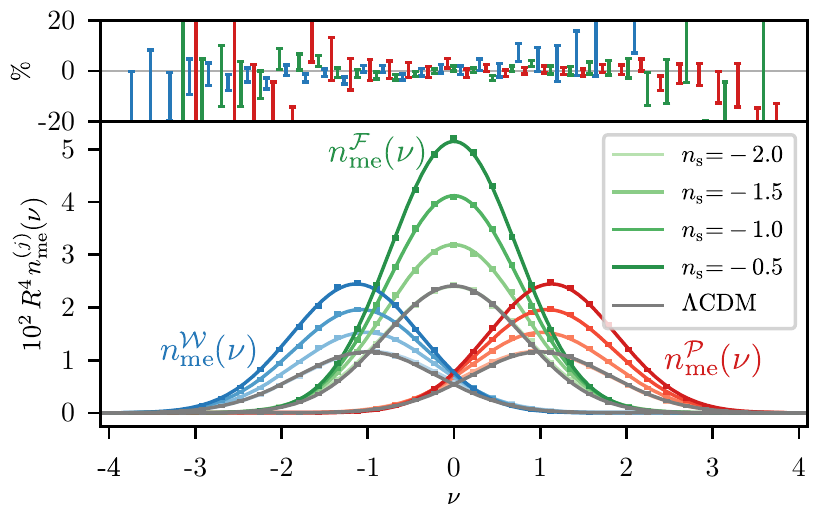}
  \caption{PDF of the critical events as a function of height in a scale invariant GRF (Gaussian Random Field) as labelled. The left bundle corresponds to wall mergers, the middle bundle to filament mergers and the right bundle to peak mergers.
  The solid curve corresponds to the theory while the error bars correspond to the error on the mean extracted from 160 simulations.
  The grey lines are the results obtained for a $\Lambda$CDM power spectrum initially smoothed over a scale of \SI{2.5}{Mpc\per\hred}.
  The top panel shows the residuals for $n_\mathrm{s}=-2$.
  The detection algorithm is still accurate in 3D.
  }
  \label{fig:crit-event-PDF-GRF}
\end{figure}

Let us now proceed to the number count at fixed density.
\Cref{fig:crit-event-PDF-GRF} shows the PDF of the critical events as a function of their height for different power-law spectra ($n_\mathrm{s}=-2,-1.5,-1,-0.5,\Lambda\mathrm{CDM}$).
The critical events have been selected at scale $\SI{2.35}{Mpc\per\hred} \leq R \leq \SI{3.01}{Mpc\per\hred}$ ($\SI{6.0}{px} \leq R \leq \SI{7.7}{px}$). The lower boundary ensures that the critical points are well separated\footnote{critical points are typically separated by $R_* \gtrsim 0.6 R$ (for $n_\mathrm{s}<0$), so $R=\SI{6}{px}$ gives a typical separation of \SI{3.6}{px} between critical points, which is larger than the number of points used to infer the curvature.}.
The upper boundary is fixed so that the smoothed cubes have consistent effective spectral parameters $\gamma_\mathrm{eff}(R)$ and $\tilde\gamma_\mathrm{eff}(R)$.
Indeed, the cubes have scale-dependent spectral parameters induced by the finiteness of the box and the discreteness of the grid \citep[see \eg\@][figure 5.1]{pierre_squelette_2011}.
Error bars have been estimated using a bootstrap method  on 400 sub-samples each made of 50 randomly chosen cubes.
Solid lines show the result of a fit of the theoretical formula to the cube data with free parameters $\hat{\gamma},\hat{\tilde{\gamma}}$.

The effective spectral index $\hat{n}_\mathrm{s}$ is fixed using $\gamma=\sqrt{(n_\mathrm{s}+3)/(n_\mathrm{s}+5)}$.
The measured values of $\gamma$ and $\tilde\gamma$ are consistent with the effective values measured directly in the cubes using \cref{eq:defgam}.
For example with $n_\mathrm{s}=-2$, the values measured in the cubes are $\gamma_\mathrm{eff}=0.62\pm 0.02, \tilde{\gamma}_\mathrm{eff} = 0.72\pm0.01$ (${n}_\mathrm{s,eff}=-1.75\pm0.13$) using \cref{eq:defgam}.
The mean values have been estimated with a sample of 100 cubes and the errors are the standard deviations of the sample.
The fitting procedure on the PDF of the critical events yields $\hat\gamma=0.621\pm0.002, \hat{\tilde{\gamma}}=0.724\pm0.003$ ($\hat{n}_\mathrm{s}=-1.75\pm 0.02$).
The relative difference between theory and measurements, presented on the upper panel of \cref{fig:crit-event-PDF-GRF}, shows no systematic deviation of the measurements and is within a few percent in the region where most of the events are.

\begin{figure}
  \centering
  \includegraphics[width=1.0\columnwidth]{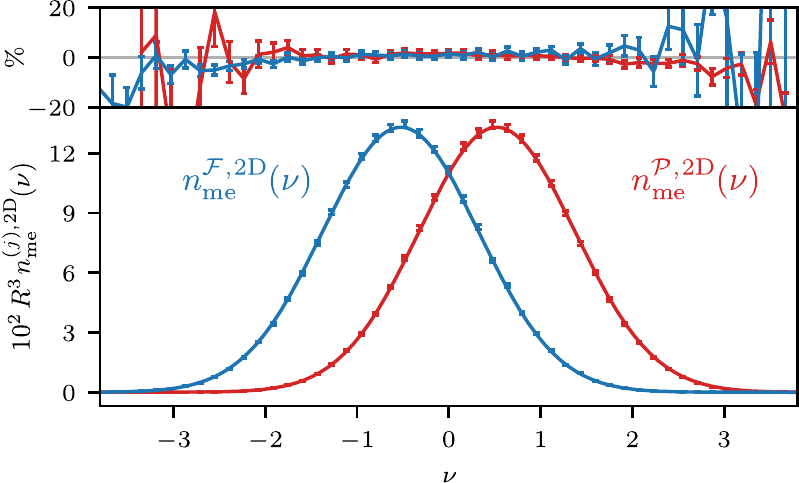}
  \caption{PDF of the critical events as a function of height in a scale invariant GRF in 2D with spectral index $n_\mathrm{s} = -1$.
    The left curve corresponds to filament mergers and the right curve to peak mergers.
  The solid curve correspond to the theory while the error bars correspond to the error on the mean extracted from 200 simulations.
    The top panel shows the residuals. The agreement between the analytic prediction
    and the measurements reflects the accuracy of the algorithm presented in
    \cref{sec:detection} in identifying critical events.
    }
  \label{fig:crit-event-PDF-GRF-2D}
\end{figure}

In order to further test the theoretical prediction, we have proceeded to the same analysis in the 2D case.
The results are presented in \cref{fig:crit-event-PDF-GRF-2D} and show that the agreement between theory and measurements is of the order of the percent.
Once again, no systematic deviation of the measurements is noted.
The results in 2+1 and 3+1D confirm the analytical formula derived in \cref{sec:critical-event-versus height} and illustrate the accuracy of the detection algorithm presented in \cref{sec:detection}.
Interestingly, since the algorithm has been designed to make no assumption on the number of dimensions, it is expected to work as well in $d$ dimensions.

\subsection{Two-point statistics}

Let us now estimate the two-point statistics of critical events using the critical events from the cubes presented above.
For each cube in the simulation, we select all critical events in a thick slice of smoothing scales ($\Delta R/R=0.3$).
The critical events are then split in two sub-samples, the first is selected at an overdensity $\nu=1$ with kind $j$ and the second at $\nu=0.7$ with kind $k$ ($j,k\in\{\mathcal{P,F,W}\}$).
The correlation functions are then computed from the number of pairs at distance $s=r/R$ in all cubes.%
The pair counting was done using a dual-tree algorithm, as described in \cite{moore_fast_2001}\footnote{See \href{https://docs.scipy.org/doc/scipy-0.19.1/reference/generated/scipy.spatial.cKDTree.count_neighbors.html}{the scipy doc} for more information.}.

\begin{figure}
  \centering
  \includegraphics{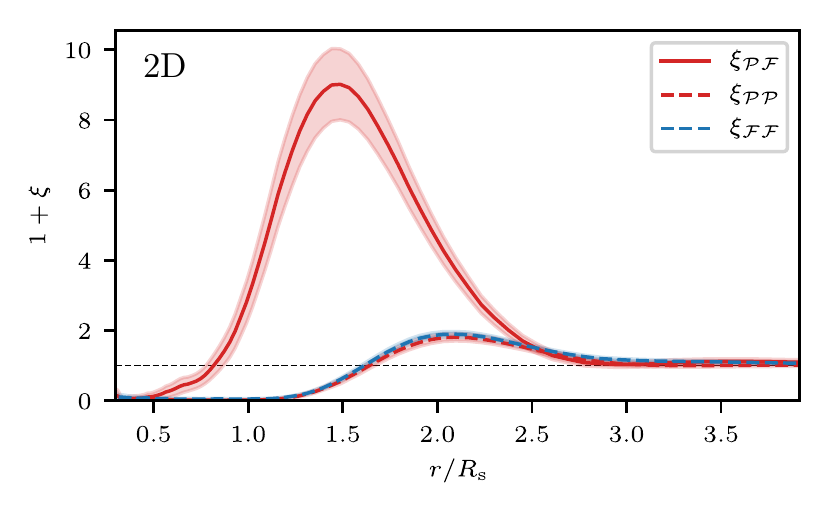} %
  \includegraphics{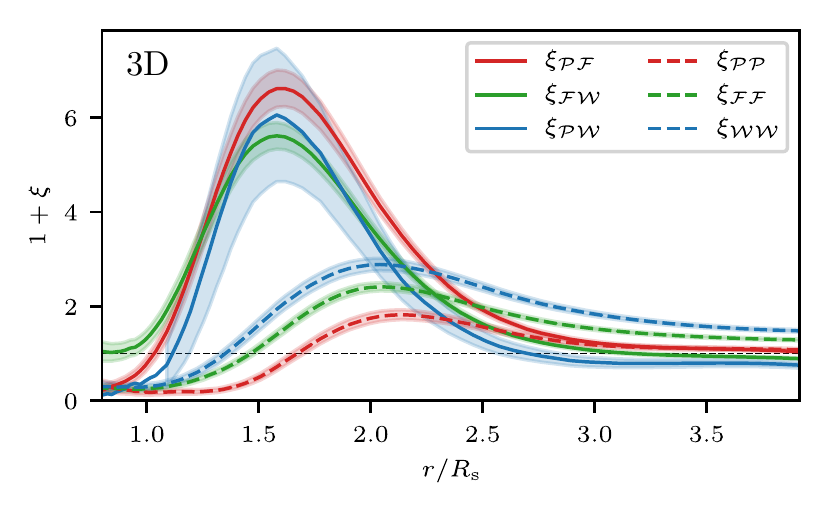}
  \caption{
    \emph{Top:} Correlation functions between critical events $\mathcal{P,F}$ in 2D at fixed smoothing scale for $n_\mathrm{s} = -1$.
    \emph{Bottom:} Correlation functions between critical events $\mathcal{P,F,W}$ in 3D at fixed smoothing scale for a $\Lambda$CDM power spectrum.
    Pairs of critical events have been selected at $\nu = 0.7$ and $\nu=1.0$.
  The correlation function of halo-merger with filament-merger, $\xi_\mathcal{PF}$, peaks at $r\sim 1.5R$ while the halo-merger auto-correlation functions $\xi_\mathcal{FF}$ peaks at $r\sim 2R$.
    This shows that halo-mergers are more likely to be followed by filament-mergers.
    The data have been filtered using a Savgol filter.
    Error bars have been estimated assuming a Poisson noise on the sample.
  }
  \label{fig:correlation-function}
\end{figure}

\Cref{fig:correlation-function} shows the measured correlation functions in 2D for a power law power spectrum with spectral index $n_\mathrm{s}=-1$ (top panel) and in 3D with a $\Lambda$CDM power spectrum smoothed at scales between $1$ and \SI{20}{Mpc\per\hred} (bottom panel).
In both cases the $\mathcal{PF}$ cross-correlation function (peak merger to filament merger correlation) peaks at $r\approx 1.5 R$ while the $\mathcal{PP}$ auto-correlation function (peak merger auto-correlation) peaks at $r\approx 2.1 R$.
This indicates that each halo merger is more likely to be followed by a filament merger compared to another halo merger.
Interestingly, peak mergers are also more likely to be followed by wall mergers.
Indeed, a halo merger induces a topological defect, as it leads to a resulting over-connected halo.
The defect is quickly corrected by a filament merger, decreasing the local connectivity of the halo back towards the cosmic average.
Doing so, another topological defect appears as a void becomes under-connected as one of its walls disappeared.
This last defect is then corrected by a last wall merger that makes the under-connected void disappear.
Note that, while the above sequence of critical events is a possible one, other sequences are possible that leave the connectivity conserved.
On average, critical events happen so that the local ratio of peak-to-filament, filament-to-wall and wall-to-void stays constant as smoothing increases, so that the global connectivity is preserved.
The link between critical events and global connectivity of the cosmic web is further discussed in \cref{sec:consistency-connectivity}.

\section{Applications and discussion}
\label{sec:discussion}

The scope of application of  the present formalism is obviously very wide.
Rather than attempting to cover it all, only a few examples will be
presented, while a more thorough investigation is left for future work.

In a cosmic framework, \cref{sec:Mzmerger} will first translate the one-point statistics presented in the previous section
into destruction rates as a function of mass and redshift.
\Cref{sec:consistency-connectivity} explains how mergers of filaments need to match that of haloes
in order to preserve the connectivity of peaks.
\Cref{sec:assembly-bias} explains how conditional merger counts in the vicinity of a filament explains
how the environment drives assembly bias.
\Cref{sec:NG} compares theoretical predictions of the destruction rates to results from $N$-body simulations and shows that the theory is able to reproduce the early non-linear stages of gravitational collapse.
Finally,  \cref{sec:other} presents an illustration of a correspondance between two critical events and mergers of walls and filaments in $N$-body simulations, 
 while
applications to other fields of research in cosmology (semi-analytical models, machine learning, intensity mapping) and beyond are discussed.

\subsection{Destruction rates as a function of mass and time}
\label{sec:Mzmerger}

The predictions in the initial Lagrangian space bear theoretical interest, yet they do not translate easily to measurable quantities.
In this section, let us show how one can map these predictions to observable quantities, and in particular destruction rates in  mass $M$, and redshift $z$, space.
Qualitatively, each critical event encodes a merger that involves three proto-structures (\eg two proto-haloes and their shared proto-filament).
In the rest of this section, we will show that it is possible to relate the mass and the destruction time of the disappearing structure\footnote{For halo mergers spotted by critical events, the disappearing halo is likely to be, but not necessarily, the less massive of the two proto-haloes.} to the density and smoothing scale of the field at the same location.

Together with the results of \cref{sec:ce_by_type}, one can then compute the destruction rates at different epochs for different object masses.
When dealing with void mergers, we will in this section use the dual interpretation of critical events from the point of view of the low density objects (see \cref{fig:scheme_critical_events_3D_dual}).
One can use the spherical collapse model to establish a mapping between collapse time of spherical regions and their initial overdensity -- high overdensity regions collapse earlier in the history of the Universe than lower densities.
At the same time, larger overdensities enclose more mass and will hence give birth to more massive structures.
These relations mathematically read
\begin{equation}
  \nu_\TH(R) = \frac{\deltac}{\sigma_\TH(R)D(z)}, \quad
  M = \frac{4\pi}{3}\bar{\rho}R^3,
  \label{eq:spherical-collapse-eq}
\end{equation}
where $\sigma_\TH(R)$ is the variance of the field smoothed by a Top-Hat filter on scale $R$, $\deltac = 1.69$ is the spherical collapse critical overdensity, $D(z)$ is the linear matter growth function and $\bar{\rho}$ is the mean matter density of the Universe.
The spherical collapse threshold can also be adapted to study the formation of voids \citep{sheth_hierarchy_2004,jennings_abundance_2013} with $\delta_\mathrm{v}=-2.7$.
Note that this simple relation holds in principle for small enough voids only ($R\lessapprox \SI{3}{Mpc\per\hred}$).

From a theoretical perspective, the action of smoothing the density field $\delta$ enables to probe the time-evolution of spherical proto-haloes by following the density evolution of peaks as the smoothing scale increases.
In order to match the results of \cref{eq:spherical-collapse-eq} with a Gaussian filter, 
one needs to establish a mapping of the smoothing scales between Top-Hat filtering and Gaussian filtering.
This can be achieved by matching the variance of the field smoothed with a Gaussian filter $\sigma_\mathrm{G}(R/\alpha) = \sigma_\mathrm{TH}(R)$, although different approaches have been used\footnote{Possible prescriptions include matching $\langle\delta_\TH\delta_\mathrm{G}\rangle = \sigma^2_{\TH}$ or matching masses $M_\mathrm{G}=M_\TH$.}.
Without loss of generality, \cref{eq:spherical-collapse-eq} becomes for a Gaussian filter and a prescription for the value of $\alpha$
\begin{equation}
  M = \frac{4\pi}{3}\bar{\rho} (\alpha R)^3.
  \label{eq:spherical-collapse-gaussian}
\end{equation}
This means that the volume associated to a Gaussian filter is equivalent to the volume associated with a Top-Hat filter (a sphere) with an effective size $\alpha$ times larger.

It is now straightforward to change variable from $R$ to $M$ and from $\nu$ to $z$ using the spherical collapse condition with a Gaussian filter (equations~\ref{eq:spherical-collapse-eq} and~\ref{eq:spherical-collapse-gaussian}),
so that for condition $c$ (peak or void)\footnote{since $\dv*{D}{z}\!=\! - D f /(1\!+\!z)$ with $f\equiv \dv*{\log D}{\log a} \sim \Omega_\mathrm{m}^{0.6}$.} the destruction rate reads
\begin{align}
\pdv{n}{\log M}{z}\Big|_{c}
    &=n_\me^{(c)}(R,\nu)\pdv{R}{\log M} \left|\pdv{\nu}{z}\right|\,,\label{eq:condmerger}\\
    &=-n_\me^{(c)}(R, \nu)\frac{|\deltac|}{3 \alpha \sigma(R) D(z)^2} \dv{D}{z} \left(\frac{3 M}{4\pi\bar\rho}\right)^{1/3},\notag
\end{align}
where $\alpha\approx 2.1$ and $\bar\rho\approx 2.8\times 10^{11}\,h^2\mathrm{M}_\odot/\mathrm{Mpc}^3\, \Omega_\mathrm{m}$ \citep[see \eg][Table A1]{Musso2018}.
From \cref{eq:nmerger_nu,eq:condmerger}, we can now count explicitly how many peaks and voids of a certain mass or within some mass range are destroyed early or late in the accretion history, \emph{via} straightforward integration.

\begin{figure*}
  \centering
    \resizebox{\textwidth}{!}{%
      \includegraphics[height=3cm]{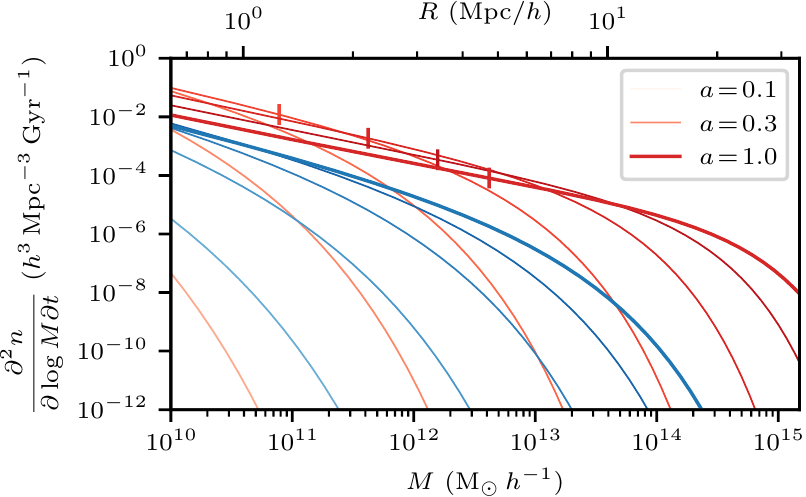}
      \includegraphics[height=2.7cm]{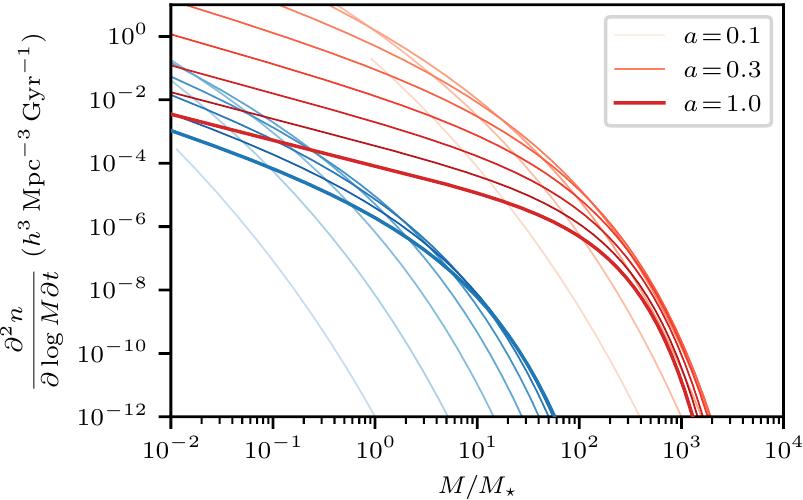}
    }
  \caption{
    Destruction rates of haloes (red lines) and voids (blue lines) from expansion factor $0.1$ (light colour) to $1.0$ (dark colour), linearly spaced, in a $\Lambda$CDM Universe as a function of object mass (\emph{left panel}) and as a function of mass relative to the non-linear mass (\emph{right panel}).
  }
  \label{fig:fig-merger-vz-mass}
\end{figure*}

\Cref{fig:fig-merger-vz-mass} shows the destruction rate of peaks and voids as a function of the object mass.
The cosmology-dependent terms of \cref{eq:condmerger} ($D(z)$, $\dv*{D}{z}$ and $\sigma$) have been computed using the code {\sc Colossus} \citep{diemer_colossus_2018} in a $\Lambda$CDM cosmology.
The power spectrum has been computed using the fitting formulas of \cite{eisenstein_BaryonicFeaturesMatter_1998}.
In order to evaluate the number density of critical events (the $n_\me$ term), we have assumed a scale-dependent equivalent power-law power spectrum\footnote{At each scale, the equivalent power-law power spectrum is given by the formula $n_\mathrm{s,eq}(R)=-3\!-\!2\dv*{\log\sigma}{\log R}$, where $\sigma$ is computed using a $\Lambda$CDM power spectrum.}.
The figure shows that for both peaks and voids, there is a cut-off mass scale above which objects are not destroyed any more.

The high-mass cut-off comes from the exponential cut-off of high $|\nu|$ objects, which suppresses massive objects (high $R$) at high redshifts.
Due to the dependence of the destruction rate to the effective spectral index of the power spectrum as well as $\sigma(R)$, the destruction rates show significant redshift evolution.
This is particularly emphasized on \cref{fig:fig-merger-vz-mass}, right panel which shows the evolution of the destruction rate with mass at different redshifts.
The evolution with redshift both depends on the rarity of the object, as encoded by $\nu$ but also on the local shape of the power spectrum, as encoded by the equivalent spectral index $n_\mathrm{s,eq}$.

Quantitatively, it should be noted that the mass scale of the cut-off and the precise value of the merger rate-mass relation will be subject to the same uncertainty in the value of $\deltac$ that also affects the halo mass function \citep{RobertsonEtal2008,Ludlow2011}.
The focus of this section is anyway to rephrase the critical event theory in astrophysical variables: the implementation of realistic merger tree models is left for future work.
It should also in principle be possible to generalise \cref{eq:condmerger} to filament mergers, but this would require the knowledge of a relation between the initial overdensity (or any other functional of the initial overdensity field) and the mass of the filament or its length, as well as a collapse condition.
\cite{ShenEtal2006} and \cite{Pogosyanetal1998} suggested this could be achieved using a spherical collapse criterion with a critical overdensity smaller than $\deltac$.

The impact of our results on filament merger rates in $M,z$ space will be done in a follow-up work.
Beyond the scope of this paper, those results could also be re-expressed in terms of the surviving structure and take into account the two objects' mass ratio, so that they can be compared to merger ratios measured in numerical simulations \citep[\eg][]{genel_halo_2009,fakhouri_merger_2010,rodriguez-gomez_merger_2015}.

\subsection{Consistency with cosmic connectivity evolution}
\label{sec:consistency-connectivity}

The properties of the initial random  field
was shown  by \cite{codis2018} to control to a large extent
the connectivity of dark matter haloes, as defined by the number of connected filaments (locally and globally)
at a given cosmic time. The upshot of this work is that the packing of peaks (\ie the `volume' they occupy, as imposed by their exclusion zone)  and saddles implies that 3-4 filaments typically dominate locally. Interestingly, the rate of filament  disappearing must match the peak merger rate, in order to preserve this number.
Beyond numerology, this rate is important because filaments  feed coherently dark matter haloes,
so their lifespan matters to understand the balance between filamentary cold gas inflow (from subsisting filaments) and environmentally-driven disruptions (from filament mergers).

Our qualitative understanding of the critical structure of Gaussian random fields remains  in close relation to packaging: each vicinity of a critical point, and with the same argument, of a critical event, must by continuity occupy a certain volume of space, as set by its eigenvalues, which puts constraints on the position of other points in the vicinity.
Indeed, critical points are found where the gradient vanishes, with some local curvature, so that the field is quadratic in each eigenvector's direction.
As a consequence, the gradient of the field is linear at non-null separation and cannot vanish, so that no other critical point can be found in the direct vicinity of another critical point or event.
At large separations the field decorrelates from its values at the critical point, so that another critical point event becomes likely.
In other words, before connecting a given peak
 to a peak of a different height, the field
must first go through a local saddle point along the ridge, which distance
is set by the `width' of that peak. 

The same reasoning applies to critical events, except that the field has a  specific third order behaviour along the ridge defined by the eigendirection of the vanishing eigenvalue (it is an inflection point in that direction).
For critical events, the process of smoothing the field will impact both the local
curvature but also the curvature of all other critical points. 
Hence, it is expected
that  smoothing will also  disconnect  neighbouring peaks as mergers occur:
the ridges are smoothed out because technically their saddle points  vanish.

\begin{figure}
\center\includegraphics[width=0.7\columnwidth]{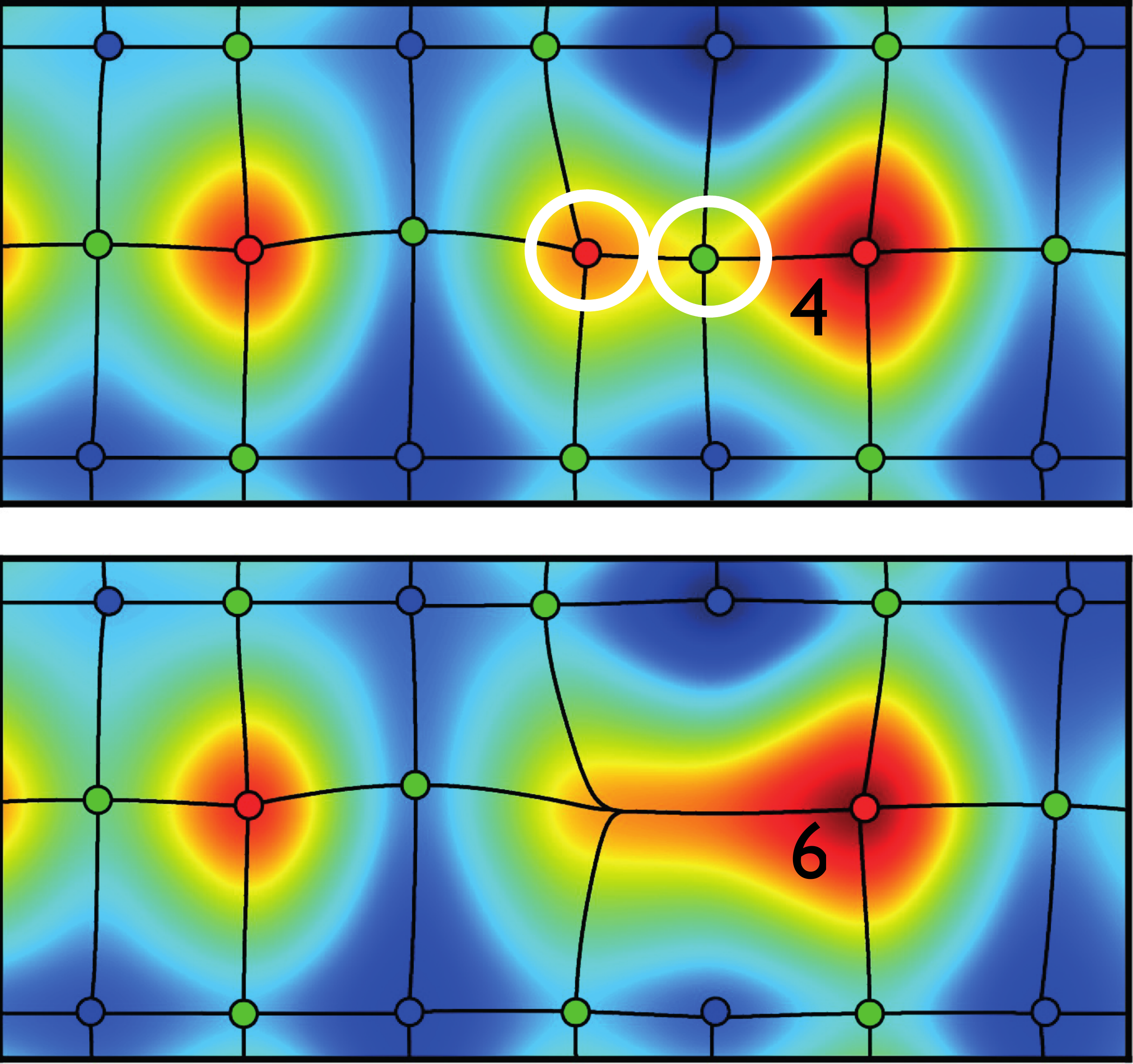}
\includegraphics[width=0.98\columnwidth]{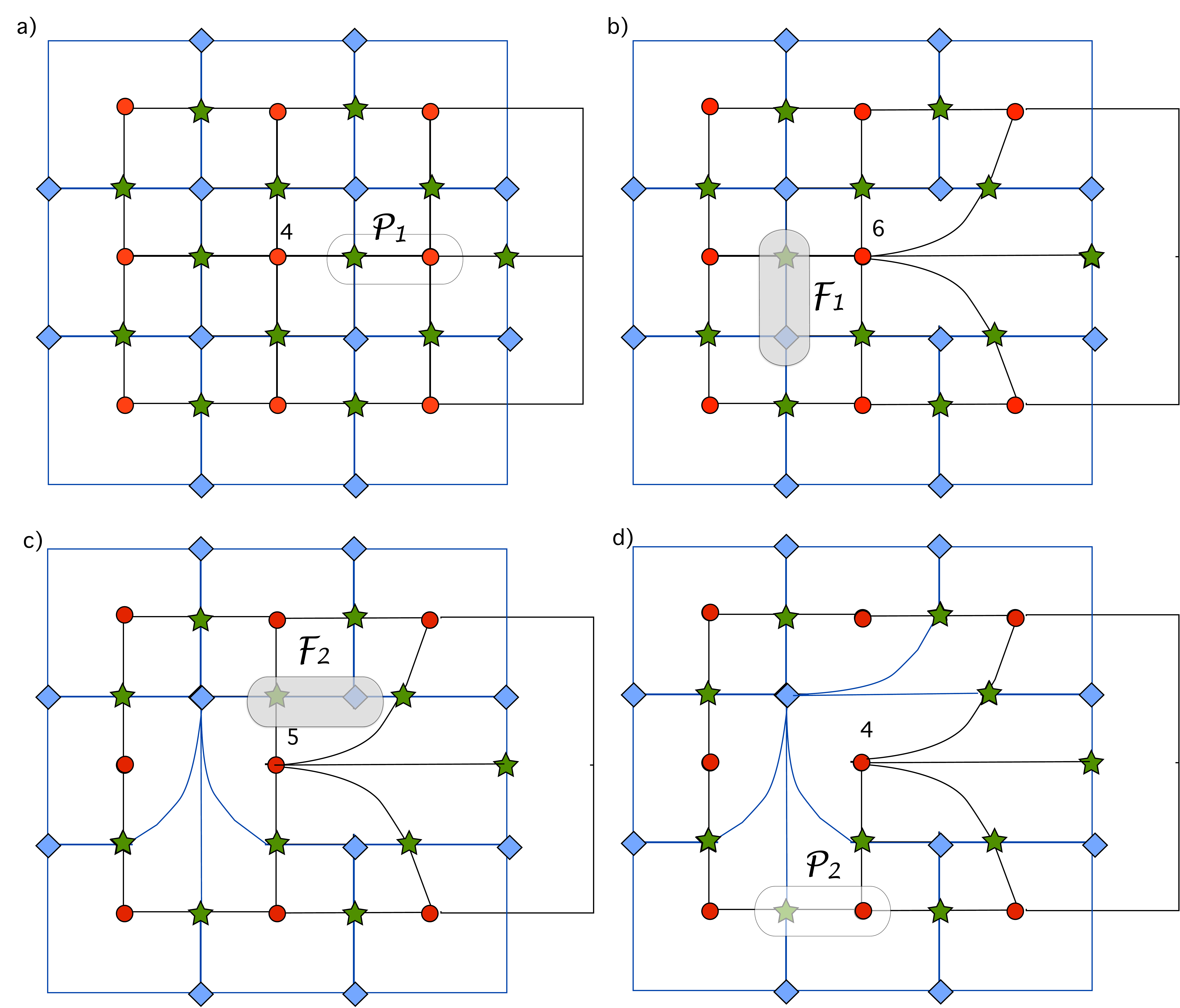}
\caption{
 {\sl Top}:
 Snapshots of the density field in 2D at two smoothing scales (colour coded from blue, low density to red high density).
 The black lines represent density ridges/troughs connecting the red peaks, and the blue voids via the green saddle points.
As the two low persistence pair
of peaks  (in white) merge the connectivity increases from 4 to 6 (as labelled).
The fate of this connectivity now depends on the nature and location of the next merger events
  \protect\cite[inspired from][]{Sousbie2011}.
 {\sl Bottom}: As labelled from a) to d) an abstraction of the merger sequence of a 2D `cosmic crystal' impacting the connectivity of
the central peak. Ridges are shown in black while troughs are shown in dark blue.
 The red circles represent the peaks, the green stars the saddles and the blue diamonds the voids.
A {${\cal P}_1$} merger (highlighted in light grey)  rises the mean connectivity of the central peak from 4 to 6,
but the next two {${\cal F}_{1,2}$} mergers   (highlighted in darker grey)  lower it back to 4.
The next {${\cal P}_2$} merger (panel d) will reduce the void's connectivity.
A more realistic representation of this process is also visible in \cref{fig:timesequence}.
\label{fig:connectix-2D}}
\end{figure}

\begin{figure}
\center\includegraphics[width=1\columnwidth]{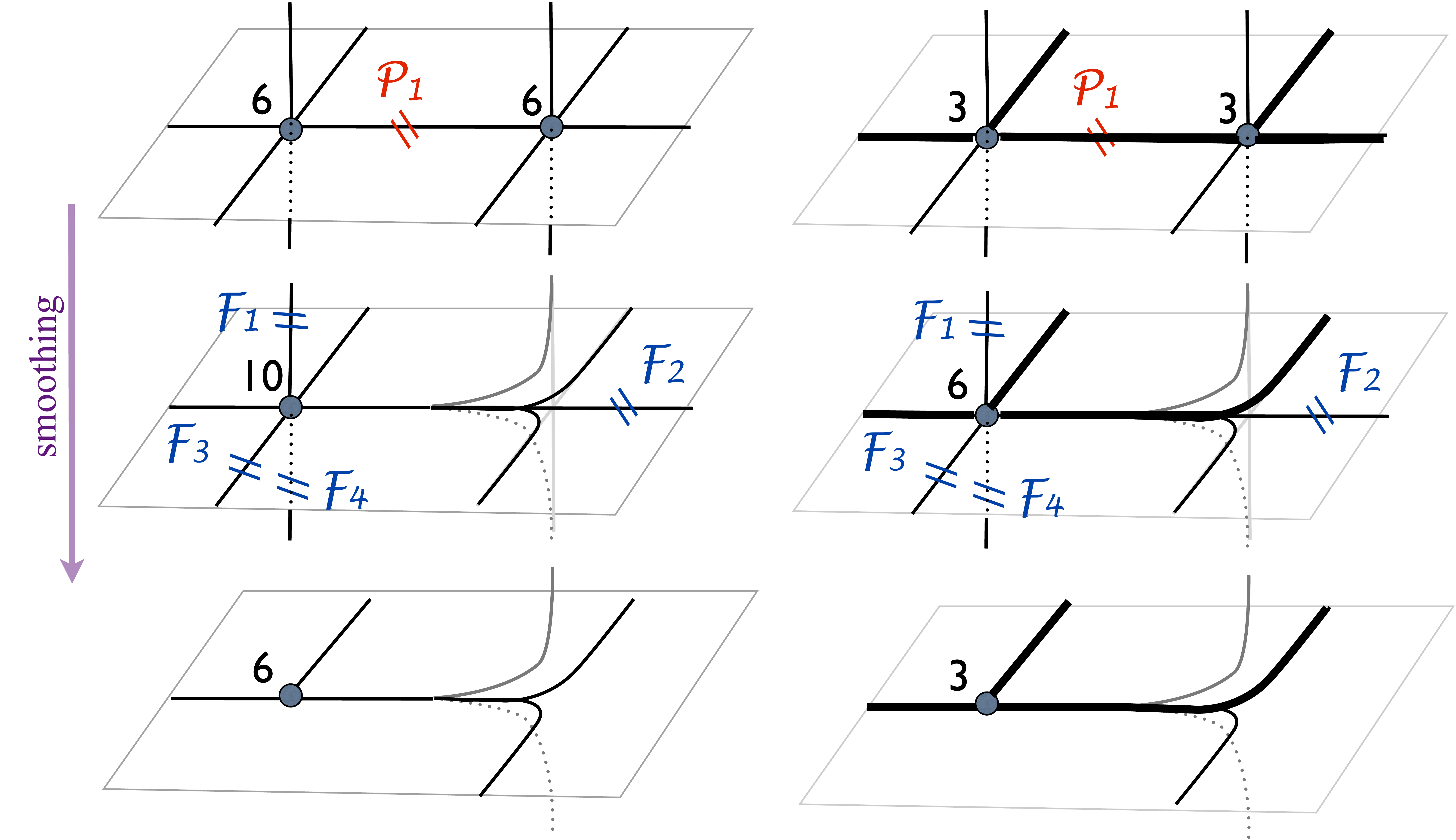}
\caption{
  Following the cartoon shown in \cref{fig:connectix-2D}, the left panel shows a smoothing sequence (from top to bottom) which would preserve the connectivity of a 3D peak.
It requires that each ${\cal P}$ merger should be followed by four  ${\cal F}$ mergers in the vicinity.
The right panel highlights how the multiplicity is preserved if one starts with 3 dominant  co-planar filaments.
\label{fig:connectix-3D}}
\end{figure}

We can  quantify this process via the two-point function of these events.
From the auto- and cross-correlations of the $\mathcal{P}$ and $\mathcal{F}$ events presented in \cref{sec:theory2pt},  we can define
the ratio of the separation at the maximum of these two correlations ($s_{ij}={\rm argmax}_s \xi_{ij}(s)$)   as a measure of the relative `proximity' of the two events.
Since this ratio $s_{\cal P\!F}/s_{\cal P\!P}\approx 3/4$ is smaller than one (see \Cref{fig:correlation-function}), it means that filament mergers are more clustered around halo mergers than halo mergers around halo mergers, so that the rate at which filaments disappear matches the merger rate and the typical number of filaments per halo remains constant through cosmic time.
As a result of this spatial clustering, the most likely sequence happening is a $\cal P\!F\!F\!P$ in 2D (one halo merger, followed by two filament mergers, followed by a halo merger), as presented on the cartoon of \cref{fig:connectix-2D}.
This sequence conserves the connectivity of peaks, and is consistent with the relative rates of events.
\Cref{fig:connectix-2D} illustrates an analogous consistent ${\cal P\!F}^4\!{\cal P}$ (one halo merger, followed by four filament mergers, followed by a halo merger) sequence in 3D.
\Cref{fig:connectix-3D} shows how the local connectivity of 3 can also be preserved, as the weaker filaments
typically lie off the main plane.

Finally, the clustering of filament disappearance impacts the connectivity of peaks as they merge as discussed in the next section (see \cref{fig:filament-crit-event-counts}, bottom right panel). This is a direct consequence of the clustering of events of the various types.

\subsection{Assembly bias in the frame of filaments}
\label{sec:assembly-bias}
\begin{figure*}
  \centering
  \begin{minipage}{.5\textwidth}
    \centering
    {\footnotesize Halo mergers ($\mathcal{P}$ events)}
    \includegraphics[width=\columnwidth]{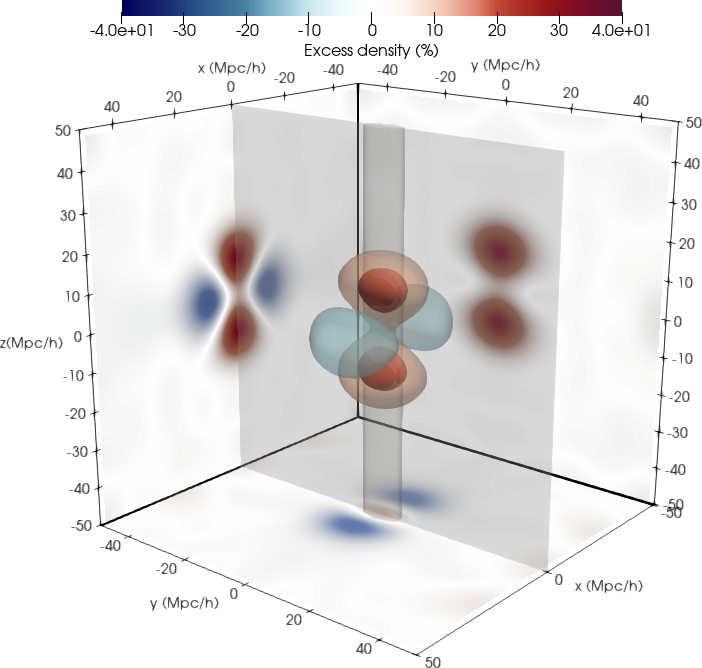}\\
  \end{minipage}%
  \begin{minipage}{.5\textwidth}
    \centering
    {\footnotesize Filament mergers ($\mathcal{F}$ events)}
    \includegraphics[width=\columnwidth]{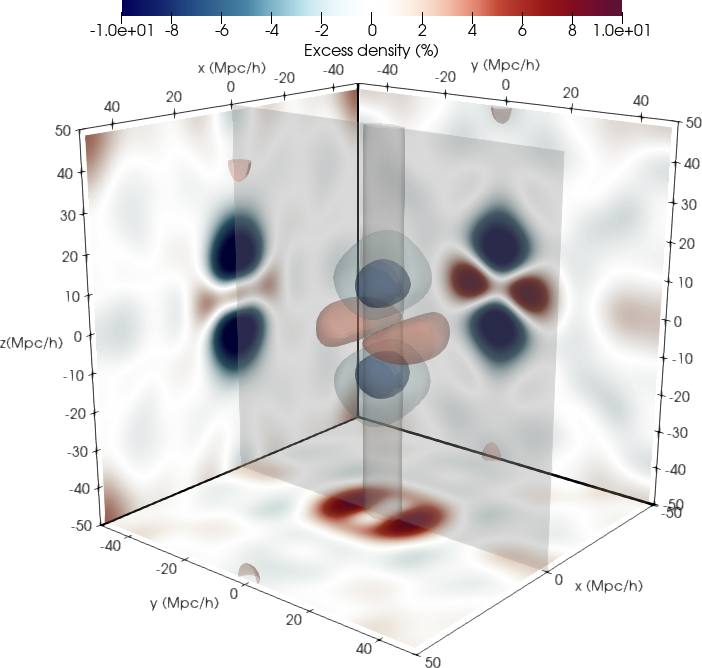}
  \end{minipage}\\
  \vspace{1em}
  \begin{minipage}{.5\textwidth}
    \centering
    {\footnotesize Wall mergers ($\mathcal{W}$ events)}
    \includegraphics[width=\columnwidth]{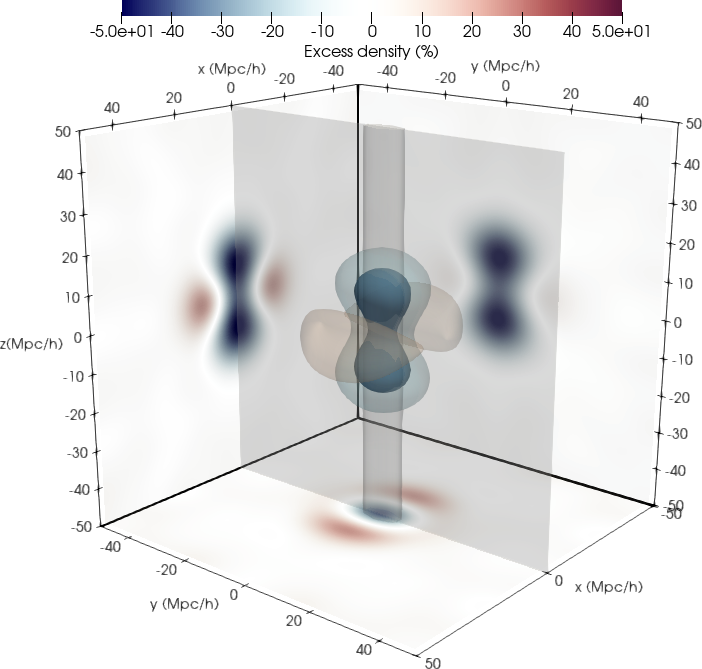}
  \end{minipage}%
  \begin{minipage}{.5\textwidth}
    \centering
    {\footnotesize Filament merger to peak merger $\mathcal{F/P}$ ratio}
    \includegraphics[width=\columnwidth]{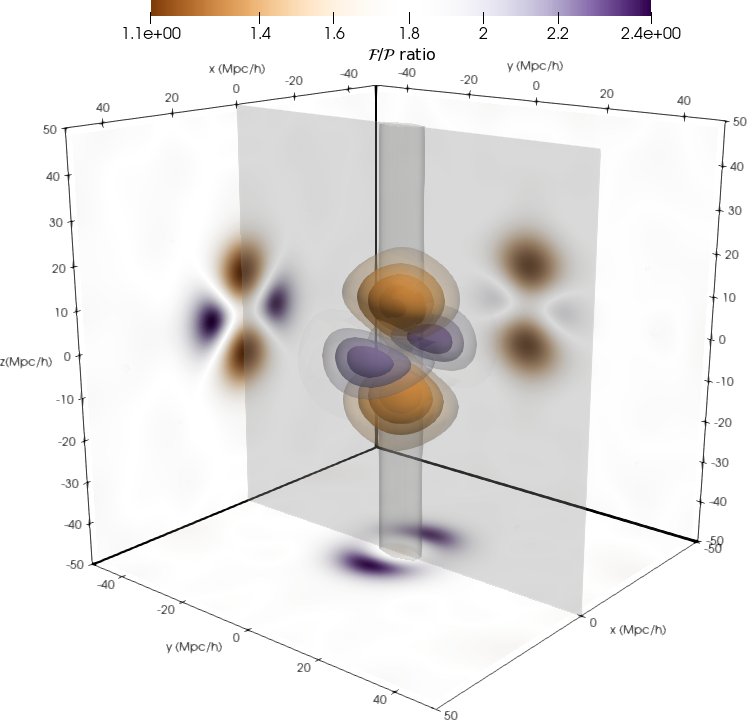}
  \end{minipage}%
  \caption{
    From left to right and top to bottom, peak-merger, filament-merger and wall-merger excess density around a large-scale proto-filament, illustrated by the vertical cylinder ($z$ direction) and the wall in which it resides, illustrated by the grey plane ($yz$ plane).
    The bottom right panel shows the local ratio of filament to peak mergers $r_\mathcal{F\!/\!P}$.
    Each side of the cube shows a slice through the centre, shifted to the side of the plot for visualisation purposes.
    Red regions have an excess of critical events while blue regions have a deficit of critical events with respect to cosmic average.
    Interactive versions of these plots can be found online for the \href{https://pub.cphyc.me/Science/3d/critical_events_peak_around_filaments.html}{halo mergers}, \href{https://pub.cphyc.me/Science/3d/critical_events_filament_around_filaments.html}{filament mergers}, \href{https://pub.cphyc.me/Science/3d/critical_events_wall_around_filaments.html}{wall mergers} and \href{https://pub.cphyc.me/Science/3d/critical_events_filament_to_peak_around_filaments.html}{filament to peak merger ratio}.
    Going from  voids to  wall, from  wall to  filament and from  filament to the nearest node (along the $z$ axis), the halo merger rate increases and the filament merger rate decreases.
    Haloes in the filament are therefore stalled: they merge less than those in the nodes.
    At the same time, the filament merger rate decreases when going from the filament towards the node so that the mean connectivity, given by the ratio of halo merger to filament merger, is expected to increase.
  }
  \label{fig:filament-crit-event-counts}
\end{figure*}

Previous works have highlighted the modulation effect induced by the environment on the assembly of dark matter haloes and the galaxies therein, which affect the secondary halo or galaxy properties, an effect often called `assembly bias'.
Let us now make use of the merger statistics to study the impact of the large scale structures on assembly bias, following \cref{sec:condcluster}.
Indeed, it is expected on theoretical ground that, at fixed mass, the typical accretion rate increases when going from filament centre towards nodes \citep{Musso2018}.
Looking at galactic properties instead, \cite{kraljic_galaxies_2018} showed that the ratio of stellar rotation-to-dispersion
($v/\sigma$) is also modulated as a function of the distance and orientation to the nearest filamentary structure.
\cite{kraljic_ImpactConnectivityCosmic_2020} suggested that galactic properties are linked to the connectivity of the halo, with more connected haloes hosting more quenched and less rotation-supported galaxies.

In this section, we show that in our framework, the connectivity of haloes increases in nodes and decreases in voids, resulting in a differential evolution of haloes depending on their spatial location in the cosmic web.
In order to do this, a suite of Gaussian random fields constrained to the presence of a proto-filament have been generated.
The proto-filament is modelled as a filament-type saddle point at the centre of the box, the exact generation procedure being described in \cref{sec:generation}.
It is defined at a scale $R=\SI{5}{Mpc\per\hred}$, is oriented along the $z$ axis and lies in a wall in the $yz$ plane.
Using the set of constrained GRFs, we compute the excess density of each kind of critical event with respect to the cosmic mean, at fixed smoothing scale (hence at fixed object mass) $2.5 \leq R \leq \SI{5}{Mpc\per\hred}$.
The results are shown in \cref{fig:filament-crit-event-counts}.

Let us first restrict ourselves to the halo merger rate (top left panel of \cref{fig:filament-crit-event-counts}).
Going from one void to the wall, from the wall to the filament and from the filament to the nearest node, the halo merger rate increases and the maximum halo merger rate is found near the location where a node is expected ($z\sim \pm \SI{10}{Mpc\per\hred}$).
At larger scales, the field becomes unconstrained so that the merger rate falls back to its cosmic mean.
We reproduce here from first principle the results of \cite{Borzyszkowski2017}, showing that haloes close to the filament centre are stalled compared to those in nodes: they do not undergo many mergers nor do they  accrete much as the local tidal fields channels all the matter towards the two surrounding nodes, bypassing the centre of the filament.
Quantitatively, haloes forming at the centre of the filament are found to have a halo merger rate close to the cosmic average, while those close to the nodes are expected to have $40\%$ more mergers.
Conversely, haloes forming in a void next to a filamentary structure are expected to have a merger rate $20\%$ smaller than the cosmic mean.

Let us now add to the emerging picture the filament coalescence rate.
Filament merger rates act locally to decrease the connectivity of haloes, as each merger will disconnect one filament from two haloes.
The top right panel of \cref{fig:filament-crit-event-counts} shows that the merger rate is maximal along the wall and minimal along the filament.
Going off the plane of the wall ($x$ direction), the filament merger rate simply decreases towards the cosmic mean.
The filament merger rate is minimal in the nodes ($-13\%$) and maximal in the wall ($+10\%$).
As a consequence, haloes forming in a filament and close to a node have a larger halo merger rate but a smaller filament merger rate.
This in turn will have an impact on the assembly of dark matter haloes and their galaxies.
In the wall where the filament merger rate is the highest, we expect filaments to merge faster than haloes, resulting in haloes with fewer connected filaments.
This can be interpreted using the results of \cref{sec:theory2D}.
Indeed, in a cosmic wall, the geometry is locally 2D so that the theoretically expected connectivity becomes 4 instead of 6.

The bottom left panel of \cref{fig:filament-crit-event-counts} shows that the wall merger rate is decreased in walls and even more strongly in filaments compared to the rate found in voids.
The minimum wall merger rate is found at the location of the node with a rate $-40\%$ smaller than the cosmic mean.
Conversely, the wall merger rate is enhanced in the two voids surrounding the wall with a rate $20\%$ above the cosmic mean.

The evolution of the connectivity with cosmic environment is summarised  by the bottom right panel of \cref{fig:filament-crit-event-counts}, which shows the ratio of halo mergers ($\mathcal{P}$ critical events) to filament  mergers ($\mathcal{F}$ critical events), for which the cosmic mean is $2.055$ (see equation~\ref{eq:crit-event-ratio-th}).
Small values of $r_\mathcal{F\!/\!P}$ indicate that haloes merge faster than their surrounding filaments, so that the connectivity increases as haloes grow.
On the contrary, large values of $r_\mathcal{F\!/\!P}$ indicate that filaments merge faster than haloes, so that the connectivity decreases as haloes grow.
The bottom right panel of \cref{fig:filament-crit-event-counts} shows that in nodes, the ratio drops to about $r_\mathcal{F\!/\!P}\approx 1.1$.
On the contrary haloes forming in voids are expected to have a ratio of about $2.4$.

We therefore expect that, at fixed final mass, haloes forming next to a node will grow an increasing number of connected filaments\footnote{
Conversely \cite{codis2015} found that when averaged over all large scale structures, connectivity increases  with mass.}.
The expected physical outcome of this process is that the streams feeding a galaxy growing next to a node will become more and more isotropic with increasing connectivity.
Assuming that an isotropic acquisition of matter leads to a smaller amount of angular momentum being transferred down to the disk, we propose that this effect prevents the formation of gaseous disks in the vicinity of nodes.
Conversely, we expect that haloes growing in the neighbouring voids see their filaments destroyed faster than they merge, so that the halo is likely to grow with steadier flows coming from a few filaments (only the dominant ones survive) \citep[see also][section 6.2.1, and 5 resp. for  similar conclusions reached via the kinematic structure of large scale flows in filaments]{codis2015,laigle2014}.

\subsection{Departure from Gaussianity at high \texorpdfstring{$z$}{z}}
\label{sec:NG}

\begin{figure}
  \centering
\hskip -0.5cm  \includegraphics[width=1.05\columnwidth]{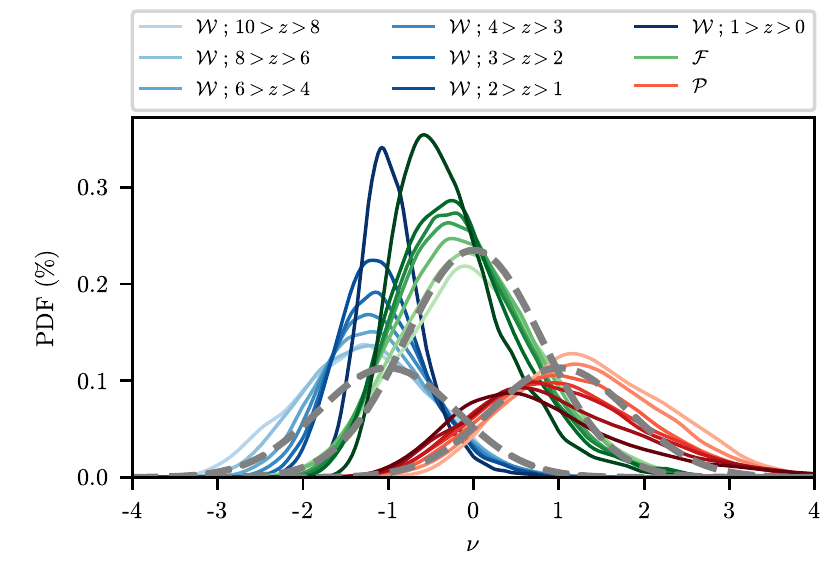}
  \caption{ %
    Critical events number counts as a function of the rarity in $\Lambda$CDM $N$-body simulations in different redshift bins as mentioned in the legend, with the same colours as \cref{fig:crit-event-PDF-GRF}.
    The curves have been normalised so that in each redshift bin, the integral of the three curves ($\mathcal{W,P,F}$) equals one.
    At high redshift, the merger rates resemble the Gaussian prediction (thick dashed grey lines, with an arbitrary normalisation).
    The skewness of the distributions increases with decreasing redshift as the field departs from Gaussianity, in qualitative agreement with the predictions of \cref{fig:prediction-critevents-product}.
  }
  \label{fig:nbody-critevents}
\end{figure}

Using the results of \cref{sec:non-gaussianity}, we detail in this section the evolution of the critical event number counts in the mildly non-linear regime, at high $z$.
Let us briefly quantify the effect first on simulations, and then compare to the proxy of \cref{sec:non-gaussianity} relying on known perturbative results.
\Cref{fig:nbody-critevents} presents the redshift evolution of critical event counts measured in 45 realisations of $\Lambda$CDM simulations in boxes of \SI{500}{Mpc\per\hred} involving $256^3$ particles evolved using {\sc Gadget} \citep{Gadget2001}. At each snapshot, the density field is sampled on a $256^3$ grid smoothed with a Gaussian filter over \SI{6}{Mpc\per\hred}.
The algorithm described in \cref{sec:detection} is used to identify and match the critical points and critical events.

At high redshift ($z\gtrapprox 10$), the measured number counts of critical events is close to the Gaussian prediction.
While we cannot make definite statements  given the level of shot noise in the measurements and existing transients at high redshifts, clear trends are seen in the counts.
In particular, at lower redshift, the $\cal P$ and $\cal F$ counts shift towards lower contrast, but resp. decrease and
increase in amplitude, while the $\cal W$ counts increase in amplitude.
Since halos in low density environments form later, it is expected that the low-$z$ counts are biased towards low densities.
Similarly, the mean density of filamentary structure decreases with increasing time, as the less dense filaments take more time to gravitationally form, so that the PDFs of the filament mergers shifts to smaller densities at low $z$.
The evolution of void structures with cosmological time mirrors that of peaks: early forming voids are the most underdense while late-time voids form out of less underdense regions.
At fixed resolution, this results in a shift of the typical density of voids towards higher densities which in turn shifts the $n^{\cal W}_\me$ towards higher densities.

Overall, the cosmic evolution of the measured event counts seems to be in fairly good  agreement with the model presented in \cref{fig:prediction-critevents-product}, suggesting that indeed, the set of critical events in the initial density field do capture the upcoming cosmic evolution of the cosmic web.
Further works beyond the scope of this paper will be necessary to better match the weakly non-Gaussian regime in more details.

\subsection{Discussion}
\label{sec:other}

\begin{figure*}
  \centering
  \includegraphics[width=\textwidth]{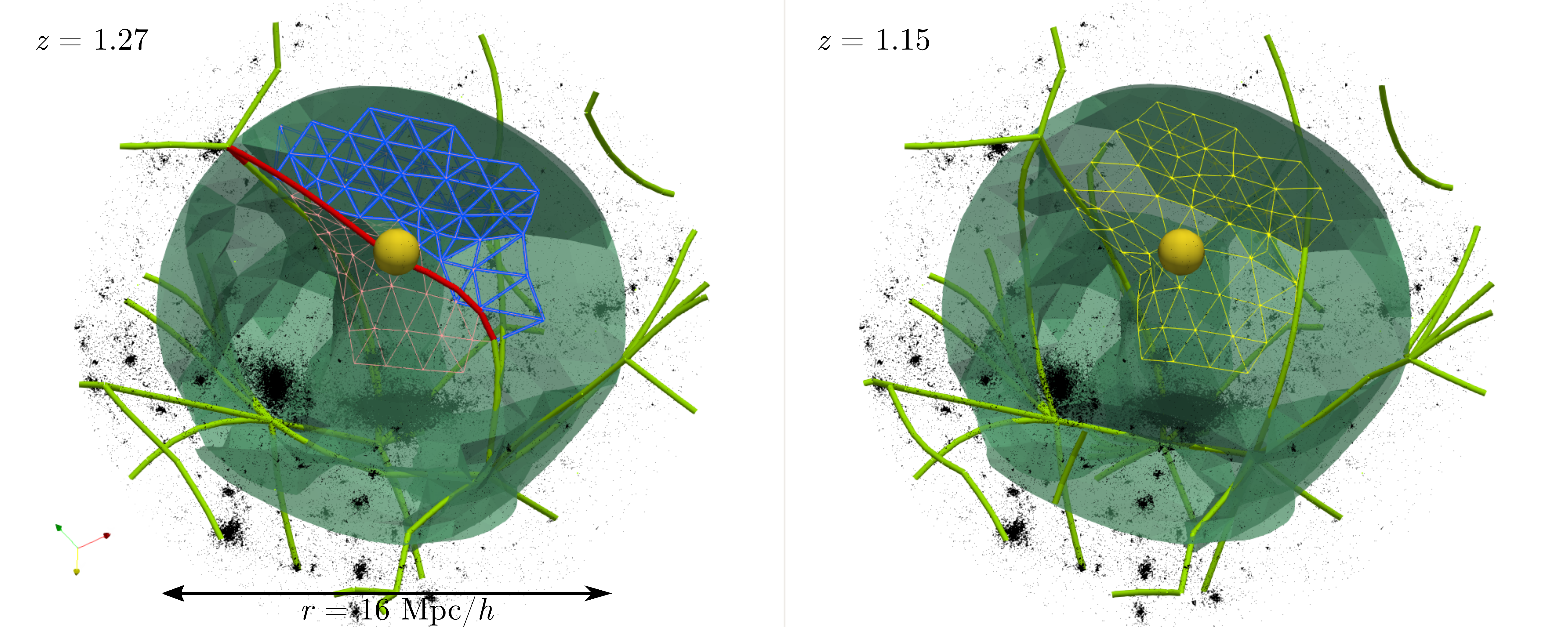}
  \includegraphics[width=\textwidth]{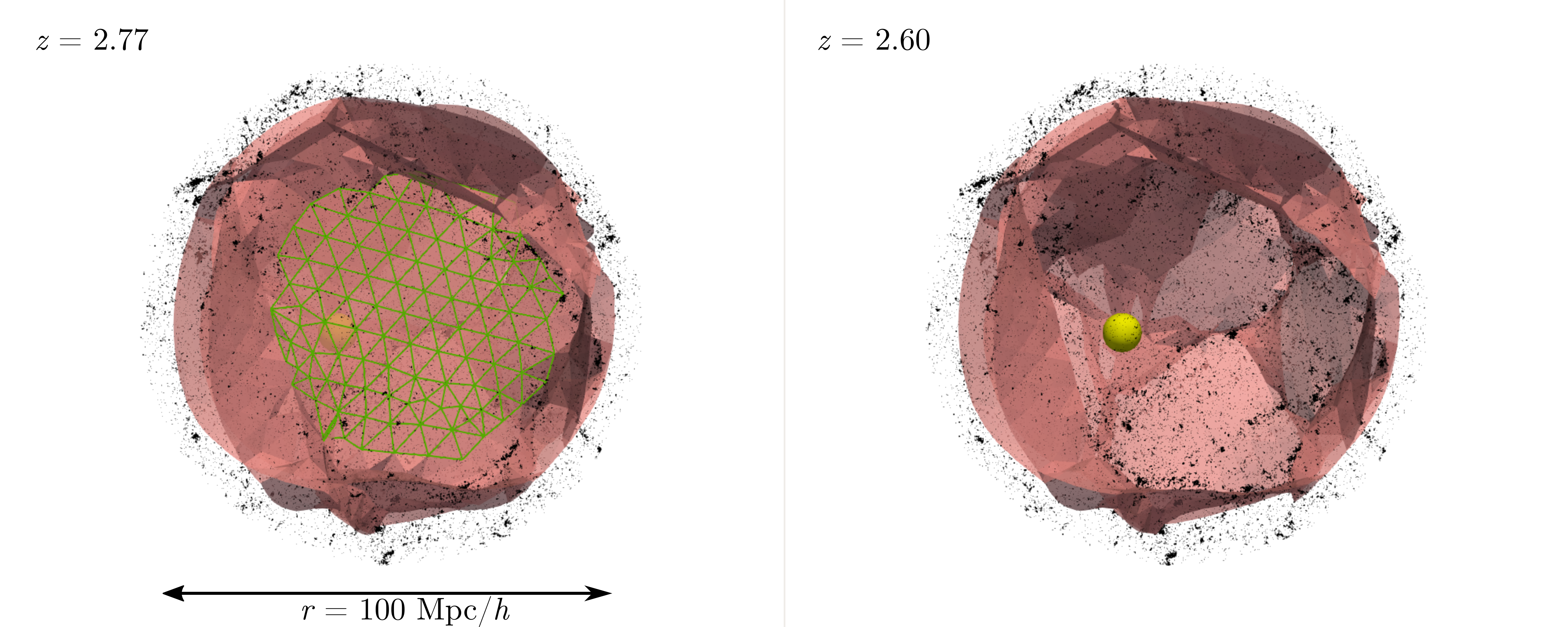}
  \caption{    
    (\emph{Top:}) Two consecutive snapshots in a \SI{100}{Mpc/h} simulation showing a $\mathcal{F}$ critical event (yellow ball) that encodes the disappearance of a filament (highlighted in red) separating two walls (red and blue wireframes).
    After the event, only one wall has survived (yellow wireframe).
    (\emph{Bottom:}) Two consecutive shapshots in a \SI{500}{Mpc/h} simulation showing a $\mathcal{W}$ critical event (yellow ball) that encodes the merger of two voids separated by a wall (green wireframe).
    After the event, only one void subsists and the wall has disappeared.
    Individual DM particles are shown as black dots.
    The skeleton (green lines) and the walls (green and red surfaces) have been extracted using \disperse.
    Both critical events can be related to merger event in the evolved Universe.
  }
  \label{fig:filament-merger-nbody}
  \label{fig:wall-merger-nbody}
\end{figure*}

There is a long tradition of relying on merger trees of dark matter haloes extracted from simulations as a means to tag the haloes with physical properties \citep[see, \eg][and references therein]{Lacey1991,White1991,Benson2010}. 
It has been suggested that galactic properties, such as spin, do not seem to be entirely encoded in the (halo) merger tree \citep{Vitvitska2002,benson_RandomwalkModelDark_2020}, 
a conclusion which could indicate that the anisotropy of the environment contributes to the spin of galaxies \citep{codis2015}.
One of the long term main motivations for the present work was to provide us with a theoretically-motivated extension to halo merger trees by adding the other two merger trees (filaments and walls).
Using the theory and the tools developed in this paper, the set of critical events that define these merger trees could be fed into semi-analytical modelling.
This would complement existing approaches by providing not only the past history of the DM halo (\emph{via} its merger tree), but also of the other substructures in the Lagrangian patch (\emph{via} the filament and wall merger trees).
To that end, the critical event theory provides an unambiguous and theoretically-motivated framework to describe and detect such events, either in the initial conditions (as a means to make predictions) or in numerical simulations (as a means to quantify the evolution).

Another possible approach would be to rely on modern machine learning techniques to identify which combination(s) of critical events are most likely to lead to galaxies of a certain type to be produced in cosmological simulations.
This strategy is likely to be efficient and rewarding, as the set of critical events is a very strong compression of the set of initial conditions, and because once the segmentation has been done, the subset of events which are in the past history of a galaxy with a given tag have physical meaning.
For instance, recent disconnect of filaments are likely to impact gas infall hence star formation and disc reformation \citep{pichonetal11,danovichetal11,2019OJAp....2E...7A}.
The set of critical events represents a useful effective  topological compression of the initial conditions which will impact the upcoming `dressed' merger tree (\ie the cosmic evolution of peaks {\sl and} their filaments and walls).
Note that the exact relative configuration of critical events in the position-smoothing space may be of relevance, and is not fully captured by the sole knowledge of the one and two-point statistics.

As an illustrative proof of concept, we have detected the critical events in the initial conditions of two $256^3$ $N$-body simulations. %
\Cref{fig:filament-merger-nbody} shows two pairs of consecutive snapshots zoomed around a $\mathcal{F}$ and a $\mathcal{W}$ critical event  (in simulations of size $100$ and \SI{500}{Mpc\per\hred} resp.).
In order to take into account the Zel'dovich flow, we have displaced the critical events from their initial Lagrangian position by the mean displacement of the neighbouring DM particles.
As expected, the $\mathcal{F}$ critical event can be related to the disappearance of a filament between two walls, while the $\mathcal{W}$ critical event encodes the disappearance of a wall between two voids. 
This figure illustrates that there exists at least a subset of critical events that can indeed be mapped to actual mergers in the evolved Universe,
but note  that the mapping between the critical events and the time of the merger in the simulation was done here by  visual inspection.
The quantitative study of the accuracy of the mapping between critical events in the initial conditions and in the corresponding
 dynamically evolved simulation will be the subject of future work.

One should note that, even if the mapping between critical events in the initial conditions and critical events in the density field evolved by the simulation could not be established uniquely, the applications highlighted above would be left unchanged as they only rely upon the detection of critical events in the \emph{evolved field}, but it would however limit the scope of theoretical predictions.

Mapping of intensity of spectral lines, for instance HI 21 cm line
\citep{Madau1997} across the sky, could also benefit from applying the present formalism to sequences of 2D maps as a function of redshift.
Existing \citep[\eg Chime,][]{Chime2014} or upcoming surveys \cite[\eg SKA,][]{SKA2015} will indeed provide both extrema and merger counts extracted from sets of maps at various redshifts.
The cosmology dependence of extrema counts is through $(R_*,\gamma)$ and the relevant cumulants, whereas the cosmology dependence of critical event counts also involve $({\tilde R},{\tilde \gamma})$ and higher order cumulants  at fixed level of non-Gaussianity (\eg involving 3rd order derivative of the field to first order as discussed in \cref{sec:non-gaussianity}).
Hence, studying both counts as a function of redshift will prove complementary.

These possible applications highlight the versatility of  critical events: they yield diagnostics  in the initial conditions, together with a theoretically-motivated description of processes driving the evolution of the cosmic web in the evolved Universe.
The theory presented in this paper provides a description of the evolution of mass infall that may play an important role in galaxy formation. 
Further efforts should be made to relate infall to the internal effects driving the formation of galaxies (star formation, feedback, turbulence, etc.).
\paragraph*{Applications beyond cosmology}
The present analysis  was mostly restricted to (quasi-)Gaussian random fields, because
of their relevance in cosmology and also because in this context
the theory can be developed in some details (as a Gaussian process defines a Morse function on a scale-by-scale basis).
But  the concept of bifurcation of critical points in a one parameter set of random fields extends beyond Gaussianity.
Any system involving  random field controlled by one parameter could in principle be
investigated with this framework in order to identify bifurcation/mergers of ridges
(though the specific role played by Gaussian smoothing would clearly generally not hold).
For instance, critical events in dust maps \citep[such as][]{Meisner_2013,PlanckDust2018} could be used as an alternative statistics to quantify the properties of the underlying turbulence. %
The theory of critical events could also find applications in fields where data are well described by their geometry, as critical events describe how this geometry changes with scale.
For example, in the context of streaming of
images, the set of critical events within a 2D image characterises its multi-scale topology.
It would therefore be of interest to send the set of critical events, starting from the ones at the largest smoothing scales, as a means of prioritising which sub-region of the image needs to be streamed first because the topology of its excursion (\ie the local parsimonious representation of the image as iso-contours) has changed.
This would allow the received image to acquire its most important topological features first.

Following the results of \cref{sec:NDevents}, our formalism could be extended to situations where the field whose evolution is investigated corresponds to realisations of probability distributions living in higher dimensions (or on more complex manifolds).
In a more abstract setting corresponding to a landscape drawn from a given probability function, a wide range of important physical  processes occur when
rare events collide, boosting detection probabilities  and passing a given threshold. For instance,  dark matter annihilation rates  (which scale like the density squared) are boosted when two substructures merges \citep{fermi2018}.
 In the context of this work this corresponds to nucleation,  or  the appearance of pairs of
critical points as one `unsmoothes' (or more generally evolves) the field.

\section{Conclusion}
\label{sec:conclusion}

As a proxy for  cosmic evolution, we computed the merger rate of critical points (peaks, saddle points and minima) as a function of smoothing scale from the primordial density field to forecast critical events (halo, filament and wall mergers) that drive the assembly of dark matter haloes and possibly galaxies.
We recovered the non-linear prediction for the net density of peak merger found by \cite{hanami} and further considered all sets of critical points coalescence,  including wall-saddle to filament-saddle (filament mergers) and wall-saddle to minima (wall or void mergers), as they modify the geometry of galactic infall, such as filament disconnection or void disappearance, thus generalising previous results that focused only on peaks.
This `critical event theory' is central to our understanding of the effect of the cosmic web on the formation of galaxies, since their evolution is the result of their past history, which is  encoded in their extended merger tree and the properties of their host halo.

The key results of the paper are the following.
\begin{enumerate}[i)]
  \item We studied critical events of all types
  and presented analytical formulas for the one-point statistics of these events in fields of dimensions up to 6 (\cref{sec:theory}), and also their clustering properties \emph{via} their two-point statistics (\cref{sec:theory2pt}).
  \item We have developed an algorithm to find critical events in numerical datasets which we used as a confirmation of the theory (\cref{sec:measurements}).
  Such algorithm could be used \eg to pre-compress streaming of images, or as input to machine learning as a means 
  to  learn galactic morphology from the initial conditions.
  We also developed an algorithm to generate Gaussian random fields subject to a given critical event.
  \item We provided a covariant formulation of the critical event theory which allowed us to also compute the two-point statistics for critical events. The two-point statistics show that halo mergers are typically followed by  filament mergers, so that the connectivity is preserved.
  \item We have shown that the critical event theory can be further extended to take into account the early stages of non-linear gravitational evolution. This has then been compared qualitatively to numerical simulations at high redshift. This extension also probes the non-Gaussianities that arise from primordial non-Gaussianities and can be used as a cosmological measurement.
\end{enumerate}

We also presented some practical applications of the theory to astrophysical problems in \cref{sec:discussion}.
We computed the destruction rate of haloes and voids as a function of mass and redshift in a $\Lambda$CDM cosmology using a simple model to assign a mass and time to critical events (\cref{sec:Mzmerger}). This can be used as a test for the critical event theory, as well as an alternative cosmological measurement.
We have established the link between critical events and connectivity. This allowed us to compute the connectivity of peaks and other critical events in arbitrary dimensions\footnote{It yields an analytical prediction of
connectivity of peak in four dimension:
$\kappa_4 ={200 \pi }/{(75 \pi \!-\!114 \!-\!100 \cot ^{-1}(2))}\approx 8.35$.
}. Physically, a duality between the evolution of the cosmic web (critical events) and its topological features (connectivity) was highlighted (\cref{sec:consistency-connectivity})
In addition, we showed that haloes forming near cosmic nodes do so by increasing their connectivity, with possible implication for the formation of their host galaxy (\cref{sec:assembly-bias}).
Finally, using $N$-body simulations, we have shown that the critical event theory statistically recovers the evolution of the merger rates of the different structures (haloes, filaments, walls) in the mildly non-linear regime at high redshift (\cref{sec:NG}).

We have only touched on practical applications for the forecasting  of special events in a multi-scale landscape.
It may prove to be a fruitful field of upcoming research in astronomy and beyond.

\section*{Acknowledgements}
This work was partially supported by the Spin(e) grant ANR-13-BS05-0005  (\href{http://cosmicorigin.org}{cosmicorigin.org}) and the Segal grant ANR-19-CE31-0017 (\href{https://www.secular-evolution.org}{www.secular-evolution.org}) of the French {\sl Agence Nationale de la Recherche}.
This project has received funding from the European Union Horizon 2020 research and innovation program under grant agreement No. 818085 GMGalaxies and  by the National Science Foundation under Grant No. PHY11-25915. SC's research is partially supported by Fondation MERAC.  This research was supported in part by the National Science Foundation under Grant No. NSF PHY-1748958.
CC thanks the Institut Lagrange de Paris for partial funding.
SC and CC thank KIAS for hospitality and financial support during the course of this project.
MM thanks the Max-Planck Institute for Astrophysics for hospitality.
CP thanks S. Colombi for pointing to Hanami's paper and acknowledges early and late discussions with  Thierry Sousbie and Pooran Memari resp.
We thank the referee, J. Primack for a constructive report. We also thank 
 A. Dekel, N. Cornuault, J. Devriendt, C.~Park and E.~Pichon for fruitful discussions and useful comments and CC thanks S. White for constructive criticisms.
This work has made use of the Horizon Cluster hosted by Institut d'Astrophysique de Paris.
We thank Stephane Rouberol for running smoothly this cluster for us and
Thierry Sousbie for \href{http://www2.iap.fr/users/sousbie/web/html/indexd41d.html}{\sc Disperse}. 

\section*{Data availability}
The data underlying this article will be shared on reasonable request to the corresponding author.

\bibliographystyle{mnras}
\bibliography{author}

\begin{thebibliography}{}
\makeatletter
\relax
\def\mn@urlcharsother{\let\do\@makeother \do\$\do\&\do\#\do\^\do\_\do\%\do\~}
\def\mn@doi{\begingroup\mn@urlcharsother \@ifnextchar [ {\mn@doi@}
  {\mn@doi@[]}}
\def\mn@doi@[#1]#2{\def\@tempa{#1}\ifx\@tempa\@empty \href
  {http://dx.doi.org/#2} {doi:#2}\else \href {http://dx.doi.org/#2} {#1}\fi
  \endgroup}
\def\mn@eprint#1#2{\mn@eprint@#1:#2::\@nil}
\def\mn@eprint@arXiv#1{\href {http://arxiv.org/abs/#1} {{\tt arXiv:#1}}}
\def\mn@eprint@dblp#1{\href {http://dblp.uni-trier.de/rec/bibtex/#1.xml}
  {dblp:#1}}
\def\mn@eprint@#1:#2:#3:#4\@nil{\def\@tempa {#1}\def\@tempb {#2}\def\@tempc
  {#3}\ifx \@tempc \@empty \let \@tempc \@tempb \let \@tempb \@tempa \fi \ifx
  \@tempb \@empty \def\@tempb {arXiv}\fi \@ifundefined
  {mn@eprint@\@tempb}{\@tempb:\@tempc}{\expandafter \expandafter \csname
  mn@eprint@\@tempb\endcsname \expandafter{\@tempc}}}

\bibitem[\protect\citeauthoryear{Agertz, Teyssier  \& Moore}{Agertz
  et~al.}{2009}]{agertz_disc_2009}
Agertz O.,  Teyssier R.,   Moore B.,  2009, \mn@doi [Monthly Notices of the
  Royal Astronomical Society: Letters] {10/bbctjs}, 397, 64

\bibitem[\protect\citeauthoryear{{Aragon Calvo}, {Neyrinck}  \& {Silk}}{{Aragon
  Calvo} et~al.}{2019}]{2019OJAp....2E...7A}
{Aragon Calvo} M.~A.,  {Neyrinck} M.~C.,   {Silk} J.,  2019, \mn@doi [The Open
  Journal of Astrophysics] {10.21105/astro.1607.07881}, \href
  {https://ui.adsabs.harvard.edu/abs/2019OJAp....2E...7A} {2, 7}

\bibitem[\protect\citeauthoryear{{Aubert} \& {Pichon}}{{Aubert} \&
  {Pichon}}{2007}]{Aubert2007}
{Aubert} D.,  {Pichon} C.,  2007, \mn@doi [\mnras]
  {10.1111/j.1365-2966.2006.11203.x}, \href
  {https://ui.adsabs.harvard.edu/abs/2007MNRAS.374..877A} {374, 877}

\bibitem[\protect\citeauthoryear{{Bardeen}, {Bond}, {Kaiser}  \&
  {Szalay}}{{Bardeen} et~al.}{1986}]{BBKS}
{Bardeen} J.~M.,  {Bond} J.~R.,  {Kaiser} N.,   {Szalay} A.~S.,  1986, \mn@doi
  [\apj] {10.1086/164143}, \href
  {http://adsabs.harvard.edu/abs/1986ApJ...304...15B} {304, 15}

\bibitem[\protect\citeauthoryear{{Benson} \& {Bower}}{{Benson} \&
  {Bower}}{2010}]{Benson2010}
{Benson} A.~J.,  {Bower} R.,  2010, \mn@doi [\mnras]
  {10.1111/j.1365-2966.2010.16592.x}, \href
  {https://ui.adsabs.harvard.edu/abs/2010MNRAS.405.1573B} {405, 1573}

\bibitem[\protect\citeauthoryear{{Benson}, {Behrens}  \& {Lu}}{{Benson}
  et~al.}{2020}]{benson_RandomwalkModelDark_2020}
{Benson} A.,  {Behrens} C.,   {Lu} Y.,  2020, arXiv e-prints, \href
  {https://ui.adsabs.harvard.edu/abs/2020arXiv200109208B} {p. arXiv:2001.09208}

\bibitem[\protect\citeauthoryear{{Bernardeau}, {Colombi}, {Gazta{\~n}aga}  \&
  {Scoccimarro}}{{Bernardeau} et~al.}{2002}]{Bernardeau2002}
{Bernardeau} F.,  {Colombi} S.,  {Gazta{\~n}aga} E.,   {Scoccimarro} R.,  2002,
  \physrep, \href
  {http://adsabs.harvard.edu/cgi-bin/nph-bib_query?bibcode=2002PhR...367....1B&db_key=AST}
  {367, 1}

\bibitem[\protect\citeauthoryear{{Blumenthal}, {Faber}, {Primack}  \&
  {Rees}}{{Blumenthal} et~al.}{1984}]{1984Natur.311..517B}
{Blumenthal} G.~R.,  {Faber} S.~M.,  {Primack} J.~R.,   {Rees} M.~J.,  1984,
  \mn@doi [\nat] {10.1038/311517a0}, \href
  {https://ui.adsabs.harvard.edu/abs/1984Natur.311..517B} {311, 517}

\bibitem[\protect\citeauthoryear{{Bond} \& {Myers}}{{Bond} \&
  {Myers}}{1996}]{bond1996}
{Bond} J.~R.,  {Myers} S.~T.,  1996, \mn@doi [\apjs] {10.1086/192267}, \href
  {https://ui.adsabs.harvard.edu/abs/1996ApJS..103....1B} {103, 1}

\bibitem[\protect\citeauthoryear{{Bond}, {Cole}, {Efstathiou}  \&
  {Kaiser}}{{Bond} et~al.}{1991}]{Bondetal1991}
{Bond} J.~R.,  {Cole} S.,  {Efstathiou} G.,   {Kaiser} N.,  1991, \mn@doi
  [\apj] {10.1086/170520}, \href
  {http://adsabs.harvard.edu/abs/1991ApJ...379..440B} {379, 440}

\bibitem[\protect\citeauthoryear{{Bond}, {Kofman}  \& {Pogosyan}}{{Bond}
  et~al.}{1996}]{bkp96}
{Bond} J.~R.,  {Kofman} L.,   {Pogosyan} D.,  1996, \mn@doi [\nat]
  {10.1038/380603a0}, \href {http://adsabs.harvard.edu/abs/1996Natur.380..603B}
  {380, 603}

\bibitem[\protect\citeauthoryear{{Borzyszkowski}, {Porciani},
  {Romano-D{\'\i}az}  \& {Garaldi}}{{Borzyszkowski}
  et~al.}{2017}]{Borzyszkowski2017}
{Borzyszkowski} M.,  {Porciani} C.,  {Romano-D{\'\i}az} E.,   {Garaldi} E.,
  2017, \mn@doi [\mnras] {10.1093/mnras/stx873}, \href
  {https://ui.adsabs.harvard.edu/abs/2017MNRAS.469..594B} {469, 594}

\bibitem[\protect\citeauthoryear{{Bournaud}, {Jog}  \& {Combes}}{{Bournaud}
  et~al.}{2007}]{2007A&A...476.1179B}
{Bournaud} F.,  {Jog} C.~J.,   {Combes} F.,  2007, \mn@doi [\aap]
  {10.1051/0004-6361:20078010}, \href
  {https://ui.adsabs.harvard.edu/abs/2007A&A...476.1179B} {476, 1179}

\bibitem[\protect\citeauthoryear{{Camera}, {Santos}  \& {Maartens}}{{Camera}
  et~al.}{2015}]{SKA2015}
{Camera} S.,  {Santos} M.~G.,   {Maartens} R.,  2015, \mn@doi [\mnras]
  {10.1093/mnras/stv040}, \href
  {http://adsabs.harvard.edu/abs/2015MNRAS.448.1035C} {448, 1035}

\bibitem[\protect\citeauthoryear{{Castorina}, {Paranjape}, {Hahn}  \&
  {Sheth}}{{Castorina} et~al.}{2016}]{2016arXiv161103619C}
{Castorina} E.,  {Paranjape} A.,  {Hahn} O.,   {Sheth} R.~K.,  2016, arXiv
  e-prints, \href {https://ui.adsabs.harvard.edu/abs/2016arXiv161103619C} {p.
  arXiv:1611.03619}

\bibitem[\protect\citeauthoryear{{Clark}, {Scott}, {Trotta}  \&
  {Lewis}}{{Clark} et~al.}{2018}]{fermi2018}
{Clark} H.~A.,  {Scott} P.,  {Trotta} R.,   {Lewis} G.~F.,  2018, \mn@doi
  [\jcap] {10.1088/1475-7516/2018/07/060}, \href
  {https://ui.adsabs.harvard.edu/abs/2018JCAP...07..060C} {2018, 060}

\bibitem[\protect\citeauthoryear{{Codis}, {Pichon}, {Pogosyan}, {Bernardeau}
  \& {Matsubara}}{{Codis} et~al.}{2013}]{Codis2013}
{Codis} S.,  {Pichon} C.,  {Pogosyan} D.,  {Bernardeau} F.,   {Matsubara} T.,
  2013, \mn@doi [\mnras] {10.1093/mnras/stt1316}, \href
  {https://ui.adsabs.harvard.edu/abs/2013MNRAS.435..531C} {435, 531}

\bibitem[\protect\citeauthoryear{{Codis}, {Pichon}  \& {Pogosyan}}{{Codis}
  et~al.}{2015}]{codis2015}
{Codis} S.,  {Pichon} C.,   {Pogosyan} D.,  2015, \mn@doi [\mnras]
  {10.1093/mnras/stv1570}, \href
  {http://adsabs.harvard.edu/abs/2015MNRAS.452.3369C} {452, 3369}

\bibitem[\protect\citeauthoryear{{Codis}, {Pogosyan}  \& {Pichon}}{{Codis}
  et~al.}{2018}]{codis2018}
{Codis} S.,  {Pogosyan} D.,   {Pichon} C.,  2018, \mn@doi [\mnras]
  {10.1093/mnras/sty1643}, \href
  {https://ui.adsabs.harvard.edu/abs/2018MNRAS.479..973C} {479, 973}

\bibitem[\protect\citeauthoryear{{Crittenden}, {Natarajan}, {Pen}  \&
  {Theuns}}{{Crittenden} et~al.}{2001}]{Cri++01}
{Crittenden} R.~G.,  {Natarajan} P.,  {Pen} U.-L.,   {Theuns} T.,  2001,
  \mn@doi [\apj] {10.1086/322370}, \href
  {http://cdsads.u-strasbg.fr/abs/2001ApJ...559..552C} {559, 552}

\bibitem[\protect\citeauthoryear{{Danovich}, {Dekel}, {Hahn}  \&
  {Teyssier}}{{Danovich} et~al.}{2012}]{danovichetal11}
{Danovich} M.,  {Dekel} A.,  {Hahn} O.,   {Teyssier} R.,  2012, \mn@doi
  [\mnras] {10.1111/j.1365-2966.2012.20751.x}, \href
  {http://cdsads.u-strasbg.fr/abs/2012MNRAS.422.1732D} {422, 1732}

\bibitem[\protect\citeauthoryear{Dekel \& Birnboim}{Dekel \&
  Birnboim}{2006}]{dekel_galaxy_2006}
Dekel A.,  Birnboim Y.,  2006, \mn@doi [Monthly Notices of the Royal
  Astronomical Society] {10/c38j76}, 368, 2

\bibitem[\protect\citeauthoryear{Diemer}{Diemer}{2018}]{diemer_colossus_2018}
Diemer B.,  2018, \mn@doi [\apjs] {10.3847/1538-4365/aaee8c}, 239, 35

\bibitem[\protect\citeauthoryear{{Doroshkevich}}{{Doroshkevich}}{1970}]{doroshkevich70}
{Doroshkevich} A.~G.,  1970, \mn@doi [Astrophysics] {10.1007/BF01001625}, \href
  {http://adsabs.harvard.edu/abs/1970Ap......6..320D} {6, 320}

\bibitem[\protect\citeauthoryear{{Dubinski}, {da Costa}, {Goldwirth}, {Lecar}
  \& {Piran}}{{Dubinski} et~al.}{1993}]{1993ApJ...410..458D}
{Dubinski} J.,  {da Costa} L.~N.,  {Goldwirth} D.~S.,  {Lecar} M.,   {Piran}
  T.,  1993, \mn@doi [\apj] {10.1086/172762}, \href
  {https://ui.adsabs.harvard.edu/abs/1993ApJ...410..458D} {410, 458}

\bibitem[\protect\citeauthoryear{{Dubois}, {Pichon}, {Haehnelt}, {Kimm},
  {Slyz}, {Devriendt}  \& {Pogosyan}}{{Dubois} et~al.}{2012}]{Dubois2011}
{Dubois} Y.,  {Pichon} C.,  {Haehnelt} M.,  {Kimm} T.,  {Slyz} A.,  {Devriendt}
  J.,   {Pogosyan} D.,  2012, \mn@doi [\mnras]
  {10.1111/j.1365-2966.2012.21160.x}, \href
  {http://cdsads.u-strasbg.fr/abs/2012MNRAS.423.3616D} {423, 3616}

\bibitem[\protect\citeauthoryear{{Dubois}, {Peirani}, {Pichon}, {Devriendt},
  {Gavazzi}, {Welker}  \& {Volonteri}}{{Dubois} et~al.}{2016}]{Dubois2016}
{Dubois} Y.,  {Peirani} S.,  {Pichon} C.,  {Devriendt} J.,  {Gavazzi} R.,
  {Welker} C.,   {Volonteri} M.,  2016, \mn@doi [\mnras]
  {10.1093/mnras/stw2265}, \href
  {https://ui.adsabs.harvard.edu/abs/2016MNRAS.463.3948D} {463, 3948}

\bibitem[\protect\citeauthoryear{Edelsbrunner, Letscher  \&
  Zomorodian}{Edelsbrunner et~al.}{2002}]{Edelsbrunner2000}
Edelsbrunner H.,  Letscher D.,   Zomorodian A.,  2002, \mn@doi [Discrete
  Comput. Geom.] {10.1007/s00454-002-2885-2}, 28, 511

\bibitem[\protect\citeauthoryear{Eisenstein \& Hu}{Eisenstein \&
  Hu}{1998}]{eisenstein_BaryonicFeaturesMatter_1998}
Eisenstein D.~J.,  Hu W.,  1998, \mn@doi [\apj] {10.1086/305424}, 496, 605

\bibitem[\protect\citeauthoryear{Eisenstein \& Hu}{Eisenstein \&
  Hu}{1999}]{eisenstein_power_1999}
Eisenstein D.~J.,  Hu W.,  1999, \mn@doi [\apj] {10.1086/305585}, 511, 5

\bibitem[\protect\citeauthoryear{Fakhouri, Ma  \& {Boylan-Kolchin}}{Fakhouri
  et~al.}{2010}]{fakhouri_merger_2010}
Fakhouri O.,  Ma C.-P.,   {Boylan-Kolchin} M.,  2010, \mn@doi [\mnras]
  {10/c2rn5b}, 406, 2267

\bibitem[\protect\citeauthoryear{Gay}{Gay}{2011}]{pierre_squelette_2011}
Gay C.,  2011, PhD thesis, Universit\'e Pierre et Marie Curie, \url
  {http://www.theses.fr/2011PA066299}

\bibitem[\protect\citeauthoryear{{Gay}, {Pichon}  \& {Pogosyan}}{{Gay}
  et~al.}{2012}]{Gay2012}
{Gay} C.,  {Pichon} C.,   {Pogosyan} D.,  2012, \mn@doi [\prd]
  {10.1103/PhysRevD.85.023011}, \href
  {https://ui.adsabs.harvard.edu/abs/2012PhRvD..85b3011G} {85, 023011}

\bibitem[\protect\citeauthoryear{Genel, Genzel, Bouch{\'e}, Naab  \&
  Sternberg}{Genel et~al.}{2009}]{genel_halo_2009}
Genel S.,  Genzel R.,  Bouch{\'e} N.,  Naab T.,   Sternberg A.,  2009, \mn@doi
  [\apj] {10/d5mvfp}, 701, 2002

\bibitem[\protect\citeauthoryear{{Hahn}, {Porciani}, {Dekel}  \&
  {Carollo}}{{Hahn} et~al.}{2009}]{Hahn2009}
{Hahn} O.,  {Porciani} C.,  {Dekel} A.,   {Carollo} C.~M.,  2009, \mn@doi
  [\mnras] {10.1111/j.1365-2966.2009.15271.x}, \href
  {https://ui.adsabs.harvard.edu/abs/2009MNRAS.398.1742H} {398, 1742}

\bibitem[\protect\citeauthoryear{{Hanami}}{{Hanami}}{2001}]{hanami}
{Hanami} H.,  2001, \mn@doi [\mnras] {10.1046/j.1365-8711.2001.04652.x}, \href
  {https://ui.adsabs.harvard.edu/abs/2001MNRAS.327..721H} {327, 721}

\bibitem[\protect\citeauthoryear{{Jedamzik}}{{Jedamzik}}{1995}]{Jedamzik95}
{Jedamzik} K.,  1995, \mn@doi [\apj] {10.1086/175936}, \href
  {https://ui.adsabs.harvard.edu/abs/1995ApJ...448....1J} {448, 1}

\bibitem[\protect\citeauthoryear{Jennings, Li  \& Hu}{Jennings
  et~al.}{2013}]{jennings_abundance_2013}
Jennings E.,  Li Y.,   Hu W.,  2013, \mn@doi [\mnras] {10/f48w3f}, 434, 2167

\bibitem[\protect\citeauthoryear{Jones, Oliphant, Peterson  et~al.}{Jones
  et~al.}{01  }]{scipy_paper}
Jones E.,  Oliphant T.,  Peterson P.,   et~al., 2001--, {SciPy}: Open source
  scientific tools for {Python}, \url {http://www.scipy.org/}

\bibitem[\protect\citeauthoryear{{Kaiser}}{{Kaiser}}{1984}]{Kaiser1984}
{Kaiser} N.,  1984, \mn@doi [\apjl] {10.1086/184341}, \href
  {http://adsabs.harvard.edu/abs/1984ApJ...284L...9K} {284, L9}

\bibitem[\protect\citeauthoryear{{Kraljic} et~al.,}{{Kraljic}
  et~al.}{2018}]{kraljic_galaxies_2018}
{Kraljic} K.,  et~al., 2018, \mn@doi [\mnras] {10.1093/mnras/stx2638}, \href
  {https://ui.adsabs.harvard.edu/abs/2018MNRAS.474..547K} {474, 547}

\bibitem[\protect\citeauthoryear{Kraljic et~al.,}{Kraljic
  et~al.}{2020}]{kraljic_ImpactConnectivityCosmic_2020}
Kraljic K.,  et~al., 2020, \mn@doi [\mnras] {10.1093/mnras/stz3319}, 491, 4294

\bibitem[\protect\citeauthoryear{{Lacey} \& {Cole}}{{Lacey} \&
  {Cole}}{1993}]{LaceyCole1993}
{Lacey} C.,  {Cole} S.,  1993, \mn@doi [\mnras] {10.1093/mnras/262.3.627},
  \href {https://ui.adsabs.harvard.edu/abs/1993MNRAS.262..627L} {262, 627}

\bibitem[\protect\citeauthoryear{{Lacey} \& {Silk}}{{Lacey} \&
  {Silk}}{1991}]{Lacey1991}
{Lacey} C.,  {Silk} J.,  1991, \mn@doi [\apj] {10.1086/170625}, \href
  {https://ui.adsabs.harvard.edu/abs/1991ApJ...381...14L} {381, 14}

\bibitem[\protect\citeauthoryear{{Laigle} et~al.,}{{Laigle}
  et~al.}{2015}]{laigle2014}
{Laigle} C.,  et~al., 2015, \mn@doi [\mnras] {10.1093/mnras/stu2289}, \href
  {http://adsabs.harvard.edu/abs/2015MNRAS.446.2744L} {446, 2744}

\bibitem[\protect\citeauthoryear{Ludlow, Porciani  \& Borzyszkowski}{Ludlow
  et~al.}{2014}]{Ludlow2011}
Ludlow A.~D.,  Porciani C.,   Borzyszkowski M.,  2014, \mn@doi [\mnras]
  {10.1093/mnras/stu2021}, 445, 4110

\bibitem[\protect\citeauthoryear{{Madau}, {Meiksin}  \& {Rees}}{{Madau}
  et~al.}{1997}]{Madau1997}
{Madau} P.,  {Meiksin} A.,   {Rees} M.~J.,  1997, \mn@doi [\apj]
  {10.1086/303549}, \href
  {https://ui.adsabs.harvard.edu/abs/1997ApJ...475..429M} {475, 429}

\bibitem[\protect\citeauthoryear{{Maggiore} \& {Riotto}}{{Maggiore} \&
  {Riotto}}{2010}]{MaggioreRiotto2009}
{Maggiore} M.,  {Riotto} A.,  2010, \mn@doi [\apj]
  {10.1088/0004-637X/711/2/907}, \href
  {https://ui.adsabs.harvard.edu/abs/2010ApJ...711..907M} {711, 907}

\bibitem[\protect\citeauthoryear{{Manrique} \& {Salvador-Sole}}{{Manrique} \&
  {Salvador-Sole}}{1995}]{Manrique1995}
{Manrique} A.,  {Salvador-Sole} E.,  1995, \mn@doi [\apj] {10.1086/176364},
  \href {https://ui.adsabs.harvard.edu/abs/1995ApJ...453....6M} {453, 6}

\bibitem[\protect\citeauthoryear{{Manrique} \& {Salvador-Sole}}{{Manrique} \&
  {Salvador-Sole}}{1996}]{Manrique1996}
{Manrique} A.,  {Salvador-Sole} E.,  1996, \mn@doi [\apj] {10.1086/177627},
  \href {https://ui.adsabs.harvard.edu/abs/1996ApJ...467..504M} {467, 504}

\bibitem[\protect\citeauthoryear{Meisner \& Finkbeiner}{Meisner \&
  Finkbeiner}{2013}]{Meisner_2013}
Meisner A.~M.,  Finkbeiner D.~P.,  2013, \mn@doi [\apj]
  {10.1088/0004-637x/781/1/5}, 781, 5

\bibitem[\protect\citeauthoryear{Moore et~al.,}{Moore
  et~al.}{2001}]{moore_fast_2001}
Moore A.~W.,  et~al., 2001, \mn@doi [Mining the Sky] {10.1007/10849171_5},
  p.~71

\bibitem[\protect\citeauthoryear{{More}, {Diemer}  \& {Kravtsov}}{{More}
  et~al.}{2015}]{More2015}
{More} S.,  {Diemer} B.,   {Kravtsov} A.~V.,  2015, \mn@doi [\apj]
  {10.1088/0004-637X/810/1/36}, \href
  {https://ui.adsabs.harvard.edu/abs/2015ApJ...810...36M} {810, 36}

\bibitem[\protect\citeauthoryear{{Musso} \& {Sheth}}{{Musso} \&
  {Sheth}}{2012}]{MussoSheth2012}
{Musso} M.,  {Sheth} R.~K.,  2012, \mn@doi [\mnras]
  {10.1111/j.1745-3933.2012.01266.x}, \href
  {http://adsabs.harvard.edu/abs/2012MNRAS.423L.102M} {423, L102}

\bibitem[\protect\citeauthoryear{{Musso} \& {Sheth}}{{Musso} \&
  {Sheth}}{2019}]{Musso2019}
{Musso} M.,  {Sheth} R.~K.,  2019, arXiv e-prints, \href
  {https://ui.adsabs.harvard.edu/abs/2019arXiv190709147M} {p. arXiv:1907.09147}

\bibitem[\protect\citeauthoryear{{Musso}, {Cadiou}, {Pichon}, {Codis},
  {Kraljic}  \& {Dubois}}{{Musso} et~al.}{2018}]{Musso2018}
{Musso} M.,  {Cadiou} C.,  {Pichon} C.,  {Codis} S.,  {Kraljic} K.,   {Dubois}
  Y.,  2018, \mn@doi [\mnras] {10.1093/mnras/sty191}, \href
  {https://ui.adsabs.harvard.edu/abs/2018MNRAS.476.4877M} {476, 4877}

\bibitem[\protect\citeauthoryear{Naab \& Burkert}{Naab \&
  Burkert}{2003}]{naab_StatisticalPropertiesCollisionless_2003}
Naab T.,  Burkert A.,  2003, \mn@doi [The Astrophysical Journal]
  {10.1086/378581}, 597, 893

\bibitem[\protect\citeauthoryear{Paranjape \& Sheth}{Paranjape \&
  Sheth}{2012}]{Paranjape2012}
Paranjape A.,  Sheth R.~K.,  2012, \mn@doi [\mnras]
  {10.1111/j.1365-2966.2012.21911.x}, 426, 2789

\bibitem[\protect\citeauthoryear{Paranjape, Sheth  \& Desjacques}{Paranjape
  et~al.}{2013}]{ESP2013}
Paranjape A.,  Sheth R.~K.,   Desjacques V.,  2013, \mn@doi [\mnras]
  {10.1093/mnras/stt267}, 431, 1503

\bibitem[\protect\citeauthoryear{{Peacock} \& {Heavens}}{{Peacock} \&
  {Heavens}}{1990}]{PH90}
{Peacock} J.~A.,  {Heavens} A.~F.,  1990, \mn@doi [\mnras]
  {10.1093/mnras/243.1.133}, \href
  {https://ui.adsabs.harvard.edu/abs/1990MNRAS.243..133P} {243, 133}

\bibitem[\protect\citeauthoryear{{Pichon}, {Pogosyan}, {Kimm}, {Slyz},
  {Devriendt}  \& {Dubois}}{{Pichon} et~al.}{2011}]{pichonetal11}
{Pichon} C.,  {Pogosyan} D.,  {Kimm} T.,  {Slyz} A.,  {Devriendt} J.,
  {Dubois} Y.,  2011, \mn@doi [\mnras] {10.1111/j.1365-2966.2011.19640.x},
  \href {http://cdsads.u-strasbg.fr/abs/2011MNRAS.tmp.1739P} {pp 2493--2507}

\bibitem[\protect\citeauthoryear{{Planck Collaboration}}{{Planck
  Collaboration}}{2018a}]{PlanckDust2018}
{Planck Collaboration} 2018a, arXiv e-prints, \href
  {https://ui.adsabs.harvard.edu/abs/2018arXiv180104945P} {p. arXiv:1801.04945}

\bibitem[\protect\citeauthoryear{{Planck Collaboration}}{{Planck
  Collaboration}}{2018b}]{Planck2018}
{Planck Collaboration} 2018b, arXiv e-prints, \href
  {https://ui-adsabs-harvard-edu.insu.bib.cnrs.fr/\#abs/2018arXiv180706209P}
  {p. arXiv:1807.06209}

\bibitem[\protect\citeauthoryear{{Pogosyan}, {Bond}, {Kofman}  \&
  {Wadsley}}{{Pogosyan} et~al.}{1998}]{Pogosyanetal1998}
{Pogosyan} D.,  {Bond} J.~R.,  {Kofman} L.,   {Wadsley} J.,  1998, in
  {S.~Colombi, Y.~Mellier, \& B.~Raban} ed., Wide Field Surveys in Cosmology.
  p.~61 (\mn@eprint {} {arXiv:astro-ph/9810072})

\bibitem[\protect\citeauthoryear{{Pogosyan}, {Gay}  \& {Pichon}}{{Pogosyan}
  et~al.}{2009a}]{pogo09b}
{Pogosyan} D.,  {Gay} C.,   {Pichon} C.,  2009a, \mn@doi [\prd]
  {10.1103/PhysRevD.80.081301}, \href
  {https://ui.adsabs.harvard.edu/abs/2009PhRvD..80h1301P} {80, 081301}

\bibitem[\protect\citeauthoryear{{Pogosyan}, {Pichon}, {Gay}, {Prunet},
  {Cardoso}, {Sousbie}  \& {Colombi}}{{Pogosyan} et~al.}{2009b}]{pogo09}
{Pogosyan} D.,  {Pichon} C.,  {Gay} C.,  {Prunet} S.,  {Cardoso} J.~F.,
  {Sousbie} T.,   {Colombi} S.,  2009b, \mn@doi [\mnras]
  {10.1111/j.1365-2966.2009.14753.x}, \href
  {http://cdsads.u-strasbg.fr/abs/2009MNRAS.396..635P} {396, 635}

\bibitem[\protect\citeauthoryear{Pranav, Edelsbrunner, van~de Weygaert, Vegter,
  Kerber, Jones  \& Wintraecken}{Pranav et~al.}{2017}]{Pranav2016}
Pranav P.,  Edelsbrunner H.,  van~de Weygaert R.,  Vegter G.,  Kerber M.,
  Jones B. J.~T.,   Wintraecken M.,  2017, \mn@doi [\mnras]
  {10.1093/mnras/stw2862}, 465, 4281

\bibitem[\protect\citeauthoryear{Press \& Schechter}{Press \&
  Schechter}{1974}]{press_formation_1974}
Press W.~H.,  Schechter P.,  1974, \mn@doi [\apj] {10.1086/152650}, 187, 425

\bibitem[\protect\citeauthoryear{{Prunet}, {Pichon}, {Aubert}, {Pogosyan},
  {Teyssier}  \& {Gottloeber}}{{Prunet} et~al.}{2008}]{prunetetal08}
{Prunet} S.,  {Pichon} C.,  {Aubert} D.,  {Pogosyan} D.,  {Teyssier} R.,
  {Gottloeber} S.,  2008, \mn@doi [\apjs] {10.1086/590370}, \href
  {http://cdsads.u-strasbg.fr/abs/2008ApJS..178..179P} {178, 179}

\bibitem[\protect\citeauthoryear{{Ramakrishnan}, {Paranjape}, {Hahn}  \&
  {Sheth}}{{Ramakrishnan} et~al.}{2019}]{Ramakrishnan2019}
{Ramakrishnan} S.,  {Paranjape} A.,  {Hahn} O.,   {Sheth} R.~K.,  2019, arXiv
  e-prints, \href {https://ui.adsabs.harvard.edu/abs/2019arXiv190302007R} {p.
  arXiv:1903.02007}

\bibitem[\protect\citeauthoryear{Robertson, Kravtsov, Tinker  \&
  Zentner}{Robertson et~al.}{2009}]{RobertsonEtal2008}
Robertson B.~E.,  Kravtsov A.~V.,  Tinker J.,   Zentner A.~R.,  2009, \mn@doi
  [Astrophys. J.] {10.1088/0004-637X/696/1/636}, 696, 636

\bibitem[\protect\citeauthoryear{{Rodriguez-Gomez} et~al.,}{{Rodriguez-Gomez}
  et~al.}{2015}]{rodriguez-gomez_merger_2015}
{Rodriguez-Gomez} V.,  et~al., 2015, \mn@doi [\mnras] {10/f7dxn4}, 449, 49

\bibitem[\protect\citeauthoryear{Rossi}{Rossi}{2013}]{Rossi:2013fk}
Rossi G.,  2013, Monthly Notices of the Royal Astronomical Society, 430, 1486

\bibitem[\protect\citeauthoryear{{Shaw}, {Sigurdson}, {Pen}, {Stebbins}  \&
  {Sitwell}}{{Shaw} et~al.}{2014}]{Chime2014}
{Shaw} J.~R.,  {Sigurdson} K.,  {Pen} U.-L.,  {Stebbins} A.,   {Sitwell} M.,
  2014, \mn@doi [\apj] {10.1088/0004-637X/781/2/57}, \href
  {https://ui.adsabs.harvard.edu/abs/2014ApJ...781...57S} {781, 57}

\bibitem[\protect\citeauthoryear{Shen, Abel, Mo  \& Sheth}{Shen
  et~al.}{2006}]{ShenEtal2006}
Shen J.,  Abel T.,  Mo H.,   Sheth R.~K.,  2006, \mn@doi [Astrophys. J.]
  {10.1086/504513}, 645, 783

\bibitem[\protect\citeauthoryear{Sheth \& {van de Weygaert}}{Sheth \& {van de
  Weygaert}}{2004}]{sheth_hierarchy_2004}
Sheth R.~K.,  {van de Weygaert} R.,  2004, \mn@doi [\mnras] {10/drkqbz}, 350,
  517

\bibitem[\protect\citeauthoryear{{Sheth}, {Mo}  \& {Tormen}}{{Sheth}
  et~al.}{2001}]{SMT01}
{Sheth} R.~K.,  {Mo} H.~J.,   {Tormen} G.,  2001, \mn@doi [\mnras]
  {10.1046/j.1365-8711.2001.04006.x}, \href
  {https://ui.adsabs.harvard.edu/abs/2001MNRAS.323....1S} {323, 1}

\bibitem[\protect\citeauthoryear{{Sousbie}}{{Sousbie}}{2011}]{Sousbie2011}
{Sousbie} T.,  2011, \mn@doi [\mnras] {10.1111/j.1365-2966.2011.18394.x}, \href
  {https://ui.adsabs.harvard.edu/abs/2011MNRAS.414..350S} {414, 350}

\bibitem[\protect\citeauthoryear{{Sousbie}, {Pichon}  \& {Kawahara}}{{Sousbie}
  et~al.}{2011}]{sousbie10}
{Sousbie} T.,  {Pichon} C.,   {Kawahara} H.,  2011, \mn@doi [\mnras]
  {10.1111/j.1365-2966.2011.18395.x}, \href
  {http://cdsads.u-strasbg.fr/abs/2011MNRAS.414..384S} {414, 384}

\bibitem[\protect\citeauthoryear{{Springel}, {Yoshida}  \& {White}}{{Springel}
  et~al.}{2001}]{Gadget2001}
{Springel} V.,  {Yoshida} N.,   {White} S.~D.~M.,  2001, New Astronomy, \href
  {http://cdsads.u-strasbg.fr/cgi-bin/nph-bib_query?bibcode=2001NewA....6...79S&db_key=AST}
  {6, 79}

\bibitem[\protect\citeauthoryear{Toomre \& Toomre}{Toomre \&
  Toomre}{1972}]{toomre_GalacticBridgesTails_1972}
Toomre A.,  Toomre J.,  1972, \mn@doi [The Astrophysical Journal]
  {10.1086/151823}, 178, 623

\bibitem[\protect\citeauthoryear{{Vitvitska}, {Klypin}, {Kravtsov}, {Wechsler},
  {Primack}  \& {Bullock}}{{Vitvitska} et~al.}{2002}]{Vitvitska2002}
{Vitvitska} M.,  {Klypin} A.~A.,  {Kravtsov} A.~V.,  {Wechsler} R.~H.,
  {Primack} J.~R.,   {Bullock} J.~S.,  2002, \mn@doi [\apj] {10.1086/344361},
  \href {https://ui.adsabs.harvard.edu/abs/2002ApJ...581..799V} {581, 799}

\bibitem[\protect\citeauthoryear{{White} \& {Frenk}}{{White} \&
  {Frenk}}{1991}]{White1991}
{White} S. D.~M.,  {Frenk} C.~S.,  1991, \mn@doi [\apj] {10.1086/170483}, \href
  {https://ui.adsabs.harvard.edu/abs/1991ApJ...379...52W} {379, 52}

\bibitem[\protect\citeauthoryear{Zhang}{Zhang}{2015}]{zhang2015}
Zhang L.,  2015, Volumes of orthogonal groups and unitary groups (\mn@eprint
  {arXiv} {1509.00537})

\bibitem[\protect\citeauthoryear{van~de Weygaert et~al.}{van~de Weygaert
  et~al.}{2011}]{vandeWeygaert2013}
van~de Weygaert R.,  et~al., 2011, Trans. Comput. Sci., 14, 60

\makeatother
\end{thebibliography}

\appendix

\section{Critical events in ND}
\label{sec:NDevents}
For the sake of completeness and possible interest in other fields of research, let us present
the one-point statistics of critical events in arbitrary dimensions.
We first generalise the spectral parameters relevant to the critical event theory in $d$ dimensions in \cref{sec:spectral-params-d}.
We then proceed to derive the joint PDFs of the field and its second derivatives in \cref{sec:field-second-deriv}, and its first and third derivatives in \cref{sec:defQprobOdd}.
These results are then used in \cref{sec:NDcounts} to derive the critical event number counts in higher dimensions.
From this, we then proceed to
provide asymptotic formulas in the high density limit (\cref{sec:asymptotics}),
compute the ratios of critical events (\cref{sec:ratio-critical-events-ND}),
and establish the connection between critical points counts and critical events in any dimension (\cref{sec:self}).
In \cref{sec:nce_versus_nsp}, we finally provide a confirmation of the net merger density derived using the number counts of critical points in 3D.

\subsection{Spectral parameters}
\label{sec:spectral-params-d}
In this section we provide definitions for the spectral parameters of a $d$ dimensional Gaussian random field.
Let us first define the variance of the $i$-th derivative of the field
\begin{equation}
 \hskip -0.2cm \sigma^2_i(R) \!=\! \frac{d}{(4\pi)^{d/2}\Gamma \left(1+\frac{d}{2}\right)} \int_0^\infty\!\!\! \!\!\dd{k} k^{d-1} P_k(k) k^{2i}W^2(kR),
\end{equation}
where $P_k(k)$ is the ND power spectrum and $W(kR) = \exp(-(kR)^2/2)$.
The characteristic scales $R_0$, $R_*$ and $\tilde{R}$ are defined by \cref{eq:defR0}
and the spectral parameters $\gamma$ and $\tilde\gamma$ are defined by \cref{eq:gammadef}.
In $d$ dimensions for a power-law power spectrum with index $n$, we have
\begin{gather}
  \nonumber
  \frac{R_0^2}{R^2} = \frac{2}{n+d}, \quad \frac{R_*^2}{R^2} = \frac{2}{n+d+2}, \quad \frac{\tilde{R}^2}{R^2} = \frac{2}{n+d+4},\\
  \gamma^2 = \frac{n+d}{n+d+2}, \quad \tilde{\gamma}^2 = \frac{n+d+2}{n+d+4}.
\end{gather}

\subsection{Joint PDF of the field and its second derivatives}
\label{sec:field-second-deriv}
From \cite{pogo09} the joint distribution function of the set of $d$ eigenvalues of the $d$ dimensional Hessian $\sigma_2\vvec{\lambda}$  and
density $\nu$ is
\begin{equation}
{ P}(\nu,\vvec{\lambda}) = \frac{1}{\mathcal N}
 \Delta(\vvec{\lambda})
  \exp\left( -\frac{1}{2}
Q_{\gamma}(\nu,\vvec{\lambda})
\right)\,,  \label{eq:defQprob}
\end{equation}
where $\vvec{\lambda}=\{\lambda_i\}_{i=1\dots d}$, $\Delta(\vvec{\lambda})=\prod_{i<j}(\lambda_j-\lambda_i)$ is the Vandermonde determinant, and
$Q_{\gamma}$ is a quadratic form in $\lambda_{i}$ and $\nu$ given by
\begin{equation}
Q_{\gamma}(\nu,\vvec{\lambda})=\nu^2+ \frac{\left(\sum_{i}\lambda_{i}+\gamma \nu\right)^2}{ (1-\gamma^2)}+
{\cal Q}_{d}(\vvec{\lambda})\,, \label{eq:Qgam}
 \end{equation}
with
\begin{align}
{\cal Q}_{d}(\vvec{\lambda})
&= \frac{d(d+2)}{ 2}\left[ \sum_i \lambda_i^2 - \frac{1}{ d}\left(\sum_i \lambda_i\right)^2\right] \nonumber \\
&= (d+2)\left[\frac{1}{2}(d-1) \sum_{i} \lambda^{2}_{i}- \sum_{i < j} \lambda_{i}\lambda_{j}\right]
\label{eq:QJ2} 
\end{align}
proportional to the Euclidean norm of the detraced Hessian matrix. Note that the expression in \cref{eq:defQprob} assumes that the eigenvalues are sorted, otherwise the 
Vandermonde determinant would come with an absolute value. Finally ${\cal N}$ is a normalisation  quantified below. 

\subsubsection{Determining the PDF normalisation}
Directly integrating \cref{eq:defQprob} is not easy due to the presence of couplings between the different variables in $Q_\gamma$. However, this integral is actually related to the integral over the joint PDF of the field and its second derivatives in an arbitrary frame. Indeed, the expression of the PDF in \cref{eq:defQprob,eq:Qgam,eq:QJ2} was obtained after a change of variables from an arbitrary frame to the Hessian eigenframe, and the presence of the Vandermonde determinant is related to the orthogonality constraint of the matrix of eigenvectors \citep{doroshkevich70,BBKS,pogo09}. More precisely, taking into account the volume of integration over all possible eigenvectors that keep the diagonalisation of the Hessian unique, we have \citep{zhang2015}:
\begin{align}
\mathcal{I} &= \int \dd{\vvec{y}} \exp({-\frac{1}{ 2}\vvec{y}^\mathrm{T}\cdot\vb{\Sigma}^{-1}\cdot \vvec{y}}) \,,\nonumber \\
 &= \frac{\mathrm{Vol}(O(d))}{2^d} \int \dd{\nu}\prod_{i=1}^d \dd{\lambda_i} \Delta(\vvec{\lambda}) \exp\left(-\frac{Q_\gamma(\vvec{\lambda})}{ 2} \right),
\label{eq:Irot}
\end{align}
where $\vvec{y}=\{\nu,x_{ij},1\leq i \leq j \leq d\}$, and the volume of the orthogonal group in dimension $d$, $O(d)$, is given by:
\begin{equation}
    \mathrm{Vol}(O({d})) = \frac{2^d \pi^{d(d+1)/4} }{ \prod_{k=1}^d \Gamma(k/2)}.
\end{equation}
Note that in \cref{eq:Irot} the integral is made over the \emph{sorted} eigenvalues $\lambda_1 < \lambda_2 <\dots <\lambda_d$. The factor $2^d$ in the denominator comes from the fact that the diagonalisation mapping is unique if we sort the eigenvalues \emph{and} we choose the sign of each eigenvector. Finally, the matrix $\vb{\Sigma}$ above is the covariance matrix of the $\vvec{y}$ vector.

We note that the first line of \cref{eq:Irot} is nothing else than a multivariate Gaussian integral over $d(d+1)/2 + 1$ variables, so that its value is
\begin{equation}
\mathcal{I}=(2\pi)^{\frac{d(d+1) + 2}{ 4}}|\vb{\Sigma}|^{1/2}~.
\end{equation}
Comparing equations~\eqref{eq:Irot} and \eqref{eq:defQprob}, the normalisation
factor $\mathcal{N}$ can then be written as a function of the determinant of $\vb{\Sigma}$:
\begin{equation}
\mathcal{N} = \frac{2^d }{ \mathrm{Vol}(O({d}))} (2\pi)^{\frac{d(d+1) + 2}{ 4}}|\vb{\Sigma}|^{1/2}.
\end{equation}

\subsubsection{Determinant of $\vb{\Sigma}$}

To evaluate $|\vb{\Sigma}|$ we recall that simple computations in Fourier space
show that \citep{BBKS,pogo09}:
\begin{equation}
\langle x_{ij}x_{kl}\rangle = \frac{1}{ d(d+2)}\left(\delta_{ij}\delta_{kl} + \delta_{ik}\delta_{jl} + \delta_{il}\delta_{jk}\right)~,
\end{equation}
while $\langle x^2\rangle=1$ and $\langle x x_{ii} \rangle=-\gamma/d$.
Therefore, the matrix $\vb{\Sigma}$ is block diagonal, with a dense sub-block $\vb{\Sigma}_{\rm diag}$ corresponding to the field $x$ itself and the diagonal
elements of the Hessian $x_{ii}$, and a diagonal sub-block $\vb{\Sigma}_{\mathrm{offdiag}}$ corresponding to the $d(d-1)/2$ independent off-diagonal elements $x_{i<j}$ of the Hessian, each having equal variance $\left(d(d+2)\right)^{-1}$. The determinant of $\vb{\Sigma}$ thus factorizes into $|\vb{\Sigma}_\mathrm{diag}|\times |\vb{\Sigma}_\mathrm{offdiag}|$, where 
the determinant of the off-diagonal Hessian elements is simply given by 
\begin{equation}
|\vb{\Sigma}_\mathrm{offdiag}| = \left[d(d+2)\right]^\frac{-d(d-1)}{ 2}.
\label{eq:detOffdiag}
\end{equation}

Let us now compute the contribution of the dense covariance matrix $\vb{\Sigma}_\mathrm{diag}$ of the field and the diagonal elements $x_{ii}$ of the Hessian.
To compute its determinant, we will diagonalise the matrix using two successive steps. Let us first change to set of variables $\{\nu,I_1,x_{11}-I_1/d,\dots,x_{(d-1)(d-1)}-I_1/d\}$. The matrix $\vb{M}$ of the variable change reads:
\begin{equation}
   \vb{M}=
    \begin{pmatrix*}[c]
    1            & 0            & \dots        & \dots       & \dots        & 0 \\
    0            & 1            & \dots        & \dots       & \dots        & 1\\ 
    \vdots       & \frac{d-1}{ d} & -\frac{1}{ d}  & \dots       & \dots        & -\frac{1}{ d} \\
    \vdots       & -\frac{1}{ d}  & \frac{d-1}{ d} & -\frac{1}{ d} & \dots        & \vdots \\
    \vdots       & \vdots       & \ddots       & \ddots      & \ddots       & \vdots \\
    0            & -\frac{1}{ d}  & \dots        & -\frac{1}{ d} & \frac{d-1}{ d} & -\frac{1}{ d}
    \end{pmatrix*}.
    \label{eq:Mmatrix}
\end{equation}
It is straightforward to show that the Jacobian determinant $|\vb{M}|=1$, \eg by first casting the trace $I_1$ as the last variable, and by using the Schur complement to the diagonal element corresponding to the trace. In this new set of variables the covariance matrix reads:
\begin{equation}
    \vb{M}\vb{\Sigma}\vb{M}^\mathrm{T} = 
    \begin{pmatrix*}[c]
    \begin{pmatrix*}
    1  & -\gamma \\ 
    -\gamma    & 1
    \end{pmatrix*} & \vb{0} \\
    \vb{0} & \vb{\Xi}
    \end{pmatrix*},
\end{equation}
where 
\begin{equation}
    \vb{\Xi} = \frac{2}{d(d+2)}\vb{I}_{d-1} - \frac{2}{ d^2(d+2)}\vvec{u} \vvec{u}^\mathrm{T},
\end{equation}
and $\vvec{u}$ is a $d-1$ column vector with all elements equal to $1$. 

At the second step we diagonalise the matrix $\vb{\Xi}$ 
which is easily accomplished by collecting orthogonal bases of $\mathrm{Span}(\vvec{u})$ and $\mathrm{Span}(\vvec{u})^\perp$ respectively. The last $d-2$ eigenvalues are equal to ${2}/{ d(d+2)}$, while the first eigenvalue is equal to ${2}/{ d(d+2)} - {2(d-1)}/{ d^2(d+2)}={2}/{ d^2(d+2)}$. 
The determinant of $\vb{\Sigma}_\mathrm{diag}$ is therefore
\begin{equation}
|\vb{\Sigma}_\mathrm{diag}| = (1-\gamma^2) \left[\frac{d(d+2)}{2}\right]^{-(d-1)}d^{-1}.
\end{equation}

Putting everything together, we get the final expression of the PDF normalisation $\mathcal{N}$ entering \cref{eq:defQprob}
\begin{equation}
  \hskip -0.25cm
  \mathcal{N} \!=\! 2^\frac{d(d+3)}{ 4}(d(d+2))^{\!-\frac{d(d+1)-2}{ 4}}\!\sqrt{\frac{\pi}{ d}}\! \sqrt{1-\gamma^2}\prod_{k=1}^d\!\Gamma(k/2). 
\end{equation}

\subsection{Joint statistics of the first and third derivatives}
\label{sec:defQprobOdd}
\subsubsection{Expression for the joint PDF}
Here, we will look into the pdf of the first and third derivatives in $d$ dimensions in order to compute the odd derivative term $C_{\rm odd}$ that enters critical event number counts in $d$ dimensions.
Let us first describe the joint distribution globally. The structure of this distribution is quite similar to that of the field and its second derivatives that were presented in appendix~\ref{sec:field-second-deriv}. We have
\begin{equation}
P(\vvec{z}) = \frac{1}{ \sqrt{\left(2\pi\right)^n |\vb{\widetilde{\Sigma}}|} } \exp(-\frac{1}{2} \vvec{z}^\mathrm{T}\vb{\widetilde{\Sigma}}^{-1}\vvec{z})\,,
\label{eq:PDFodd}
\end{equation}
where $\vvec{z}=\{x_i,x_{ijk}, 1\leq i\leq j\leq k\leq d\}$ is the set of the first and third derivative tensors, and $\vb{\widetilde{\Sigma}}$ is the covariance matrix of $\vvec{z}$, and $n=d(d+1)(d+2)/6+1$ is the number of non-redundant terms in $\vvec{z}$.
As before, the first derivatives get coupled to the third derivatives only via traces of the latter. Indeed, it can be shown that the quadratic form of the Gaussian PDF reads \citep{pogo09}
\begin{align}
\vvec{z}^\mathrm{T}\vb{\widetilde{\Sigma}}^{-1}\vvec{z} &= d\, \mathrm{Tr} \left\{ \begin{bmatrix} 1 & -\tilde{\gamma} \\ -\tilde{\gamma} & 1\end{bmatrix}^{-1} \begin{bmatrix} x_i x_i & x_i x_{iaa} \\ x_i x_{ibb} & x_{icc}x_{idd} \end{bmatrix}\right\} \nonumber\\
& + \frac{d(d+2)(d+4)}{6} \bar{x}_{jkl}\bar{x}_{jkl},
\label{eq:QprobOdd_quad}
\end{align}
where summation is assumed over repeated indices, and $\bar{x}_{ijk} = x_{ijk} - 3x_{aa(i}\delta_{jk)}/(d+2)$ is the traceless part of $x_{ijk}$. 
Given that we have $d$ traces of
$x_{ijk}$, and that the space of traceless symmetric tensors of order
$3$ is of dimension
$w_3=d(d+1)(d+2)/6-d=d(d-1)(d+4)/6$, one expects the determinant of
$\vb{\widetilde{\Sigma}}$ to show factors $(1-\tilde{\gamma})^d
\left(d(d+2)(d+4)\right)^{-w_3}$.
Indeed, calculations similar to (but more cumbersome than) those of
appendix~\ref{sec:field-second-deriv} yield:
\begin{equation}
|\vb{\widetilde{\Sigma}}| = 3^d 2^{d(d-1)} (1-\tilde{\gamma})^d \left(d(d+2)(d+4)\right)^{-w_3} (d^2(d+2))^{-d}.
\label{eq:QprobOdd_det} \notag
\end{equation}
Together with equations~\ref{eq:PDFodd} and \ref{eq:QprobOdd_quad},
this fully describes the joint distribution of first and third
derivatives of the field. In the following, we will compute the
conditional statistics needed for the term $C_{\mathrm{odd}}$.

\subsubsection{Conditional statistics needed for $C_{\mathrm{odd}}$}

First, let us note that the first derivatives are Gaussian distributed with individual variance $\left\langle x_{i}^{2} \right\rangle=1/d$ so that the probability density of first derivatives 
near the configuration when they all vanish is 
\begin{equation}
\label{eq:grad0}
    { P}(x_i=0)=\left(\frac d {2\pi}\right)^{d/2}.
\end{equation}

Now let us specify the different statistics of the third derivatives.
By symmetry, one can note that
\begin{equation}
\left\langle \left(\sum_{i}x_{1ii}\right)^{2} \right\rangle=\frac 1 d,
\end{equation}
since the third derivatives are rescaled by $\sigma_{3}$,
and
\begin{equation}
\left\langle x_{1jj}^{2} \right\rangle=\left\langle x_{111}x_{1jj} \right\rangle=\frac 1 5 \left\langle x_{111}^{2} \right\rangle= 3 \left\langle x_{1jj} x_{1kk} \right\rangle \quad \forall j\neq k\neq 1.\notag
\end{equation}
Therefore,
\begin{multline}
\frac 1 d = \left\langle x_{111}^{2} \right\rangle+(d-1) \left\langle x_{1jj}^{2} \right\rangle+2 (d-1)\left\langle x_{111}x_{1jj} \right\rangle\\
+(d-1)(d-2)\left\langle x_{1kk}x_{1jj} \right\rangle \quad \forall j\neq k\neq 1 \notag
\end{multline}
implies that $\left\langle x_{iii}^{2} \right\rangle={15}/{d(d+2)(d+4)}$
and all terms of the third derivatives covariance have been specified.

However, we are interested in statistics subject to a zero gradient constraint, in particular the three quantities of interest are
(choosing the last dimension $d$ as the degenerate one and assuming an implicit summation on the $i$ indices)
\begin{align}
\left\langle x_{ddd}^{2} |x_{d}=0\right\rangle
&=
\left\langle x_{ddd}^{2} \right\rangle
-\frac{\left\langle x_{dii}x_{d}\right\rangle^{2}}{\left\langle x_{d}^{2} \right\rangle},\label{eq:tmp-b10}\\
\left\langle\left(x_{dii}\right)^{2}  |x_{d}=0\right\rangle&=\left\langle \left(x_{dii}\right)^{2} \right\rangle-\frac{\left\langle x_{ddd}x_{d} \right\rangle^{2}}{\left\langle x_{d}^{2} \right\rangle}, \label{eq:tmp-b11}\\
\left\langle x_{dii}x_{ddd}  |x_{d}=0\right\rangle&=\left\langle x_{dii}x_{ddd}  \right\rangle-\frac{\left\langle x_{d} x_{ddd}\right\rangle\left\langle x_{d} x_{dii} \right\rangle}{\left\langle x_{d}^{2} \right\rangle},
\end{align}
which can easily be computed thanks to the additional relation $\left\langle x_{ii}^{2} \right\rangle=3/d(d+2)$,
\begin{align}
\label{eq:oddapprox}
\left\langle x_{ddd}^{2} |x_{d}=0\right\rangle
&=\frac 3 {d(d+2)}\left[\frac{5}{d+4}-\frac{3\tilde\gamma^{2}}{d+2}\right],\\
\left\langle\left(x_{dii}\right)^{2}  |x_{d}=0\right\rangle&=\frac{1-\tilde\gamma^{2}}{d},\\
\left\langle x_{dii}x_{ddd}  |x_{d}=0\right\rangle&=\frac 3 {d(d+2)}(1-\tilde\gamma^{2}).
\end{align}

\subsection{Critical event number counts in ND}
\label{sec:NDcounts}
It now follows that the critical event  number counts of type $j$ at height $\nu$ in dimension $d$ reads 
\begin{equation}
\hskip -0.15cm
n_{d,\ce}^{(j)}(\nu)= \frac{ R \, 
 }{{\tilde R}^2 R_*^d  }C^d_{{\rm odd}}C^d_{j,{\rm even}}(\nu) ~. 
\label{eq:NDdiff}
\end{equation}

The contribution from the odd part of the distribution function, $C_{d,{\rm odd}}$ depends on
whether we consider total critical point count or net merger events, but can be obtained in
a closed analytical form for arbitrary $d$ in both cases. 

To count net merger events as defined in \cref{eq:dndRmaintext} we evaluate $C_{d,{\rm odd}}$ as
\begin{equation}
C^d_{{\rm odd}}= \left\langle  {\sum_i x_{jii}}\,{x_{jjj}} \dirac^{(d)}(x_i) \right\rangle\,. \label{eq:NDdiffodd}
\end{equation}
where the expectation in \cref{eq:NDdiffodd} should be computed using the results for odd-order derivatives given in \cref{sec:defQprobOdd}.
Note that due to symmetries, the result does not depend on $j$.
Using  \cref{eq:grad0,eq:oddapprox}, we get
\begin{equation}
\label{eq:Coddd}
C^d_{{\rm odd}} = \frac 3 {d(d+2)}\left(\frac d {2\pi}\right)^{d/2}(1-\tilde\gamma^{2}),
\end{equation}
which is analogous to \cref{eq:C1even_nu} in $d$ dimensions.

If we are counting total density of critical events instead (\cref{eq:eventcount_covariant} in $d$ dimensions),
one is led to introduce
\begin{equation}
 C^{\ce,d}_{{\rm odd}} = \left\langle   \!\abs{\sum_i x_{jii}}\abs{x_{jjj}} \dirac^{(d)}(x_i) \right\rangle\,, \label{eq:NDdiffoddce}
\end{equation}
where once again the final results does not depend on $j$.
After a bit of algebra,
\begin{align}
C^{\ce,d}_{{\rm odd}}&= \Big(\frac{d}{2 \pi}\Big)^{\tfrac{d}{2}}
\frac{2 \sqrt{6}}{\pi } 
{ \sqrt{\frac{(d-1)
   \left(1-\tilde{\gamma
   }^2\right)}{d^2 (d+2)^2
   (d+4)}}}+\\
 & \hskip -0.5cm  \Big( \frac{d}{2 \pi}\Big)^{\tfrac{d}{2}} \frac{6
 \left(1\!-\!\tilde{\gamma }^2\right)}{\pi  d
   (d+2)}{
   \tan
   ^{-1}\!\left(\!\sqrt{\frac{3}{2}}\frac{ \sqrt{d\!+\!4}\sqrt{1-{\tilde\gamma}^2}}
   {\sqrt{d\!-\!1}}\!\right)}\notag \,.
\end{align}

The contribution from the even, density threshold dependent term, $C^d_{j,{\rm even}}(\nu)$ is given by
\begin{equation}
\label{eq:Cevend}
C^d_{j,{\rm even}}(\nu)=\!\left\langle \!  {\dirac(x-\nu)} \dirac(\lambda_{j}) 
\left| \prod_{i\neq j}\!  \lambda_i \right|\!\right\rangle, 
\end{equation}
where the condition of critical point of type $j$ refers to the vanishing eigenvalue in
the ordered list $\lambda_1 \le \lambda_2 \le \ldots \le \lambda_d$, $j=d$
corresponds to peak-filament mergers.
The expectation value in \cref{eq:Cevend} is computed using the distribution function in \cref{eq:defQprob}.

\begin{figure}
  \includegraphics[width=\columnwidth]{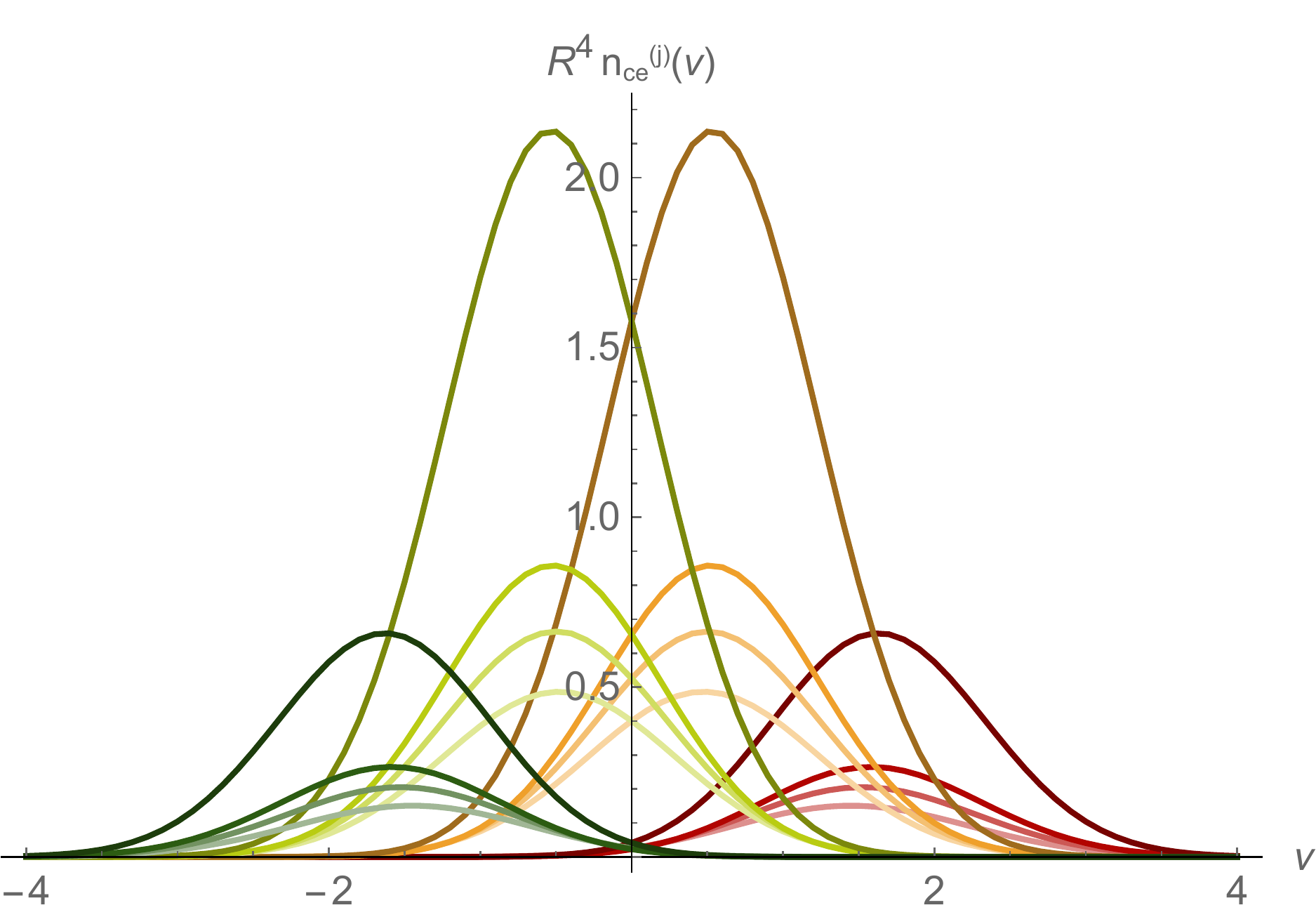}
  \includegraphics[width=\columnwidth]{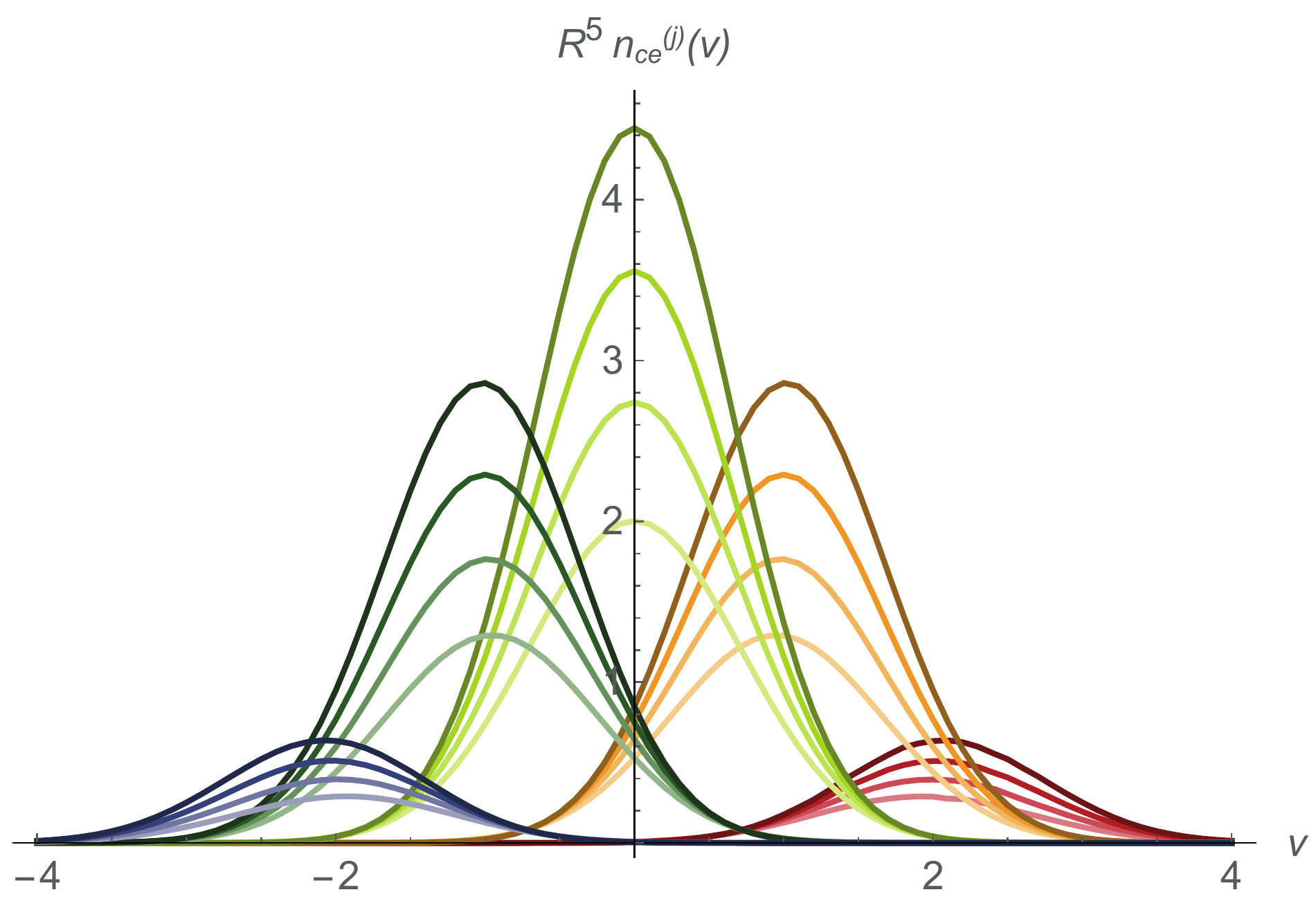}
    \includegraphics[width=\columnwidth]{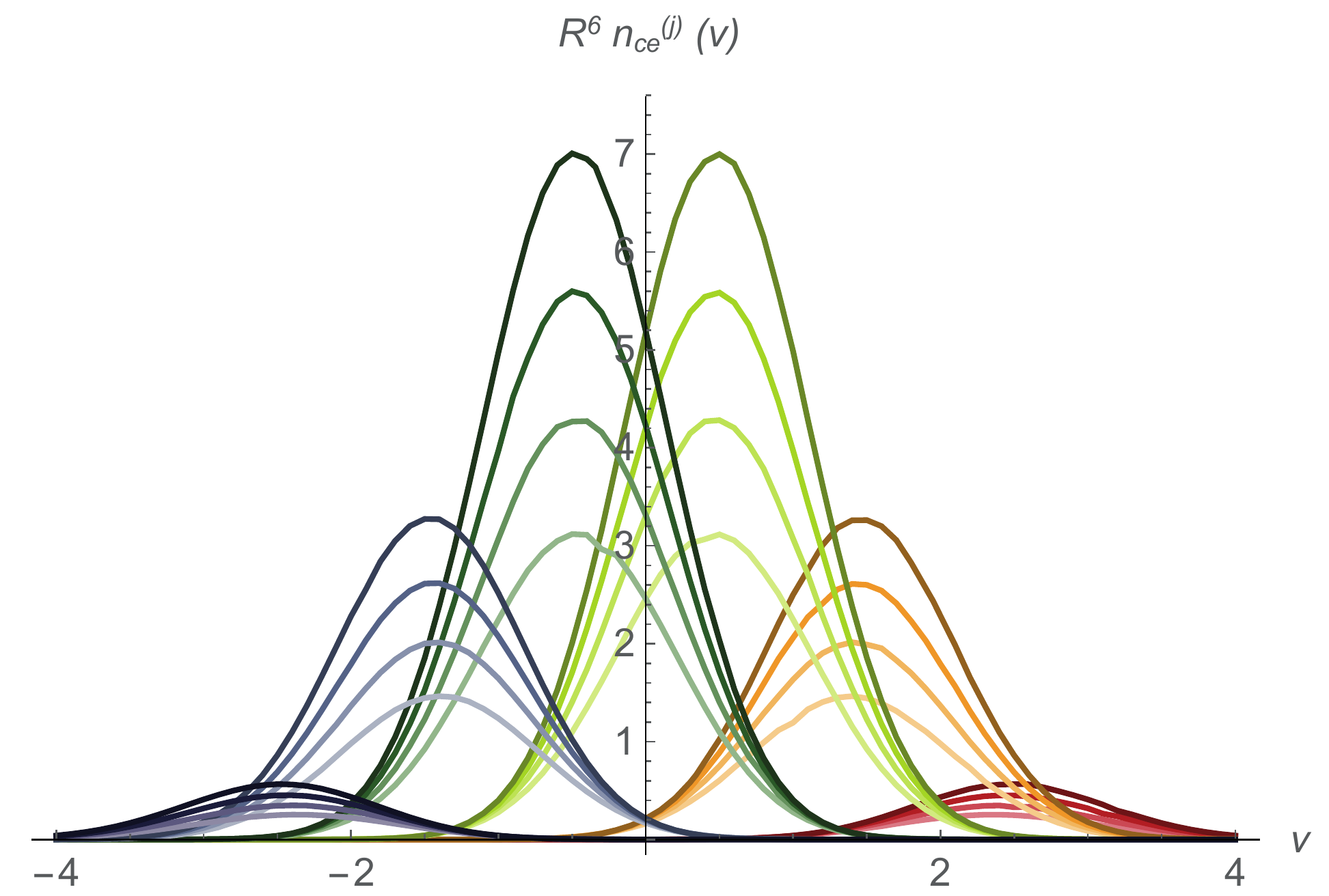}
  \caption{ The PDF of critical events of the various types (${\cal P,F},{\cal W}_1, {\cal W}_2$) in 3+1D (\emph{top}), in 4+1D (\emph{middle}) and 5+1D (\emph{bottom})
  for $n_\mathrm{s}=-2,-3/2,-1,-1/2$ from light to dark.}
  \label{fig:PDF-4D-5D}
\end{figure}

$C^d_{j,{\rm even}}(\nu)$ is a non-trivial function of $\nu$ because of the correlation between $\nu$ and $\sum_i \lambda_i$ seen in \cref{eq:Qgam}.
It does not allow for an exact analytical form, however we can obtain the
asymptotic behaviour of $C^d_{j,{\rm even}}(\nu)$ at large overdensities $\nu$, as will be shown below .
The PDFs of total critical events in 3+1D, 4+1D and 5+1D can be obtained numerically using \cref{eq:defQprob,eq:NDdiff,eq:Coddd}, and are shown in \cref{fig:PDF-4D-5D}. Note that the intermediate signature events dominate in number over the extreme ones, in accordance with the relative number of critical points.

\subsection{Asymptotics}
\label{sec:asymptotics}
In the large $\nu$ limit, the number density of peak-filament mergers in $d$ dimensions will now be shown  to scale like
\begin{equation}
\label{eq:Cevend-largenu} \hskip -0.01cm
C^d_{j,{\rm even}}(\nu) \!\!\underset{\gamma\nu \to \infty}{\propto}\!\! \left(\gamma\nu\right)^{2(d-1)}\exp\left(-\frac 1 2 \frac{\nu^{2}}{1- \displaystyle\frac{d+2}{3d}\gamma^2}\right).
\end{equation}

To get to \cref{eq:Cevend-largenu},  first note 
that  in $d$ dimensions  the average over the full range of eigenvalues of any monomial $\prod_i \lambda_i^{n_i}$
behaves as
\begin{equation}
\left\langle \prod_i \lambda_i^{n_i} \right\rangle \propto (\gamma \nu)^{\sum_i \! n_i} e^{-\nu^2/2} \,,
\label{eq:high_nu_all_lambda}
\end{equation}
in the high-$\nu$ limit.
This follows from rewriting the exponential argument in \cref{eq:Qgam,eq:QJ2} of the
distribution in \cref{eq:defQprob}
 in terms of uncorrelated d-dimensional Hessian invariants \citep{pogo09b}
$J_1^{(d)} = \sum_i \lambda_i$ and
$J_{2}^{(d)}=\left(\sum_i \lambda_i \right)^{2}-\displaystyle \frac{2d}{d-1}\sum_{i<j} \lambda_i \lambda_j$ as
\vskip -0.3cm
\begin{equation}
 Q_{d}(\nu,\vvec{\lambda})=\frac{1}{2}(d+2)(d-1) J_2^{(d)} ~,
\label{eq:QdJ}
\end{equation}
so that 
\begin{equation}
\hskip -0.2cm Q_{\gamma}(\nu,\vvec{\lambda})=\nu^2+ \frac{\left(J_1^{(d)}+\gamma \nu\right)^2}{ (1-\gamma^2)}+\frac{1}{2}(d+2)(d-1) J_2^{(d)} ~,
\label{eq:QgamJ}
\end{equation}
where $J_2^{(d)}$ is also uncorrelated with the overdensity $\nu$.
In the limit $\nu \to \infty$, $J_1^{(d)} \to -\infty$ and if all eigenvalues 
$\lambda_i$ are unrestricted,
the exact boundaries
of integration in $\lambda_i$ space become irrelevant. The average over 
$J_1^{(d)}$ and $J_2^{(d)}$ gives a power law in $\nu$, while
the factored out $\nu^2$ term in $Q_{\gamma}$ is responsible for the
exponential `controlling factor' of the asymptotic behaviour
$e^{-\nu^2/2}$. A classical example of this
situation is found in the peak counts \citep{BBKS}. 
For high peaks, all eigenvalues 
tend to be large and negative, and asymptotically yield \cref{eq:high_nu_all_lambda}.

The situation changes when one or more eigenvalues are restricted to 
remain small and/or positive.  This is the case for critical events where $\lambda_j = 0$, and thus $\lambda_i \ge 0$ for all
$ i \ge j$. In  the  $\nu \to \infty$, $J_1^{(d)} \to -\infty$ limit, only
the subset of $j-1$ eigenvalues $\lambda_{i<j}$  becomes large and negative,
so the average over the Hessian terms is effectively restricted to a subspace
of dimension $j-1$. This affects the asymptotics, since $J_1^{(d)}$ and
$J_2^{(d)}$ are  correlated when projected to a lower dimensional
hypersurface. Instead, we need to rewrite the PDF using combinations of $J_1^{(j-1)}$ and
$J_2^{(j-1)}$. 

In the case of peak filament critical events $j=d$. Setting $\lambda_d=0$ leads to the following transformation
 \[
J_1^{(d)} \to J_1^{(d-1)}\,\,, \mbox{but}\,\,\,
J_2^{(d)} \to \frac{1}{(d-1)^2} J_1^{(d-1)} + \frac{d (d-2)}{(d-1)^2} J_2^{(d-1)}\,,
\] which displays a coupling to $J_1$. 
Closing the square term for $J_1^{(d-1)}$ in \cref{eq:QgamJ}
now gives
\begin{multline}
{\cal Q}_{\gamma}(\{\lambda_{d-1}\},0)=
\frac{3d}{3d-(d+2) \gamma^2} \nu^2
+\frac{d(d-2)(d+2)}{2(d-1)}J_{2}^{(d-1)}
\\
+\left(J_{1}^{(d-1)}+\frac{2(d-1)\gamma\nu}{3d-(d+2)\gamma^{2}}\right)^{2}\frac{3d-(d+2)\gamma^{2}}{2(d-1)(1-\gamma^{2})}.
\label{eq:Qd-1}
\end{multline}
which yields a new coefficient in front of $\nu^2$. 
The averaging in \cref{eq:Cevend}
leaves this term as  controlling  the exponential factor of
the $\nu \to \infty$ asymptote, and yields a polynomial in $\gamma\nu$ 
scaling like
$\propto (\gamma \nu)^{2(d-1)}$, as stated in \cref{eq:Cevend-largenu}, given  that the Dirac in $\lambda_d$ 
changes the measure, $\prod_{i<j\le d} (\lambda_i-\lambda_j)$ to $\prod_{i< j \le d-1} (\lambda_i-\lambda_j)\times\prod_{i< d} \lambda_i$,
hence the extra factor $(\gamma \nu)^{(d-1)}$.

Incidentally, a similar situation arises when computing the number density of filamentary
saddle points \citep{Gay2012},  where the largest eigenvalue, though not zero, 
is still restricted to positive values, leading to an effective change of
dimension by one, and asymptotes with the same exponential behaviour as \cref{eq:Cevend-largenu}.

\subsection{Ratios of critical events}
\label{sec:ratio-critical-events-ND}
From \cref{eq:defQprob}, the integration over $\nu$ yields the marginal
probability of  $\vvec{\lambda}$, which, up to a multiplicative constant, reads:
\begin{equation}
  \prod_{i\le {d}}\!  d \lambda_i \!
 \prod_{i<j} (\lambda_j\!-\!\lambda_i)
  \exp\left(\! -\frac{1}{2}
{\cal Q}_{d}(\vvec{\lambda})-\!\frac{1}{2}\!\! \left(\!\sum_i \lambda_i \!\right)\!^2
\right). \label{eq:defQprob2}
\end{equation}
Finally, the $d$ dimensional  ratio of critical events of type $j$ and $k$ is simply given by
\begin{equation}
r _{j/k}\! = \!\left\langle \! \!  \dirac(\lambda_{j})  \prod_{i\neq j} \big|\! \lambda_i \big|\! \right\rangle\!\!\Big/\!
\!\left\langle \!   \dirac(\lambda_{k}) \prod_{i\neq k} \big| \! \lambda_i \big|\! \right\rangle
\notag\label{eq:defratioND}
\end{equation}
where the PDF to evaluate this expectation is given by \cref{eq:defQprob2}.
Note that these counts correspond to the area below each curve shown in \cref{fig:PDF-4D-5D}.
In 2+1D, we recover the ratio presented in the main text.
In 3+1D the ratio is analytic and reads ${2 (57\!+\!25 \pi\!-\!50 \cot^{-1}(3))}\big/({75 \pi\!-\!2(57\!+\!50 \cot^{-1}(2))})\approx 3.17$.
More generally,
\begin{align}
d=2:&\quad r_{\cal F\!/\!W}= 1 \,,\notag\\
d=3:&\quad r_{\cal F\!/\!P}= 2.06 \,,\notag\\
d=4:&\quad r_{\cal F\!/\!P}= 3.17 \,,\quad r_{\cal W\!/\!P} =3.17\,,\notag\\
d=5:&\quad r_{\cal F\!/\!P}= 4.36 \,,\quad r_{{\cal W}_1\!/{\cal P}} =6.72\,, \quad r_{{\cal W}_2\!/{\cal P}} =4.36 \,,\notag\\
d=6:&\quad r_{\cal F\!/\!P}= 5.67 \,,\quad r_{{\cal W}_1\!/{\cal P}} =11.97\,,\!\! \quad r_{{\cal W}_2\!/{\cal P}} =11.97, \notag
\end{align}
and $\quad r_{{\cal W}_3\!/{\cal P}} =5.67$.
Note that these ratios are pure numbers and do not depend on the detailed shape of the underlying power spectrum.

\subsection{Self-consistency links with critical points counts} \label{sec:self}
The results of this paper can be used to derive the connectivity as defined in \cite{codis2018}.
Indeed, let us formally write $n_\cp^{(i)}$ the number density of critical point of kind $i$ in $d$ dimensions and %
$n_{\me}^{(i)}$
the net number density of critical event of kind $d-i+1$. The evolution of $n_\cp^{(i)}$ is given by
\begin{equation}
  \frac{\mathrm{d} n_\cp^{(i)}}{\mathrm{d} R} = -\begin{cases}
    n_{\me}^{(1)} & \text{if $i=0$,} \\
    n_{\me}^{(i)} + n_{\me}^{(i+1)} & \text{if $0<i<d-1$,} \\
    n_{\me}^{(d)} & \text{if $i=d-1$.}
  \end{cases}
  \label{eq:crit-pt-crit-event-relation}
\end{equation}
For Gaussian random fields, the number density of critical point can be formally written as
\begin{equation*}
  n_\cp^{(i)} = \frac{1}{R_*^d} \underbrace{\Big\langle \Big|\prod_j \lambda_j\Big|\Big\rangle\Big\langle\dirac^{(3)}(x_i)\Big\rangle}_{C_i},
\end{equation*}
where the PDF to evaluate the left part of the r.h.s.\ is given by \cref{eq:defQprob2}.
Here $C_i$ is a number common to all power spectra.
The derivative of $n_\cp^{(i)}$ with respect to the smoothing scale is then
\begin{equation}
  \frac{\mathrm{d} n_\cp^{(i)}}{\mathrm{d} R} = -n_\cp^{(i)} \times d \dv{\log R_*}{R}.
  \label{eq:crit-pt-number-dens-derivative2}
\end{equation}
  Using \cref{eq:crit-pt-crit-event-relation} and \cref{eq:crit-pt-number-dens-derivative2} yields a simple relation between the number density of critical points and the number density of critical events
\begin{equation}
  n_\cp^{(i)}= \frac{1}{d \times \dv*{\log R_*}{R}}\begin{cases}
    n_\me^{(1)} & \text{if $i=0$,} \\
    n_\me^{(i)} + n_\me^{(i+1)} & \text{if $0<i<d-1$,} \\
    n_\me^{(d)} & \text{if $i=d-1$.}
  \end{cases} \notag
\end{equation}
  For Gaussian random fields, one has the property that $n_\cp^{(i)}= n_\cp^{(d-i)}$ (with $i\in\{0,\dots,d\}$), and $n_\me^{(i)} = n_\me^{(d-i+1)}$ (with $i\in\{1,\dots,d\}$).
This provides us with simple way to compute the ratio of critical events as a function of the ratio of the critical points. For any $d$, the ratio of filament to peak is connected to the ratio of ${\cal F}$ to ${\cal P}$ critical events
\begin{equation}
  \frac{n_\cp^{\rm f}}{n_\cp^{\rm p}} = \frac{n_\me^{\cal P} + n_\me^{\cal F}}{n_\me^{\cal F}} %
  = 1 + r_{\cal F\!/\!P}.
\end{equation}
As an example, let use derive the ratio of other critical points in dimensions up to 6D.
For $d=4$,
\begin{align*}
  \frac{n_\cp^{\rm f}}{n_\cp^{\rm p}} &= \frac{n_\cp^{(1)}}{n_\cp^{(0)}} = 1 + r_{\cal F\!/\!P} \approx 4.17,\\
  \frac{n_\cp^{(2)}}{n_\cp^{(1)}} &= \frac{n_\me^{(1)} + n_\me^{(2)}}{n_\me^{(0)} + n_\me^{(1)}} = \frac{n_\me^{(0)} + n_\me^{(1)}}{n_\me^{(0)} + n_\me^{(1)}} = 1.
\end{align*}
For $d=5$,
\begin{align*}
  \frac{n_\cp^{\rm f}}{n_\cp^{\rm p}} &= \frac{n_\cp^{(1)}}{n_\cp^{(0)}} = 1 + r_{\cal F\!/\!P} \approx 5.36,\\
  \frac{n_\cp^{(2)}}{n_\cp^{(1)}} &= \frac{n_\cp^{(2)}}{n_\cp^{(3)}} = \frac{n_\me^{(1)} + n_\me^{(2)}}{n_\me^{(0)} + n_\me^{(1)}} = \frac{r_{\cal F\!/\!P} + r_{\mathcal{W}_1\!/\!\mathcal{P}}}{1+r_{\cal F\!/\!P}} \approx 2.07.
\end{align*}
For $d=6$,
\begin{align*}
  \frac{n_\cp^{\rm f}}{n_\cp^{\rm p}} &= \frac{n_\cp^{(1)}}{n_\cp^{(0)}} = 1 + r_{\cal F\!/\!P} \approx 6.67,\\
  \frac{n_\cp^{(2)}}{n_\cp^{(1)}} &= \frac{n_\cp^{(3)}}{n_\cp^{(4)}} = \frac{n_\me^{(1)} + n_\me^{(2)}}{n_\me^{(0)} + n_\me^{(1)}} = \frac{r_{\cal F\!/\!P} + r_{\mathcal{W}_1\!/\!\mathcal{P}}}{1+r_{\cal F\!/\!P}}  \approx 2.64,\\
  \frac{n_\cp^{(3)}}{n_\cp^{(2)}} & = 1.
\end{align*}
Note that in the previous expressions we have used the following substitutions $n_\cp^{\rm p} \equiv n_\cp^{(d)}$  and $n_\cp^{\rm f} \equiv n_\cp^{(d-1)}$
and the fact that $n_\cp^{(d)}=n_\cp^{(0)}$ and $n_\cp^{(d-1)}= n_\cp^{(1)}$
Given that \cite{codis2018} provides an asymptotic limit for the global connectivity $\kappa\equiv 2{n_\cp^{\rm f}}/{n_\cp^{\rm p}}$,
we can re-express it in terms of the ratio of critical events as 
\begin{equation}
\frac{n_\cp^{\rm f}}{n_\cp^{\rm p}} = \frac{n_\cp^{(1)}}{n_\cp^{(0)}} = 1 + r_{\cal F\!/\!P} \sim  d+\frac{1}{2}\left((2d-4)/7\right)^{7/4}\,,
\end{equation}
which, in the large $d$ limit, asymptotes approximately to
\begin{equation}
 r_{\cal F\!/\!P} \stackrel{d \to \infty}{\sim} \frac{1}{2}\left(\frac 2 7\right)^{7/4}\, d^{7/4}\approx \frac 1 {17}d^{7/4}\,.
\end{equation}
\begin{figure}
  \includegraphics[width=0.98\columnwidth]{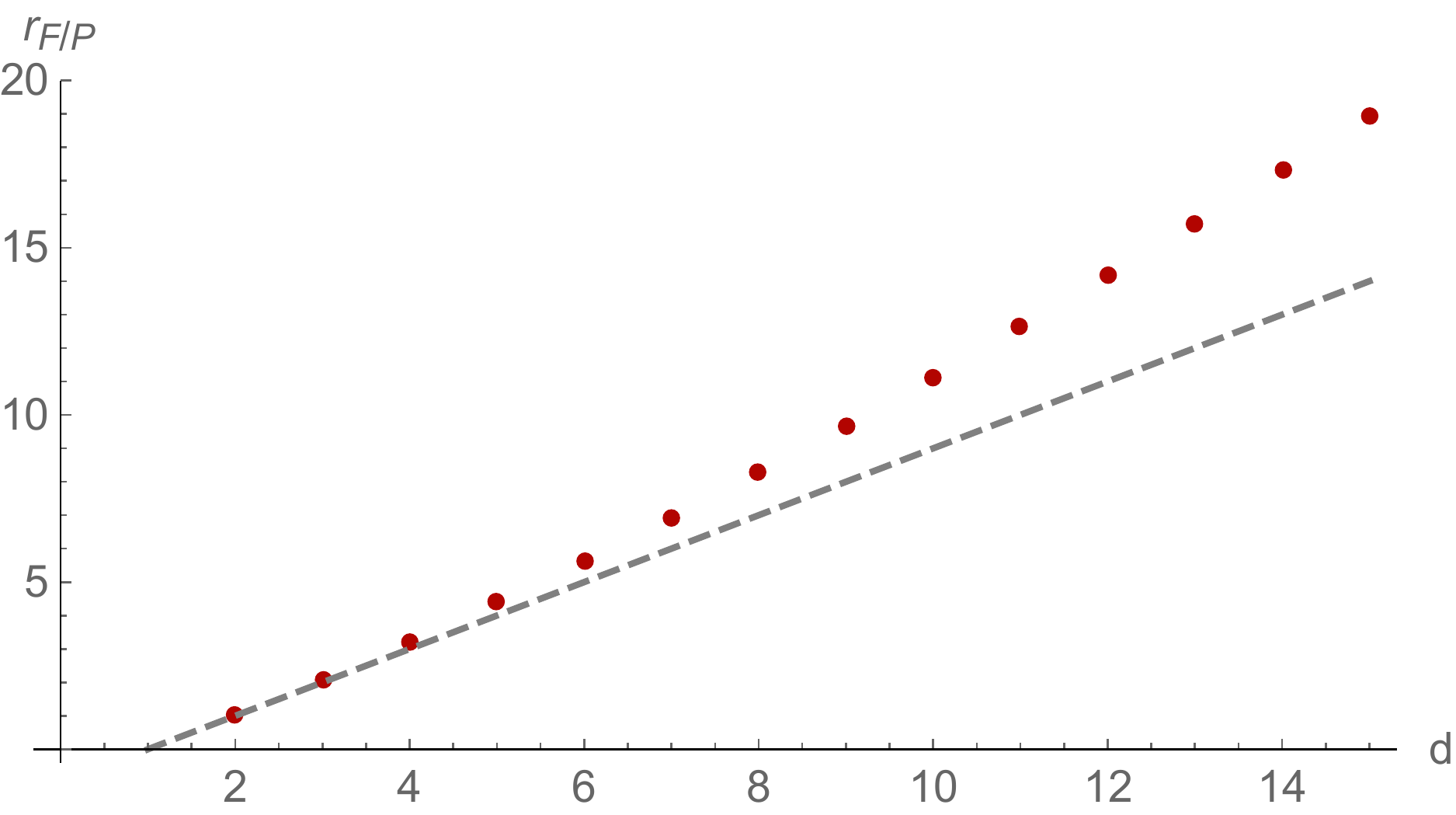}
  \caption{ The ratio of peak to filament merger as a function of $d$.
  For reference, the first diagonal is shown as a dashed grey line as well.
  The ratio is approximately fitted as $d\!-\!1\!+\!\left((2d-4)/7\right)^{7/4}\!/2$ and shown as red  dots.
  The dashed line is the identity.
  }
  \label{fig:ratio-d}
\end{figure}

\subsection{Testing the link between critical points and events counts}
\label{sec:nce_versus_nsp}
From \cref{eq:crit-pt-number-dens-derivative2}
and because for a Gaussian filter, we have
\begin{equation*}
\frac{\dd \sigma_{i}^{2}}{\dd R^{2}}=-\sigma_{i+1}^{2},
\end{equation*}
one can easily derive
\begin{equation}
  \frac{\mathrm{d} n_\cp^{(i)}}{\mathrm{d} R} = -n_\cp^{(i)} \times d \frac{R}{R_{\star}^{2}}\frac{1-\tilde \gamma^{2}}{\tilde \gamma^{2}}.
  \label{eq:crit-pt-number-dens-derivative}
\end{equation}
  which in $d=3$ for peaks reads
\begin{align}
 - \frac{\mathrm{d} n_\cp^{\rm p}}{\mathrm{d} R} &= 3 n_\cp^{\rm p} \frac{R}{R_{\star}^{2}}\frac{1-\tilde \gamma^{2}}{\tilde \gamma^{2}}\\
 &=
 \frac{3R}{R_{\star}^{3}\tilde R^{2}}
 (1-\tilde \gamma^{2})
 \frac{29\sqrt {15} -18 \sqrt{10}}{1800\pi^{2}},
\end{align}
which is equal to the net merger rate of peak type from \cref{eq:eventcount} with $C_\mathrm{even}$
and $C_{3,\mathrm{odd}}$ given by \cref{eq:C2even3D,eq:alpha-term}, respectively.
\section{Rate of change with smoothing}
\label{sec:net_mergers}

Let us show how the 3+1D number density of critical events is related to the rate of change of the 3D density of critical points with $R$,
$\dv*{n_\cp}{R}$. The 3D density of critical points is defined as
\begin{equation}
  n_\cp (R) \equiv
  \bigg\langle \sum_{\cp}\dirac(\vvec{r} - \vvec{r}_\cp) \bigg\rangle,
\label{eq:3Ddef1}
\end{equation}
where the sum runs over the solutions $\vvec{r}_\cp$ of the equation $\grad\delta=0$, $H$ is the Hessian determinant and brackets designate spatial
averages on a 3D slice, $\langle\,\dots \rangle \equiv \frac{1}{V} \int_V \dots \dd[3]{\vvec{r}}  $.
Critical points of a given kind (peak, saddle or minimum) can be defined by further imposing the signs of the eigenvalues of the Hessian.

Outside the critical events, the trajectory of each critical point in the extended 3+1D
space obtained by stacking spatial slices at different smoothing scales
can be parametrised by $R$ to yield the 3+1D coordinates $(\vvec{r}_\cp(R),R)$.
The equation for $\vvec{r}_\cp(R)$ is obtained by requiring that the field gradient
$\grad\delta(\vvec{r}=\vvec{r}_\cp(R),R)$ be constant, which gives
\begin{equation}
   \frac{\dd\vvec{r}_{\cp}}{\dd R} =
  - \pd_R \grad \delta \cdot \mathbf{H}^{-1} \bigg|_{\vvec{r}=\vvec{r}_\cp} .
  \label{eq:trackeq}
\end{equation}

If one considers a single critical point, its  contribution to the integral count
is seemingly preserved along the track at one,  $\int \dd[3]{\vvec{r}} \dirac(\vvec{r} - \vvec{r}_\cp) = 1$. Thus, if this was valid for
all trajectories at every $R$,
we would obtain a puzzling and incorrect conclusion that the number of critical points is conserved with varying smoothing.
However, only trajectories that do not encounter a critical event
can be continuously parametrised with $R$ everywhere.
At the critical event a 3+1D geometrical line tracking the critical point turns around and continues back in reverse direction
in $R$ with a change of sign in $H$, and the $R$ parametrisation breaks.
Equivalently, the lines of two critical points of different types with opposite signs of $H$ (now both taken in the same $R$ direction) meet and terminate. It can be shown that the merging of two branches is smooth to first order, but when parametrised in $R$, $\dd \vvec{r}_\cp/\dd R$ diverges
at the critical event and has opposite signs on the two branches (see \cref{sec:nucl}).
This clearly demonstrates why it is the critical events that are responsible for critical point number changes with smoothing.

To resolve this difficulty, %
we shall consider counting only half of the critical points, \eg the ones with positive $H$ (\ie minima and filamentary saddle points in 3D).
The other half, with negative $H$, has the same average number density due to the null Euler characteristic of the space, so that the total density is twice that of critical points with $H>0$.
Since the two sides of each merging pair of tracks have opposite $H$ signs, this leaves us with only one of the two branches terminating at any critical event. So we have tracks that
go forever, and tracks that terminate at critical events, but along all of them $R$ is a suitable parameter, since there is no backwinding.

Thus, we can compute the change with $R$ of the density of critical points as
\begin{equation}
  \frac{\mathrm{d} n_{\cp}}{\mathrm{d} R} = 2\frac{\mathrm{d}}{\mathrm{d} R}
  \bigg\langle \sum_\cp \dirac^{(3)}(\vvec{r}-\vvec{r}_{\cp})
  \heaviside(H(\vvec{r}_\cp,R))\bigg\rangle\,,
\label{eq:rate}
\end{equation}
and differentiating under the averaging operation we find
\begin{equation}
  \frac{\mathrm{d} n_{\cp}}{\mathrm{d} R} =
  2\bigg\langle \sum_\cp \dirac^{(3)}(\vvec{r}-\vvec{r}_{\cp})
  \frac{\mathrm{d}}{\mathrm{d} R}
  \heaviside\left(H(\vvec{r}_\cp,R)\right)\bigg\rangle\,.
\end{equation}
Note that the contribution from Dirac's delta function vanishes, since $\dd \dirac^{(3)}(\vvec{r}-\vvec{r}_{\cp})/\dd R = - \dd\vvec{r}_\cp/\dd R \cdot\grad_{\vvec{r}} \dirac^{(3)}(\vvec{r}-\vvec{r}_{\cp})$,
and there is no $\vvec{r}$ dependence left for the gradient to act on after integrating by parts.

Next we express the full derivative $\mathrm{d} H(\vvec{r}_\cp(R),R) / \mathrm{d} R$ via field variables using \cref{eq:trackeq}, and
use the representation $\sum_\cp\dirac^{(3)}(\vvec{r}-\vvec{r}_{\cp}) = |H| \dirac(\grad \delta)$ on a fixed $R$ slice, to obtain
\begin{equation}
\frac{\mathrm{d} n_{\cp}}{\mathrm{d} R} = 2
\left\langle |H| (\pd_R H \!-\! \pd_R \grad \delta \cdot \mathrm{\mathbf{H}}^{-1} \!\cdot \grad H) \dirac^{(3)}(\grad \delta)\dirac(H) \right\rangle,\notag
\end{equation}
having replaced volume averaging by ensemble averaging over the field distribution.
Here the expression is understood as the $H \to 0^+$ limit, \ie approaching the critical events along the positive $H$ tracks.
This allows us to replace the absolute value $|H|$ by $H$ itself.\footnote{Using tracks with negative $H$ would lead to the same result due to
the minus sign after differentiating $\heaviside(-H)$ and confirming that
$\lim_{H \to 0^-} (-|H|\mathrm{\mathbf{H}}^{-1}) = \lim_{H \to 0^+} |H|\mathrm{\mathbf{H}}^{-1} = \lim_{H \to 0} H \mathrm{\mathbf{H}}^{-1}$.}
In the term $H(\pd_R H - \pd_R \grad \delta \cdot \mathrm{\mathbf{H}}^{-1}\cdot \grad H)$ we recognise
the 3+1D Jacobian of \cref{eq:jacobian}, and finally obtain
\begin{equation}
\label{eq:dndR}
\frac{\mathrm{d} n_{\cp}}{\mathrm{d} R} = 2 \left\langle J\; \dirac^{(3)}(\grad \delta)\dirac(H) \right\rangle
~.
\end{equation}
In this expression, the factor of two reflects the fact that each critical event affects two critical points; the appearance of $J$, rather than its absolute value $|J|$, the fact that different critical events change the number of critical points according to the sign of $J$. Critical points are created at a critical event if $J$ is positive, and destroyed if $J$ is negative. Averaging over all $J$'s in \cref{eq:dndR} counts the balance of sources and sinks. %

It is interesting to notice the analogy of \cref{eq:dndR} with the Press-Schechter expression for the crossing rate of random walks through a threshold. Here the threshold is $H = 0$ rather than $\delta=\deltac$, and the random walks follow the critical point lines, but the crossing rate is still the total derivative of the probability of being above threshold, as in \cref{eq:rate}, and it equals the expectation value of the derivative $\dd H/\dd R$ at $H=0$ over all possible trajectories, that is $n_\ce^{\rm 3D}$. The upcrossing probability (a better approximation to the first crossing rate, where up is meant towards smaller scales) is on the other hand the expectation value over trajectories with negative derivative only, and is therefore analogous to $n_{\ce,-}^{\rm 3D}$.

\section{Duality  in  events ranking}
\label{sec:dual-interpretation}

\begin{figure}
  \centering
  \includegraphics[width=\columnwidth]{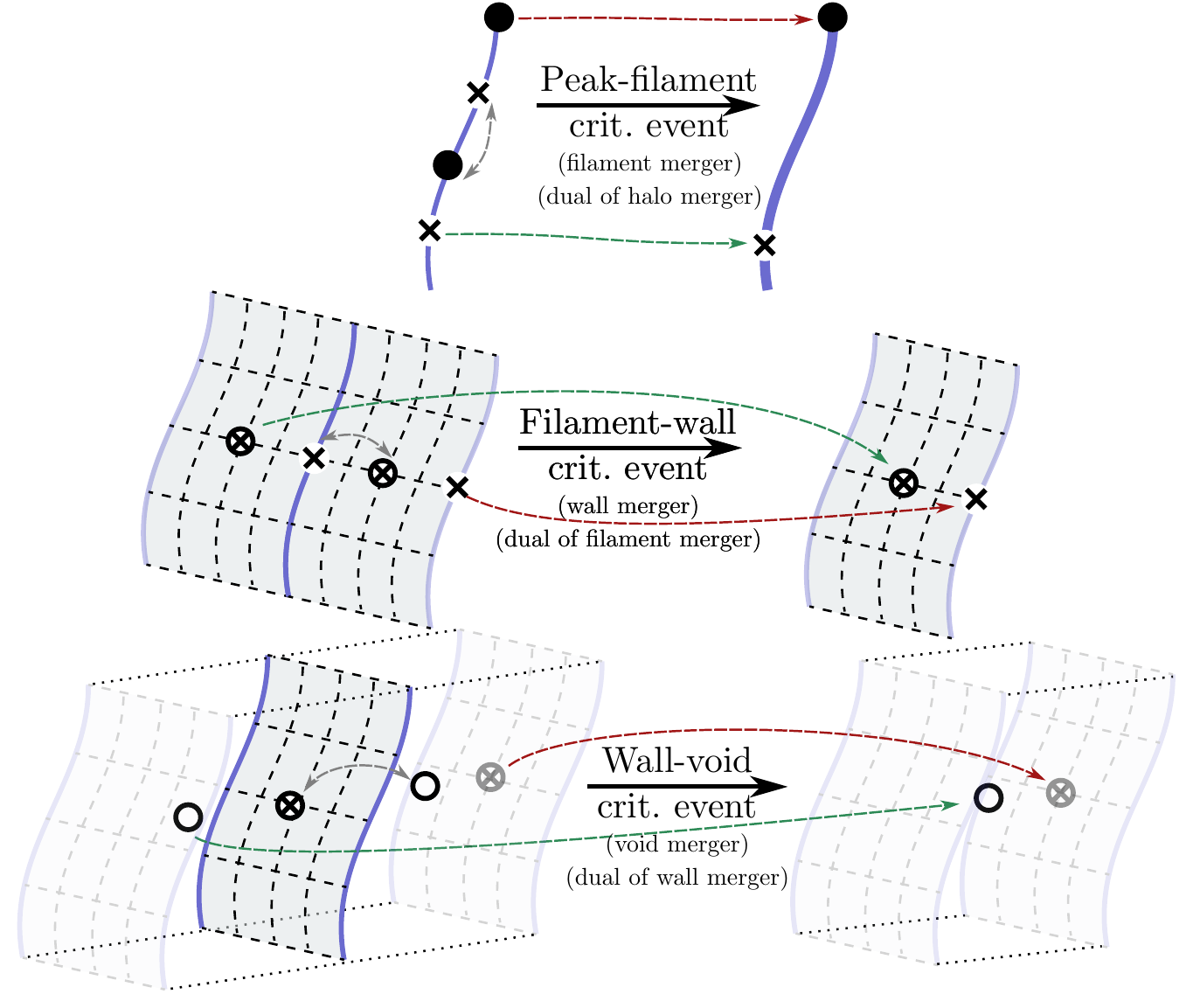}
  \caption{Same as \cref{fig:scheme_critical_events_3D} but interpreted from the point of view of the least massive structure.
  $\newmoon$ symbols are peaks, $\times$ symbols are filament-type saddle points (filament centres), $\otimes$ symbols are wall-type saddle points (wall centres) and $\fullmoon$ symbols are minima (void centres).
  Each critical event can be interpreted as the destruction of a pair of critical points (grey arrows) with a surviving structure.
  In the \cref{fig:scheme_critical_events_3D} description, the merger is associated to the densest surviving structure (peak, filament and wall mergers, red arrows) while in a dual description, the merger is associated to the least dense surviving structure (filament, wall and void mergers, green arrows).
  }
  \label{fig:scheme_critical_events_3D_dual}
\end{figure}

In the paper and unless stated otherwise, the physical interpretation of critical events was done from the perspective of the densest structure.
From this point of view, ${\cal P, F, W}$ critical events are interpreted as peak (proto-halo) mergers, filament and wall mergers respectively.
It is however also possible to interpret critical events from the perspective of the least dense structure, in which case ${\cal P, F, W}$ critical events are interpreted as 
filament, wall and void mergers respectively.
In order to illustrate this, let us focus on the central panel of \cref{fig:scheme_critical_events_3D_dual}, which illustrates a ${\cal F}$ critical event.
Before the critical event, the topology of the field is described, from left to right, by a wall-type saddle point ($\mathrm{W}_1$), a filament-type saddle point ($\mathrm{F}_1$), a wall-type saddle point ($\mathrm{W}_2$) and a filament-type saddle point ($\mathrm{F}_2$).
The critical event records the merger of $\mathrm{F}_1$ with $\mathrm{W}_2$.
Now, in order to interpret the critical event in astrophysical terms, one is left with a choice of associating the merger to the surviving wall ($\mathrm{W}_1$) or the surviving filament ($\mathrm{F}_2$).
In the former case, the merger is interpreted as a wall merger while in the latter, it is interpreted as a filament merger.
Note that if one interprets the critical event from the perspective of the disappearing structures instead, \eg to compute disappearing rates (as is done in \cref{sec:Mzmerger}), one faces the same dual interpretation as the critical event records the merging of two critical events of different kinds (here, $\mathrm{F}_1$ and $\mathrm{W}_2$).

Wall critical events ($\cal W$) share a similar dual interpretation. From the point of view of the densest structures -- the disappearing wall or the surviving one -- the critical event is interpreted as a wall merger, where the void between two walls is crushed.
This is illustrated by the red arrow, bottom panel of \cref{fig:scheme_critical_events_3D_dual}.
Conversely, the critical event can be interpreted from the point of view of the least dense structure, \ie the two voids.
In this interpretation, the critical event records a void merger where the surviving void (green arrow) is the result of the central wall being `swallowed' into the disappearing void (grey arrow).

It is worth noting that this duality  follows from  self-consistency relations between  critical events with connectivity.
Indeed, after a critical event the densest surviving structure (\eg $\mathrm{F}_2$, or a peak after a ${\cal P}$ critical event) becomes connected to the least dense surviving structure on the other side of the critical event (\eg $\mathrm{W}_1$ or a filament after a ${\cal P}$ critical event), each of these two structures being equally valid candidates as the `astrophysical outcome' of the merger.

Finally, the dual interpretation also reflects the fact that, apart from extrema, all critical points have two channels of destruction.
They can merge with a critical point of either the next or the previous kind. For example, a filament can be destroyed in a $\cal P$ critical event (where a filament between two peaks disappears) or in a $\cal F$ critical event (where a filament between two walls disappears).
This can be mathematically expressed by relating the rate of change of the number density of critical points of a given kind to the number density of critical events (see \cref{sec:self}).

\section{Local analysis of nucleation}
\label{sec:nucl}
Let us consider the problem of merging or  nucleation of
a pair of critical points near the critical event $\grad \delta = 0$,
$H=0$
as one changes the smoothing radius $R$ by $\Delta R$ (either positive in
case of a nucleation or negative in case of a merger). Smoothing is assumed to be Gaussian. Calculations are done in 2D but
are easily generalised to higher dimensions. The 1D case is
a special case, with separate conclusions as discussed briefly at the end of this section.

\subsection{Probing local vicinity of critical events}
The idea is to start with a particular, but sufficiently general
configuration of the field at smoothing $R$ and see how it changes with
smoothing. We take the field at smoothing $R$ to be described by the form
\begin{equation}
\delta_R \!=\! \delta_0 \!+\! \frac{\delta_{yy}}{2} y^2\! \!+\! \frac{\delta_{xxx}}{6} x^3\! \!+\! \frac{\delta_{yyx}}{2} y^2 x \!+\! \frac{\delta_{xxy}}{2} x^2 y \!+\! \frac{\delta_{yyy}}{6} y^3,
\label{eq:field}
\end{equation}
where $\delta_R$ is the field at smoothing $R$, position $x,y$ and $\delta_0$ is the field at smoothing $R$ at the origin.
Here we have used the short-hand notation $\delta_{x} = \partial{\delta}/\partial x$ (and similarly for higher derivatives).
This can be viewed as
one specific realisation of the random field, or
as the terms of Taylor expansion up to cubic order near the critical point
$x=0, y=0$, where,
with our choice of coordinates
\begin{equation}
\delta_x= \delta_y = 0, \quad \delta_{xx} = 0\,.
\end{equation}

We want to find out under which conditions
the shift by $\Delta R$ out of a critical event
will create two critical points. This can be done by solving the problem
perturbatively in $\Delta R$, to the lowest order in $\Delta R$.
For this purpose, the terms used in the expression (\ref{eq:field}) are sufficient, higher order terms do not modify the result.
Given our choice of filtering, the evolution of the field with $R$  is given
by the diffusion \cref{eq:diffusion}.
Hence, to first order in $\Delta R$,
\begin{equation}
\delta_{R+\Delta R}\sim \delta_R + R \Delta R  \laplacian \delta_R  ~,
\label{eq:fr+dr}
\end{equation}
with the Laplacian, 
\begin{equation}
\laplacian \delta_R = \delta_{yy} + ( \delta_{xxx} + \delta_{yyx}) x + ( \delta_{yyy} + \delta_{xxy} ) y\,,
\end{equation}
which is to be substituted in \cref{eq:fr+dr} for the final form of the field
configuration at the shifted smoothing.
The shifted field will have extrema where the gradient is zero
\begin{align}
\frac{\partial \delta_{R+\Delta R}}{\partial x} =&
\frac{1}{2} \delta_{xxx} x^2 + \delta_{xxy} x y + \frac{1}{2} \delta_{yyx} y^2 + \nonumber \\
& R \Delta R ( \delta_{xxx} + \delta_{yyx} ) = 0, \label{eq:dx}\\
\frac{\partial \delta_{R+\Delta R}}{\partial y} =& \delta_{yy} y +
\frac{1}{2} \delta_{xxy} x^2 + \delta_{yyx} x y + \frac{1}{2} \delta_{yyy} y^2 + \nonumber \\
& R \Delta R ( \delta_{xxy} + \delta_{yyy} ) = 0.\label{eq:dy}
\end{align}
Since we are looking for appearance of critical points near the critical event, \ie at
position $x,y$ close to  $0$ as $ \Delta R \to 0$, we should solve the system of equations (\ref{eq:dx},~\ref{eq:dy})
perturbatively in $\Delta R$.

We start with the $y$-derivative, \cref{eq:dy}. At leading order  in $y$, the terms $y^2$ and $x y$ can be dropped, leaving us with the following relation
\begin{equation}
\delta_{yy} y + \frac{1}{2} \delta_{xxy} x^2 \sim -R \Delta R ( \delta_{xxy} + \delta_{yyy} ).
\label{eq:dyleading}
\end{equation}
There are two viable possibilities.  Either $y \sim \Delta R$, or $y \sim x^2$.
Let us now check \cref{eq:dx} for these possibilities.
If $y \sim \Delta R$ we see that the
$x y$ and $y^2$ terms are subdominant w.r.t. the linear $R \Delta R$  term,
but if $ y \sim x^2 $ these same terms are subdominant to $x^2$.
Thus, the $x y$ and $y^2$ terms can be always neglected, and we find that
\begin{equation}
\frac{1}{2} \delta_{xxx} x^2 \sim -R \Delta R ( \delta_{xxx} + \delta_{yyx} ),
\label{eq:dxleading}
\end{equation}
\ie we always have a parabolic $x \sim \Delta R^{1/2}$ and $y \sim \Delta R$ behaviour (see \cref{fig:nucleation-destruction}).
Note that it is not possible to have $y$ subdominant to $x^2$ or $\Delta R$,
since in this case  \cref{eq:dyleading,eq:dxleading} 
will be in general inconsistent.
The solutions to \cref{eq:dyleading,eq:dxleading} are two
points $(x_+,y)$ and $(x_-,y) $, where
\begin{eqnarray}
\!\!x_{\pm}\! &=& \!\!\pm \sqrt{ \frac{-2 R \Delta R {(\delta_{xxx} + \delta_{yyx})}}{\delta_{xxx}}}\,, \\
\!\!y\!  &=&\!\! -R \Delta R\frac{ \delta_{yyx} \delta_{xxy} + \delta_{yyy}\delta_{xxx} }{\delta_{xxx} \delta_{yy}}\,.
\end{eqnarray}
For dimensions higher than two, this standard linear dependence appears for all regular directions in which the second derivative of the field
at the critical point does not vanish.
The only condition for the existence of a pair of extrema near the critical event 
now arises from requiring that
the square root argument in the expression for $x_\pm$  be positive
\begin{equation}
-2 R \Delta R \frac{\delta_{xxx} + \delta_{yyx}}{\delta_{xxx}} > 0\,.
\end{equation}
The type of critical points created or merged at a critical event is determined by the signs of the eigenvalues of
the Hessian at the critical point locations. The Hessian of the smoothed field is given by
\begin{align}
\label{eq:dxx}
\frac{\partial^2 \delta_{R+\Delta R}}{\partial x^2} &= \delta_{xxx} x + \delta_{xxy} y,  \\
\label{eq:dyy}
\frac{\partial^2 \delta_{R+\Delta R}}{\partial y^2} &= \delta_{yy} +
\delta_{yyx} x + \delta_{yyy} y, \\
\label{eq:dxy}
\frac{\partial^2 \delta_{R+\Delta R}}{\partial x \partial y} &=
\delta_{xxy} x + \delta_{yyx} y.
\end{align}
Conversely, it is easy to show that to  leading order in $\Delta R$ the eigenvalues
of the Hessian at the critical points are
\begin{align}
\lambda_1 &= \delta_{xxx} x_{\pm}
= \pm \sqrt{ -2 R \Delta R \frac{(\delta_{xxx} + \delta_{yyx})} {\delta_{xxx}}}, \\
\lambda_2 &= \delta_{yy}.
\end{align}
This explicitly demonstrates that two merging or created critical points in a pair
differ in nature with the sign of one eigenvalue. If $\delta_{yy} < 0$ (as well as the rest of  the
eigenvalues in higher dimensional case),  the process describes interaction of
one maximum and one filamentary saddle. If $\delta_{yy} > 0$ the process describes
interaction of a minimum and a saddle. In multi-dimensional case
it is the set of signs of all non-zero eigenvalues that determine the type
of interaction. In 3D, we have three cases: when  both eigenvalues are negative
$\delta_{yy} < 0, \delta_{zz} <0 $, it describes maxima and filamentary saddle coalescence,
one negative and one positive, $\delta_{yy} < 0, \delta_{zz} > 0 $, corresponds to
filamentary and wall-like saddles interacting, and both positive $\delta_{yy} > 0, \delta_{zz} > 0 $
correspond to a wall and a void coalescence.  Note that in this discussion we do not consider eigenvalues
as sorted, so the first (degenerate) direction is arbitrary.

\subsection{Discussion of the existence condition}
The merging of a critical point pair corresponds to the situation when two critical points
disappeared as smoothing
reached $R$. Thus, two critical points existed for $\Delta R < 0$.  This happens when
\begin{equation}
\frac{\delta_{xxx} + \delta_{yyx}}{\delta_{xxx}} > 0.
\label{eq:merg}
\end{equation}
Conversely, if the solution exists for $\Delta R > 0$ then
two critical points appear out of a critical event as smoothing increases from $R$ value.
We see that this happens when
\begin{equation}
\frac{\delta_{xxx} + \delta_{yyx}}{\delta_{xxx}} < 0.
\label{eq:nuc_cond}
\end{equation}
This analysis thus proves that the nucleation process is in general possible --
even for Gaussian smoothing -- if the number of dimensions exceeds one.
 
The condition on merging or nucleation that we have derived by this local analysis
is equivalent to the condition on the sign of the Jacobian defined in
\cref{eq:jacobian} presented in the main text ($J > 0$ for nucleation and $J < 0$ for
merging events), since in our local coordinate
representation the sign dependent part of this
Jacobian is exactly $ J \propto - \delta_{xxx} \sum_i \delta_{iix}$.
Thus, we conclude that
the regions in the space of third derivatives
with negative Jacobian describe the merging (disappearance) of peak/saddle pairs,
while regions with positive Jacobian describe the creation of peak/saddle pairs.

Note finally that in 1D we do not have $y$ or higher directions, so all mixed derivatives
vanish. A solution for finding extrema
pair requires simply $x_{\pm}^2 = - 2 R \Delta R > 0$.
This solution exists therefore only for negative
$\Delta R$, so in 1D extrema pairs can only merge and never be created at a
critical event if the field is smoothed with Gaussian filters. 

\begin{figure}
  \begin{flushright}
    \includegraphics[width=0.98\columnwidth]{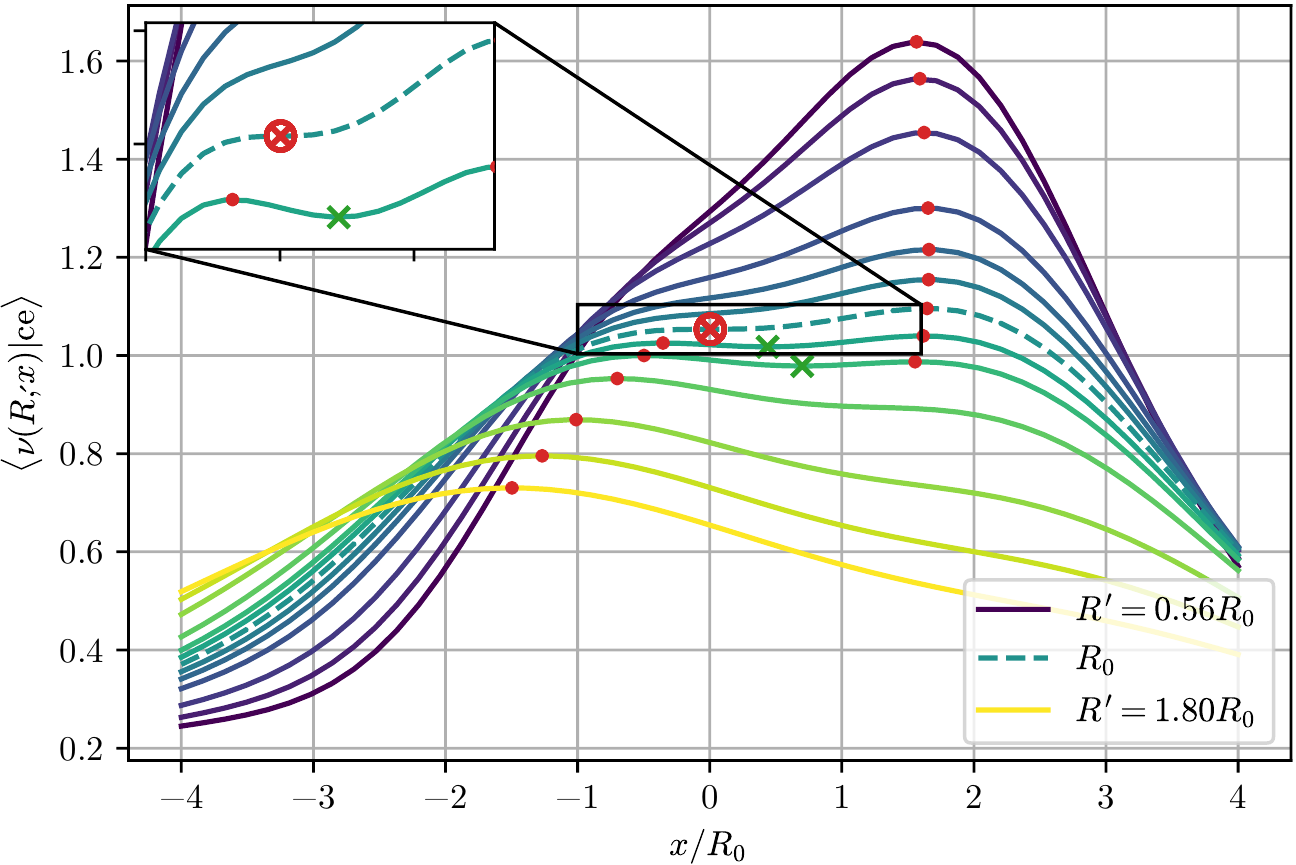}
    \includegraphics[width=\columnwidth]{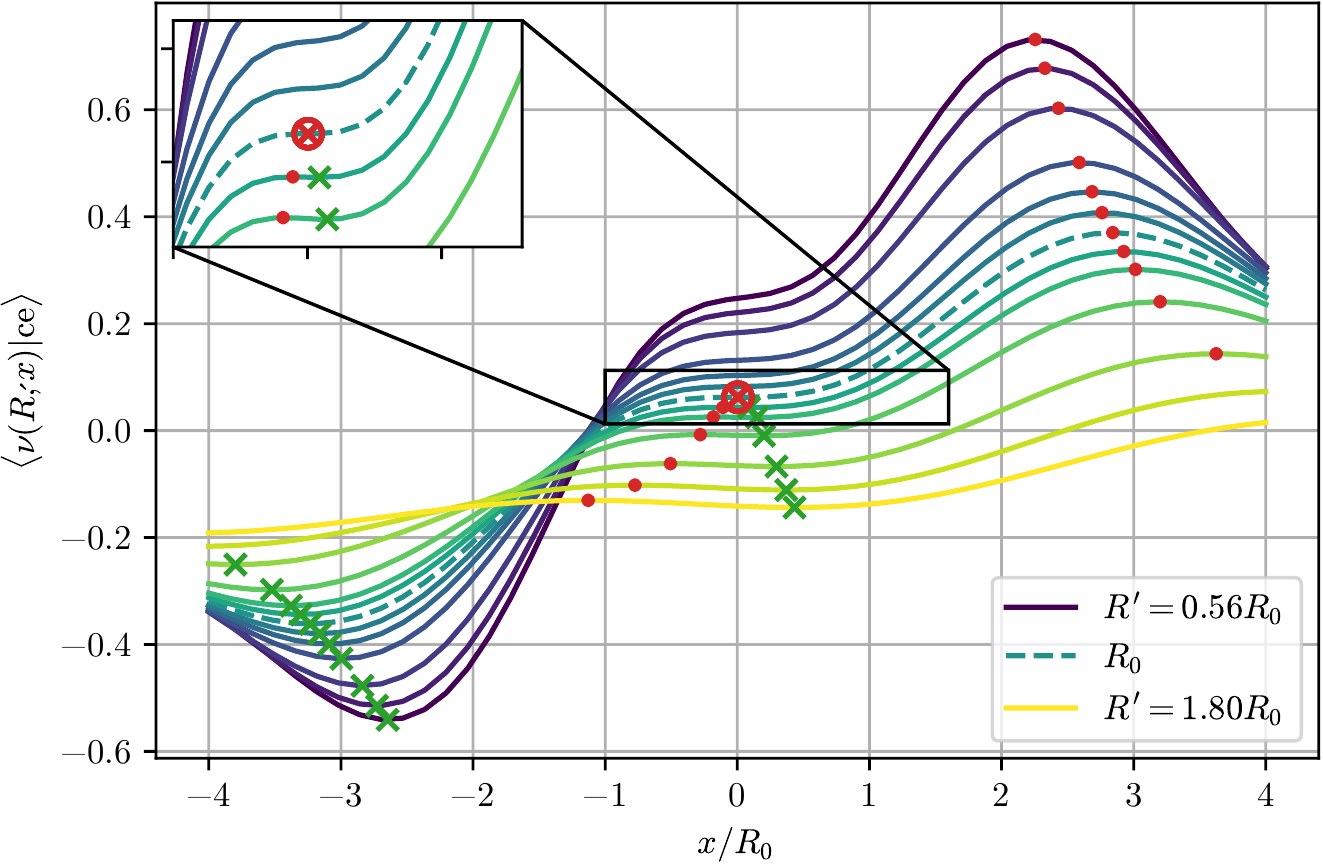}
  \end{flushright}
  \caption{
    Multiple 1D slices of the conditional mean density field in 3+1D at different smoothing scales (from $R'=0.56R_0$ to $R'=1.80R_0$, equally spaced) around a nucleation critical event (red $\bigotimes$ symbol), defined at scale $R_0$, with peaks (red $\newmoon$ symbols) and filaments (green $\times$ symbols) of the 1D slice.
    The nucleation critical event creates a pair of peak-filament that is either shortly destroyed (top panel) or long-lived (bottom panel).
    The fate of the pair depends on the particular values taken by the field and its derivatives at the critical event.
  }
  \label{fig:nucleation-slices}
\end{figure}

Two example of nucleation are presented on \cref{fig:nucleation-slices}, that shows successive slices of the density field around a nucleation critical event at different smoothing scales.
For both plots, the value of the density, its Hessian and third derivative are drawn from a Gaussian PDF until a $\cal P$ nucleation critical event is found in direction $x$. These values are then used to constrain the density field at finite distance and different smoothing scale.
The slice direction is oriented parallel to the critical event, so that peaks and minima in each 1D slice coincide at first-order with peaks and filaments of the 3D density field.\footnote{As critical points mostly slide along ridges of the skeleton.}
The figure illustrates that pairs of critical points emerging from a nucleation critical event are either long- or short-lived, and an investigation using multiple constrained field showed that the latter is the most common type.
Interestingly, it seems that pairs created from a nucleation critical event are very unlikely to annihilate, at least in this somehow contrived setup.
A likely astrophysical counterpart to  peak nucleation  event may be splashback haloes \citep[\ie the temporary reappearance of a  sub halo which as only recently been accreted][and for filament nucleation, the 
temporary re-appearance of an enclosed  wall as two filaments merge etc.]{Aubert2007,More2015}. More work will however be required to astrophysically interpret them and study their properties in the initial density field.

\section{Event generation algorithm}
\label{sec:generation}

\subsection{Constrained field -- peak constraint}
\begin{figure}
  \centering
  \includegraphics[width=\columnwidth]{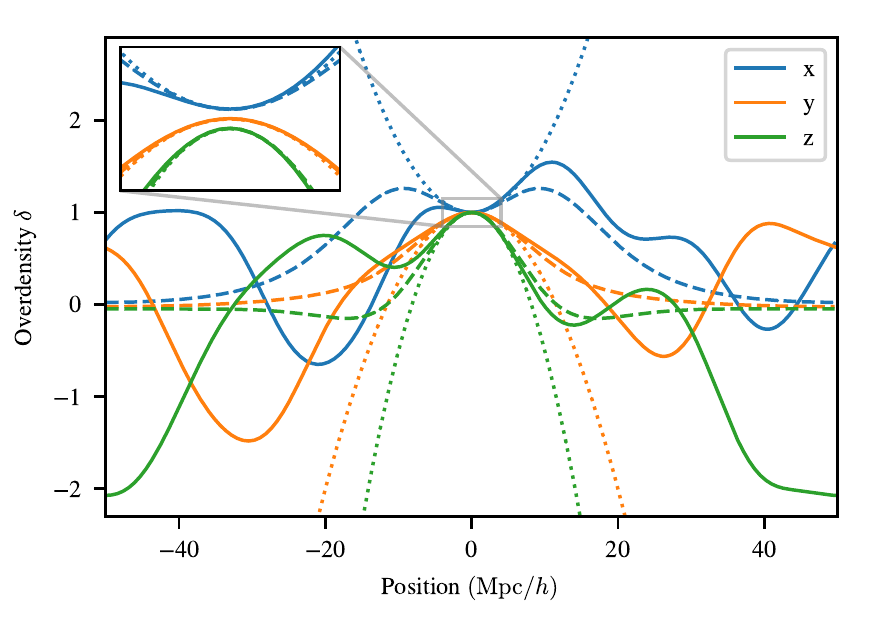}
  \caption{
    Density profile of a random field constrained to a density $\delta=1$, null gradient and a Hessian with eigenvalues $\sigma_2/2,-\sigma_2/2,-\sigma_2$ in directions $x,y,z$ at the centre of the box, assuming periodic boundary conditions.
    The expectation of the field is shown in dashed lines and the value of the field in one realisation is shown in solid lines.
    Dotted lines show the second order Taylor series of the field around the constrained point.
    The inset shows a zoom on the constrained zone. For the sake of clarity, each curve have been shifted by 0.02.
    At small distances from the constraint, the field resembles its mean and its Taylor expansion.
    }
    \label{fig:constrained_field}
\end{figure}

We have used  {\sc ConstrField} coupled with {\sc MPgrafic} from \cite{prunetetal08} to generate constrained realisations of a Gaussian random field.
We generate an unsmoothed Gaussian random field, constrained to have a filament-type saddle point of height $\delta=1$ ($\nu=1.17$) at smoothing scale $R=\SI{5}{Mpc\per\hred}$.
The eigenvalues of the Hessian are constrained to be $\{\lambda_1, \lambda_2, \lambda_3\} = \sigma_2\{-1/2,-1/2, -1\}$ with eigenvectors $\{\hat x, \hat y, \hat z\}$.
  \cref{fig:constrained_field} shows the mean density profiles as well as one realisation. As expected, the density is locally entirely set by the constraints and has a parabola-like shape.
At larger scales, the field decouples from the constraints resulting in large fluctuations around the mean value.

\subsection{Constrained field -- higher order constraints}
\label{sec:generation-higher-order}

We developed a code that is able to numerically compute the covariance matrix between any derivative of the field up to third order or any anti-derivative of the field up to second order (potential), smoothed by any filter function and at any separation.
The code relies on the numerical integration of the correlation function between any two functionals of the field.
Formally, let us define a linear functional $F$ and its Fourier representation
\begin{equation}
  \tilde{F}[\delta](\vvec{k}) = \int \dd[3]{\vvec{r}} e^{-i\vvec{k}\vdot\vvec{r}} F[\delta](\vvec{r}).
\end{equation}
Functionals that can be written as a convolution with a distribution, which includes notably derivation operators and smoothing operators, can be further simplified as $\tilde{F}[\delta](\vvec{k}) = \delta(\vvec{k})\tilde{F}(\vvec{k})$, where $\tilde{F}$ is now a function of $\vvec{k}$ only.%
Common operators take a simple form in Fourier space; for example the third derivative operator in direction $i,j,k$ reads $(-i)^3k_ik_jk_l$, the shift operator (that shifts the field by $\Delta \vvec{x}$) reads $\exp(i\vvec{k}\vdot \Delta\vvec{x})$ and the Gaussian filter has its usual form $\exp(-k^2R^2/2)$.
The covariance between two linear functionals of the field then simply reads
\begin{equation}
  \label{eq:any-correlation}
  \left\langle F[\delta] G[\delta]\right\rangle = \frac{1}{(2\pi)^3}\int\dd[3]{\vvec{k}}P_k(k)\tilde{F}(\vvec{k}) \tilde{G}^\star\!(\vvec{k}),
\end{equation}
where the star symbol denotes here the complex conjugate.
As a worked example, the covariance between the field smoothed by a Gaussian filter at scale $R_1$ at the origin and the field smoothed by a Gaussian filter at scale $R_2$, position $\vvec{r}$ is given by \cref{eq:any-correlation} with $\tilde{F}(\vvec{k}) = \exp(-k^2R_1^2/2)$ and $\tilde{G}(\vvec{k}) = \exp(-k^2R_2^2/2+i\vvec{k}\vdot\vvec{r})$.

Let us write $\vvec{X} = \{\vvec{X}_1, \vvec{X}_2\}$ where $\vvec{X}_1$ is the density field sampled at $p$ different locations and $\vvec{X}_2$ contains the $q$ values (the field and/or its (anti-)derivatives) that will later be constrained to the value $\vvec{a}$.
For example, a critical event constraint (at fixed scale) could be represented by $\vvec{X}_2 = \{x,x_{1},x_{2},x_{3},x_{11},x_{111},x_{221},x_{331}\}$ subject to the constraint $\vvec{a} = \{\nu,0,0,0,0,\alpha_1,\alpha_2,\alpha_3\}$.
The conditional mean $\bar{\vvec{\mu}}$ and covariance $\bar{\mathbf{C}}$ of the field are then obtained from the full mean $\vvec{\mu} = \langle \vvec{X} \rangle$ and covariance $\mathbf{C}=\langle\vvec{X}^\mathrm{T}\vvec{X}\rangle$, computed using \cref{eq:any-correlation}, by simple arithmetic
\begin{equation}
  \bar{\vvec{\mu}} = \vvec{\mu}_1 + \mathbf{C}_{12} \mathbf{C}_{22}^{-1}(\vvec{a}-\vvec{\mu}_2), \quad
  \bar{\mathbf{C}} = \mathbf{C}_{11} - \mathbf{C}_{12}\mathbf{C}_{22}^{-1}\mathbf{C}_{12}^\mathrm{T},
\end{equation}
where we assumed here that the covariance is decomposed as $\mathbf{C} = \begin{pmatrix}
  \mathbf{C}_{11} & \mathbf{C}_{12} \\ \mathbf{C}_{12}^\mathrm{T} & \mathbf{C}_{22}
\end{pmatrix}$, with sizes $p\times p$, $p\times q$ and $q\times q$ for $\mathbf{C}_{11}, \mathbf{C}_{12}, \mathbf{C}_{22}$ respectively and similarly for the mean.
One can then easily draw samples from the conditional multivariate distribution using $\bar{\vvec{\mu}}, \bar{\mathbf{C}}$.

\section{Pair destruction \& creation counts}
\label{sec:crit-event-numb-different-def}

Three different definitions of the number count have been discussed in the text and presented in \cref{eq:eventcount_covariant,eq:nce+-_def,eq:nme_def}.
In this section, we present the results obtained in three dimensions for a Gaussian random field using these three definitions.
For a Gaussian random field, the expectation of the even-derivatives is left unchanged but the odd part is modified.

Using the total merger density definition of \cref{eq:eventcount_covariant}, the odd part reads
\begin{align}
  \label{eq:Codd-|J|}
  C_\mathrm{odd}^\ce & = \frac{1}{5}\left({\frac{3}{2\pi}}\right)^{3/2} (1-\tilde{\gamma}^2) \times \nonumber \\
  & \times \frac{2}{\pi} \left(\frac{2}{\sqrt{21(1-\tilde{\gamma}^2)}}+\tan^{-1}\frac{\sqrt{21(1-\tilde{\gamma}^2)}}{2}\right).
\end{align}
Using the pair destruction and pair creation definition of \cref{eq:nce+-_def}, the odd part reads
\begin{align}
  \label{eq:Codd-+=}
  C_\mathrm{odd}^{\ce,-} & = \frac{1}{5}\left({\frac{3}{2\pi}}\right)^{3/2} (1-\tilde{\gamma}^2) \times \\
& \times \frac{1}{\pi} \left(\frac{2}{\sqrt{21(1-\tilde{\gamma}^2)}}+\frac{\pi}{2}+\tan^{-1}\frac{\sqrt{21(1-\tilde{\gamma}^2)}}{2}\right) ~, \nonumber \\
  C_\mathrm{odd}^{\ce,+} & = \frac{1}{5}\left({\frac{3}{2\pi}}\right)^{3/2} (1-\tilde{\gamma}^2) \times\\
& \times \frac{1}{\pi} \left(\frac{2}{\sqrt{21(1-\tilde{\gamma}^2)}}-\frac{\pi}{2}+\tan^{-1}\frac{\sqrt{21(1-\tilde{\gamma}^2)}}{2}\right) ~,  \nonumber 
\end{align}
respectively.
For the sake of completeness, let us reproduce here the result, already presented in \cref{eq:Codd}, using the net merger rate definition of \cref{eq:nme_def}
\begin{equation}
  C_\mathrm{odd}=\frac{1}{5}\left({\frac{3}{2\pi}}\right)^{3/2} (1-\tilde{\gamma}^2)\,.
\end{equation}

\begin{figure}
  \includegraphics[width=0.98\columnwidth]{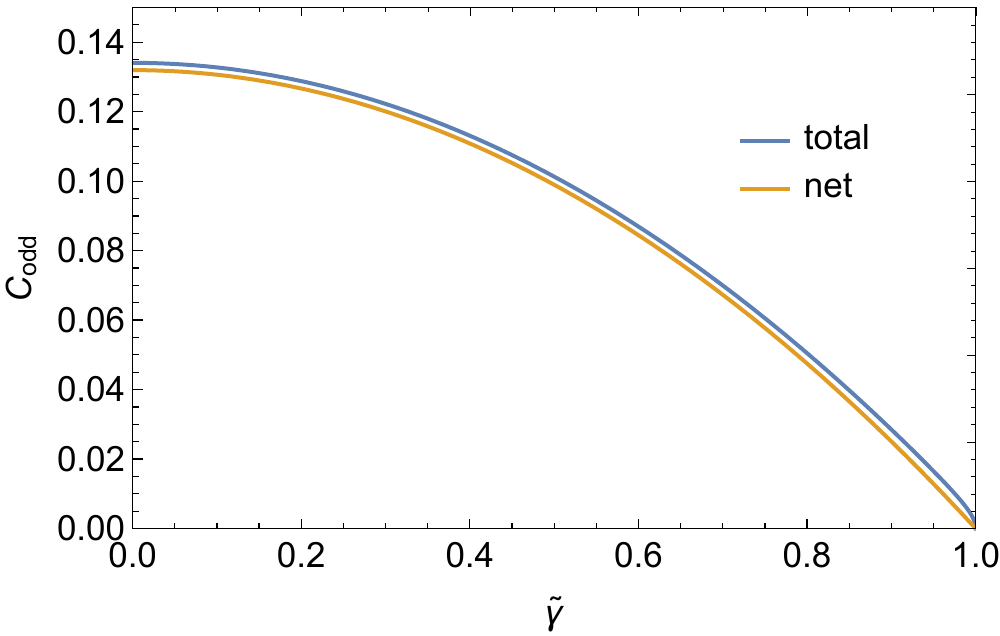}
  \caption{ $C_\mathrm{odd}$  as a function of $\tilde \gamma$ when the definition for the total or the net merger density is used. The difference between the two curves is at the percent level, at least for relatively small values of 
 of $\tilde \gamma$. At higher $\tilde \gamma$ (typically above 0.8 i.e for a spectral index above -1), both nucleations and destructions become rarer and their ratio tend towards unity.}
  \label{fig:total-vs-net}
\end{figure}

A comparison of $C_\mathrm{odd}$ between total and net merger density is given in \cref{fig:total-vs-net} and is shown to be at a few percent level only. For $\tilde \gamma$ above 0.8 (\ie for a spectral index above $-1$), there are at least 30 times fewer nucleations than destructions (this ratio is an increasing function of $\tilde \gamma$).

\section{Joint PDFs}
\label{sec:jointpdf}

Let us present here the PDF of the field and its (up to 3rd) derivative which will allow us to compute the expectations
involved in the main text.
\subsection{One-point PDFs}
Since the odd and even-order derivatives of Gaussian random fields do not correlate,
let us write the joint PDF as $ { P}_{\rm G}={P}_{0}(x,x_{kl}) {P_{1}}(x_{i},x_{ijk})$.
The  expression for ${P}_{0}(x,x_{kl})$ for the Gaussian field
was first given by \cite{BBKS}. Introducing the variables
\begin{align}
u  \equiv - \laplacian x &= -(x_{11}+x_{22}+x_{33})\,,\\
w  &\equiv\frac{1}{2}  (x_{11}-x_{33})\,, \\
v  &\equiv   \frac{1}{2}(2 x_{22}-x_{11} - x_{33})\,,
\end{align}
in place of diagonal elements of the Hessian $(x_{11},x_{22},x_{33})$
one finds that $u,v,w,x_{12} ,x_{13},x_{23}$ are
uncorrelated. Importantly, the field, $x$ is only correlated with
$u$ and
\begin{equation}
\langle x u \rangle= \gamma, \quad \langle x v \rangle =0, \quad
\langle x w \rangle=0, \quad \langle x x_{kl} \rangle=0, \ k\neq l,\nonumber
\end{equation}
where $\gamma$ is the same quantity as in \cref{eq:gammadef}.
The full expression of ${P}_{0}(x,x_{kl})$ is then
\begin{equation}
  P_{0}(x,x_{kl})  =
  \frac{5^{1/2}15^2}{(2\pi)^{7/2}({1-\gamma^2})^{1/2}}
  \exp\left[-\frac{Q_0 + Q_2}{2}\right]
 \,,
\end{equation}
with the quadratic forms $Q_0$ and $Q_2$ given by
\begin{align}
\label{eq:quadformdimadd}
  Q_0 &= x^2 + \frac{(u -\gamma x)^2}{(1-\gamma^2)}\,,\nonumber\\
  Q_2 &= 5 v^{2} + 15  (w^2 + x_{12}^{2}+x_{13}^{2}+x_{23}^{2}) \,\nonumber\\
  &=  \frac{15}2\,
  \overline x_{ab}
  \overline x_{ab}\,,
\end{align}
where the last identity is demonstrated in \cite{pogo09} and involves the detraced tensors:
\begin{align}\label{eq:detrace.trois.quatre}
\overline t_{ij} &=t_{ij} - \frac13 t_{aa} \delta_{ij}\,,\\
  \overline t_{ijk} &= t_{ijk} - \frac35 \, t_{aa(j} \delta_{kl)}\,,
 \end{align}
with an implicit summation over repeated indices and symmetrization
between parenthesised indices (for instance: %
$ t_{aa(j} \delta_{kl)} =  [t_{aaj} \delta_{kl} + t_{aak}
\delta_{lj} + t_{aal} \delta_{jk}]/3$ and so on).
\Cref{eq:quadformdimadd} depends only on a single correlation parameter: $\gamma$.
A similar procedure can be performed for the joint probability of the
first and third derivatives of the fields, ${P_{1}}(x_{i},x_{ijk})$ by
defining the following nine parameters  \citep[see also][]{hanami}
\begin{align}
u_i &\equiv \nabla_{i} u,
\quad v_i \equiv \frac{1}{2} \epsilon^{ijk} \nabla_{i }
\left( \nabla_{j} \nabla_{j}- \nabla_{k} \nabla_{k}\right)x \,,\,\,\,
{\rm with}\,\,\, j<k\,,\nonumber \\
w_i &\equiv \sqrt\frac{5}{12}\nabla_{i}
\left( \nabla_{i} \nabla_{i}  -\frac{3}{5}\laplacian \right)x\,,
\end{align}
and replacing the variables $(x_{i11},x_{i22},x_{i33})$ with $(u_i,v_i,w_j)$.
In that case, the only cross-correlations in the vector
$(x_1,x_2,x_3,u_1,v_1,w_1,u_2,v_2,w_2,u_3,v_3,w_3,x_{123})$ which
do not vanish are between the same components of the gradient and
the gradient of the Laplacian of the field:
\begin{equation}
\langle x_i u_i \rangle = {\tilde \gamma}/3,\qquad i=1,2,3,
\end{equation}
where $\tilde \gamma$ was defined in \cref{eq:gammadef}.
This allows us to write:
\begin{equation}
{P_{1}}(x_{i},x_{ijk})\! =\!
\frac{105^{7/2} 3^{3}  }{(2\pi)^{13/2}\!(1-{\tilde \gamma}^2)^{3/2}}\! \exp\left[-\frac{Q_1  +Q_3}{2}\right]
, \label{eq:defp13D}
\end{equation}
with the quadratic forms:
\begin{align}
Q_1 &= 3 \sum_i \left( \frac{(u_i -{\tilde \gamma} x_i)^2}{(1-{\tilde \gamma}^2)} +  x_i^2 \right) \,\nonumber ,\\
Q_3 &= 105 \left( x_{123}^{2}+\sum_{i=1}^{3}(v_{i}^{2}+w_{i}^{2})\right)  \,
, \nonumber\\
&= \frac{35}2\
  \overline x_{ijk}
  \overline x_{ijk} \,.\label{eq:defp1Q3D}
\end{align}
\subsection{Two-point PDFs}  \label{sec:twptPDF}
Calling $\vvec{x}=(x,x_i,x_{ij},x_{ijk})$ and  $\vvec{y}=(y,y_i,y_{ij},y_{ijk})$, the joint PDF reads
\begin{equation}
{P}_2(\vvec{x},\vvec{y})= \frac{\exp
\left[{\displaystyle-\frac{1}{2}}
\left(\begin{array}{c}
 \vvec{x}
\\
\vvec{y}
 \\
 \end{array} \right)^{\rm T}
 \vdot
  \mathbf{C}^{-1}\vdot \left(\begin{array}{c}
 \vvec{x}
\\
\vvec{y}
  \\
\end{array} \right) \right]
}
{{\rm det}|\mathbf{C}|^{1/2} \left(2\pi\right)^{\rm 15 }} \,,
\label{eq:defPDF}
\end{equation}
where $\mathbf{C}$ is the covariance matrix which depends on the separation vectors only because of homogeneity
\begin{equation}
\mathbf{C}=\left(\begin{array}{cc}
\mathbf{C}_{\mathbf{ xx}}&\mathbf{C}_{\mathbf{ xy}}\\
 \mathbf{C}_{\mathbf{xy}}^{\rm T}  &\mathbf{C}_{\mathbf{yy}}\\
\end{array}
\right)\,.
\end{equation}
Note that $ \vvec{x}^{\rm T}\vdot \mathbf{C}_{\mathbf{xx}}^{-1} \vdot  \vvec{x} $ is given by $Q_0(x)+Q_2(x)+Q_1(x)+Q_3(x)$,
where the $Q_i$ are given by \cref{eq:quadformdimadd,eq:defp1Q3D}.
The cross terms will involve correlations of all components of $\vvec{x}$ and
 $\vvec{y}$
\begin{equation}
\mathbf{C}_{\mathbf {xy}} =\langle \vvec{x} \vdot  \vvec{y}^{\rm T}  \rangle \,.
\label{eq:correl}
\end{equation}
The correlation length of the various components of $\mathbf{C}_{\mathbf {xy}} $
differ, as higher derivatives decorrelate faster.
Note that the separations are measured in units of $R$, whereas
the $Q_i$ are independent of $R$.

\section{Detection algorithms}
\label{sec:detection}

The source code of the implementation
is available upon request.
It is based on {\tt Python} and the {\tt Scipy} stack \citep{scipy_paper}.

\subsection{Critical points detection}
\label{sec:critical_points_detection}

This section presents the algorithm used to find the critical points in a $N$-dimensional field.
Let $F$, $F_i$ and $F_{ij}$ be a field evaluated on a grid, its derivative and its Hessian. For any point $\vvec{x}$ on the grid, we have the following relation
\begin{equation}
  F_j(\vvec{x}) = F_j(\vvec{x_{\rm c}}) + (x_i-x_{{\rm c},i}) F_{ij}(\vvec{x}) + \mathcal{O}(\Delta x_i^2).
\end{equation}
Critical points are found where $F'_j=0$ by solving the linear system of equation
\begin{equation}
  \label{eq:deltaxi}
  \Delta x_i F_{ij} = - F_j,
\end{equation}
where $\vvec{\Delta x} = \vvec{x}-\vvec{x}_{{\rm c}}$. The algorithm works as follows:
\begin{enumerate}
\item Solve \cref{eq:deltaxi} for each cell on the grid. We then get a set of points $(\vvec{x}_{\vvec{\rm c}}^{i}, \vvec{x}^{i})$, where the former is the cell centre and the latter the closest critical point.
\item Remove all critical points found at $|\vvec{x}_{\vvec{\rm c}}^{i}, \vvec{x}^{i}|_\infty \geq \Delta x$, where $\Delta x$ is the grid spacing.
\item For all critical point, compute the value of the Hessian by interpolating linearly from the $2N$ (4 in 2D, 6 in 3D) neighbouring cells.
\item Compute the eigenvalues of the Hessians and the type of the critical point (maximum, saddle point(s) or minimum).
\item Merge all critical points of the same kind closer than $\Delta x$. To do this, we first build a KD-Tree of the critical points and find all the pairs located at a distance $d_{ij} = |\vvec{x}^i-\vvec{x}^j|_\infty \leq \Delta x$. For each pair, we keep only the point that is the closest to its associated cell.
\end{enumerate}

\subsection{Critical event detection}
\label{sec:critical_events_detection}

The algorithm is based on the idea that each critical event has two predecessors at the previous smaller smoothing scale (two critical points). Conversely, each critical point has either a critical point successor of the \emph{same kind} at the next (larger) smoothing scale or a critical event. Therefore, a way to detect critical events is to find critical points that do not have a successor. These points will be referred to as `heads' as they are the tip of a continuous line of critical points in the smoothing scale direction. Critical events are then found between pairs of heads of kind $k$ and $k+1$ (\eg a peak and a filament).

Following this idea, the algorithm can be decomposed in two steps: compute the heads of each kind, than find pairs of heads to detect critical events.
In the rest of the section, let us call $R_0$ (resp. $R_1$) the smallest (resp. largest) scale at which the field is smoothed.
Let ${C}_{R,k} = \{\vvec{r}_i, R\}_{i=1,\dots,N}$ be the set of the $N$ critical points of kind $k$ at scale $R$.
The whole detection algorithm reads
\begin{algorithmic}[1]
\Procedure{FindCritEvents}{$C_{R,k}, \alpha$}
  \State $E\gets \{\}$ \Comment{All critical events}
  \For{$k$ \textbf{in} $1,\dots,d$} \Comment{Find heads of critical points}
    \State $H_k \gets$ \Call{BuildHeads}{$k$, $\Delta\log R$}
    \label{alg:buildheads}
  \EndFor
  \State $R\gets R_0$
  \While{$R \leq R_1$} \Comment{Find pairs of heads (crit. events)}
    \State $\Delta R \gets R \times \Delta \log R $ \Comment{}
    \State $E \gets E + $\Call{FindHeadPairs}{$H_1,\dots,H_d, R, \alpha \Delta R$}
    \label{alg:pairheads}
    \State $R\gets R + \Delta R$
  \EndWhile
  \State\Return $E$
\EndProcedure
\end{algorithmic}
  The parameter $\alpha$ controls how far heads can be in the smoothing scale direction, in units of $\log R$. A value of 1 looks for pairs of heads at the same scale, a value of 2 looks for pairs of heads at scales $R,R+\Delta R$.

The first step (\cref{alg:buildheads}) of the algorithm builds the set of heads $H_k$. It works as follows
\begin{algorithmic}[1]
\Procedure{BuildHeads}{$k$, $\Delta\log R$}\Comment{Build heads of kind $k$}
    \State $H_k \gets C_{R_1,k}$ \Comment{Initialise heads}
    \State $P_k \gets H_k$ \Comment{Initialise progenitors}
  \State $R\gets R_1$
  \While{$R\geq R_0$}
    \State $P'_k\gets \{\}$\Comment{Initialise new progenitors at $R$}
    \For{$p,c,d$ \textbf{in} \Call{SortedPairs}{$P_k, C_{R,k}, R$}}
      \If{$c \not\in P'_k$}
        \State $P'_k\gets P'_k + \{p, c\}$ \Comment{Found new progenitor}
      \EndIf
    \EndFor
    \State $P_k \gets P'_k$
    \For{$c$ \textbf{in} $C_{R,k}$} \Comment{Loop over crit. points}
      \If{$c \not\in P'_k$} \Comment{Keep only unpaired ones\dots}
        \State $H_k\gets H_k + \{c\}$ \Comment{\dots and add them to heads}
        \State $P_k\gets P_k + \{c\}$
      \EndIf
    \EndFor
    \State $R \gets R(1-\Delta\log R)$
  \EndWhile
  \State \Return $H_k$ \Comment{Heads are points with no successors at larger $R$}
\EndProcedure
\end{algorithmic}
Here, $\mathrm{SortedPairs}(X, Y, R_\mathrm{max})$ returns $(x,y,d)$, where $x,y$ are points in $X, Y$ and $d\leq R_\mathrm{max}$ is their relative distance (in $(\vec{r},R)$ space).
  The tuples are sorted by increasing distance. This can be efficiently implemented using a KD-tree with periodic boundary conditions.
$\mathrm{BuildHeads}$ builds all heads by using a watershed approach.
Starting from the largest smoothing scales, it finds and discards all critical events that are progenitors of a head at any larger scale. The remaining points have no successor (they are the progenitor of nothing) and are hence heads.

Once the heads have been computed, the second step of the algorithm pairs them (\cref{alg:pairheads})
\begin{algorithmic}[1]
\Procedure{FindHeadPairs}{$H_1, \dots, H_{d}, R, \Delta R$} \Comment{Find pairs of heads (crit. events)}
  \State $H_{R,k}\gets \{c \in H_k\ |\ R\leq c.R < R+\Delta R\}$ \Comment{Keep heads at scale $R$}
  \State $P\gets \{\}$ \Comment{Head pair list}
  \For{$k$ \textbf{in} $1,\dots,d-1$}
    \State $P\gets P + $\Call{SortedPairs}{$H_{R,k}, H_{R,k+1}, R$}
    \label{alg:pairheads-symmetrize}
    \State $P\gets P + $\Call{SortedPairs}{$H_{R,k+1}, H_{R,k}, R$}
    \label{alg:pairheads-symmetrize-end}
  \EndFor
  \State $P\gets$ \Call{SortByDistance}{$P$}
  \State $P'\gets \{\}$ \Comment{Pairs with no double counts}
  \For{$c_1, c_2, d$ \textbf{in} $P$}
    \If{$c_1 \not\in P'$ \textbf{and} $c_2 \not\in P'$}
      \State $P'\gets P' + \{c_1, c_2\}$
    \EndIf
  \EndFor
  \State $E\gets \{\}$ \Comment{Critical events}
  \For{$c_1, c_2$ \textbf{in} $P'$}
    \State $E\gets E + $ \Call{CritEventData}{$c_1, c_2$}
  \EndFor
  \State \Return $E$
\EndProcedure
\end{algorithmic}
Lines~\ref{alg:pairheads-symmetrize}-\ref{alg:pairheads-symmetrize-end} ensure that the detection method is invariant by permutation of $k\gets d-k+1$.
$\mathrm{CritEventData}(c_1, c_2)$ computes the properties (position, kind, gradient, \ldots) of the critical events given two critical points.
  $\mathrm{FindHeadPairs}$ works as follows.
It first finds all pairs of heads separated by less than a smoothing scale.
It then loops over all pairs (sorted by increasing distance) and greedily consumes heads.
Each head can only be paired once, to its closest not-yet-paired head of either the previous or next kind.
  This prevents for example $F$ critical points from being paired to a $P$ and a $W$ critical point, which would result in a double count.
Note that this procedure may leave some heads unpaired (\eg critical points at the largest smoothing scale do not merge but have no successor).
In practice the unpaired heads typically account for less than a percent (\SI{0.5}{\percent} for $\Delta R = \alpha R \Delta \log R$ with $\alpha=2$) of the total number of heads.

\end{document}